\documentclass[a4paper,11pt]{article}
\usepackage{jheppub-tpajero}
\usepackage[nottoc]{tocbibind}

\usepackage[section]{placeins}

\usepackage{ifthen} 
\newboolean{uprightparticles}
\setboolean{uprightparticles}{false} 

\usepackage{bm}
\makeatletter
\g@addto@macro\bfseries{\boldmath}
\makeatother

\usepackage{xspace} 
\usepackage{upgreek}

\def\lhcb   {\mbox{LHCb}\xspace}

\def\belletwo {\mbox{Belle~II}\xspace}

\def\cdf    {\mbox{CDF}\xspace}

\def\lhc    {\mbox{LHC}\xspace}

\def\upgradeone {\mbox{Upgrade~I}\xspace}
\def\upgradetwo {\mbox{Upgrade~II}\xspace}

\def\MagUp {\mbox{\em Mag\kern -0.05em Up}\xspace}

\ifthenelse{\boolean{uprightparticles}}%
{

 \def\Peta        {\ensuremath{\upeta}\xspace}

 \def\Pmu         {\ensuremath{\upmu}\xspace}                 
 \def\Pnu         {\ensuremath{\upnu}\xspace}                 
                  
 \def\Ppi         {\ensuremath{\uppi}\xspace}

 \def\Pphi        {\ensuremath{\upphi}\xspace}

 \def\PDelta      {\ensuremath{\Delta}\xspace}                 
 \def\PXi         {\ensuremath{\Xi}\xspace}                 
 \def\PLambda     {\ensuremath{\Lambda}\xspace}                 
 \def\PSigma      {\ensuremath{\Sigma}\xspace}                 
 \def\POmega      {\ensuremath{\Omega}\xspace}                 
 \def\PUpsilon    {\ensuremath{\Upsilon}\xspace}

 \def\PB      {\ensuremath{\mathrm{B}}\xspace}                 
                  
 \def\PD      {\ensuremath{\mathrm{D}}\xspace}

 \def\PK      {\ensuremath{\mathrm{K}}\xspace}

 \def\PP      {\ensuremath{\mathrm{P}}\xspace}

 \def\PW      {\ensuremath{\mathrm{W}}\xspace}

 \def\Pb      {\ensuremath{\mathrm{b}}\xspace}                 
 \def\Pc      {\ensuremath{\mathrm{c}}\xspace}                 
 \def\Pd      {\ensuremath{\mathrm{d}}\xspace}                 
 \def\Pe      {\ensuremath{\mathrm{e}}\xspace}

 \def\Ph      {\ensuremath{\mathrm{h}}\xspace}                 
 \def\Pi      {\ensuremath{\mathrm{i}}\xspace}

 \def\Pp      {\ensuremath{\mathrm{p}}\xspace}

 \def\Ps      {\ensuremath{\mathrm{s}}\xspace}                 
                  
 \def\Pu      {\ensuremath{\mathrm{u}}\xspace}

 \def\thebaroffset{0.0em}
}
{

 \def\Peta        {\ensuremath{\eta}\xspace}

 \def\Pmu         {\ensuremath{\mu}\xspace}                 
 \def\Pnu         {\ensuremath{\nu}\xspace}                 
                  
 \def\Ppi         {\ensuremath{\pi}\xspace}

 \def\Pphi        {\ensuremath{\phi}\xspace}

 \mathchardef\PDelta="7101
 \mathchardef\PXi="7104
 \mathchardef\PLambda="7103
 \mathchardef\PSigma="7106
 \mathchardef\POmega="710A
 \mathchardef\PUpsilon="7107
                  
 \def\PB      {\ensuremath{B}\xspace}                 
                  
 \def\PD      {\ensuremath{D}\xspace}

 \def\PK      {\ensuremath{K}\xspace}

 \def\PP      {\ensuremath{P}\xspace}

 \def\PW      {\ensuremath{W}\xspace}

 \def\Pb      {\ensuremath{b}\xspace}                 
 \def\Pc      {\ensuremath{c}\xspace}                 
 \def\Pd      {\ensuremath{d}\xspace}                 
 \def\Pe      {\ensuremath{e}\xspace}

 \def\Ph      {\ensuremath{h}\xspace}                 
 \def\Pi      {\ensuremath{i}\xspace}

 \def\Pp      {\ensuremath{p}\xspace}

 \def\Ps      {\ensuremath{s}\xspace}                 
                  
 \def\Pu      {\ensuremath{u}\xspace}

 \def\thebaroffset{0.18em}
}
\newcommand{\offsetoverline}[2][\thebaroffset]{\kern #1\overline{\kern -#1 #2}}%

\makeatletter
\newcommand{\miniscule}{\@setfontsize\miniscule{4}{5}}
\makeatother

\DeclareRobustCommand{\optbar}[1]{\shortstack{{\miniscule (\rule[.5ex]{1.25em}{.18mm})}
  \\ [-.7ex] $#1$}}

\def\en         {{\ensuremath{\Pe^-}}\xspace}   
\def\ep         {{\ensuremath{\Pe^+}}\xspace}

\def\mup        {{\ensuremath{\Pmu^+}}\xspace}
\def\mun        {{\ensuremath{\Pmu^-}}\xspace} 

\def\ellp       {{\ensuremath{\ell^+}}\xspace}

\def\neu        {{\ensuremath{\Pnu}}\xspace}

\def\neul       {{\ensuremath{\neu_\ell}}\xspace}

\def\W      {{\ensuremath{\PW}}\xspace}

\def\uquark    {{\ensuremath{\Pu}}\xspace}
\def\uquarkbar {{\ensuremath{\overline \uquark}}\xspace}

\def\dquark    {{\ensuremath{\Pd}}\xspace}
\def\dquarkbar {{\ensuremath{\overline \dquark}}\xspace}

\def\squark    {{\ensuremath{\Ps}}\xspace}
\def\squarkbar {{\ensuremath{\overline \squark}}\xspace}

\def\cquark    {{\ensuremath{\Pc}}\xspace}
\def\cquarkbar {{\ensuremath{\overline \cquark}}\xspace}

\def\bquark    {{\ensuremath{\Pb}}\xspace}

\def\hadron {{\ensuremath{\Ph}}\xspace}
\def\pion   {{\ensuremath{\Ppi}}\xspace}
\def\piz    {{\ensuremath{\pion^0}}\xspace}
\def\pip    {{\ensuremath{\pion^+}}\xspace}
\def\pim    {{\ensuremath{\pion^-}}\xspace}
\def\pipm   {{\ensuremath{\pion^\pm}}\xspace}
\def\pimp   {{\ensuremath{\pion^\mp}}\xspace}

\def\kaon    {{\ensuremath{\PK}}\xspace}

\def\KorKbar {\kern \thebaroffset\optbar{\kern -\thebaroffset \PK}{}\xspace}
\def\Kz      {{\ensuremath{\kaon^0}}\xspace}

\def\Kp      {{\ensuremath{\kaon^+}}\xspace}
\def\Km      {{\ensuremath{\kaon^-}}\xspace}
\def\Kpm     {{\ensuremath{\kaon^\pm}}\xspace}
\def\Kmp     {{\ensuremath{\kaon^\mp}}\xspace}
\def\KS      {{\ensuremath{\kaon^0_{\mathrm{S}}}}\xspace}

\def\KL      {{\ensuremath{\kaon^0_{\mathrm{L}}}}\xspace}

\newcommand{\etaz}{\ensuremath{\Peta}\xspace}
\newcommand{\etapr}{\ensuremath{\Peta^{\prime}}\xspace}
\newcommand{\phiz}{\ensuremath{\Pphi}\xspace}

\def\Dbar    {{\ensuremath{\offsetoverline{\PD}}}\xspace}
\def\D       {{\ensuremath{\PD}}\xspace}
\def\Db      {{\ensuremath{\Dbar}}\xspace}
\def\DorDbar {\kern \thebaroffset\optbar{\kern -\thebaroffset \PD}\xspace}
\def\Dz      {{\ensuremath{\D^0}}\xspace}
\def\Dzb     {{\ensuremath{\Dbar{}^0}}\xspace}
\def\Dp      {{\ensuremath{\D^+}}\xspace}
\def\Dm      {{\ensuremath{\D^-}}\xspace}

\def\DpDm    {\ensuremath{\Dp {\kern -0.16em \Dm}}\xspace}

\def\Dstarp  {{\ensuremath{\D^{*+}}}\xspace}
\def\Dstarm  {{\ensuremath{\D^{*-}}}\xspace}

\def\theDstarp{{\ensuremath{\D^{*}(2010)^{+}}}\xspace}

\def\Dsp     {{\ensuremath{\D^+_\squark}}\xspace}

\def\B       {{\ensuremath{\PB}}\xspace}

\def\BorBbar {\kern \thebaroffset\optbar{\kern -\thebaroffset \PB}\xspace}
\def\Bz      {{\ensuremath{\B^0}}\xspace}

\def\Bd      {{\ensuremath{\B^0}}\xspace}

\def\BdorBdbar {\kern \thebaroffset\optbar{\kern -\thebaroffset \Bd}\xspace}
\def\Bu      {{\ensuremath{\B^+}}\xspace}

\def\Bp      {{\ensuremath{\Bu}}\xspace}

\def\Bs      {{\ensuremath{\B^0_\squark}}\xspace}

\def\BsorBsbar {\kern \thebaroffset\optbar{\kern -\thebaroffset \Bs}\xspace}

\def\Y#1S{\ensuremath{\PUpsilon{(#1S)}}\xspace}

\def\proton      {{\ensuremath{\Pp}}\xspace}

\def\Lz          {{\ensuremath{\PLambda}}\xspace}

\def\LorLbar     {\kern \thebaroffset\optbar{\kern -\thebaroffset \PLambda}\xspace}

\def\Lc          {{\ensuremath{\Lz^+_\cquark}}\xspace}

\def\BF         {{\ensuremath{\mathcal{B}}}\xspace}

\newcommand{\decay}[2]{\ensuremath{#1\!\to #2}\xspace} 

\def\to                 {\ensuremath{\rightarrow}\xspace}

\newcommand{\tauDz}{{\ensuremath{\tau_{\Dz}}}\xspace}

\def\order   {{\ensuremath{\mathcal{O}}}\xspace}

\newcommand{\as}{{\ensuremath{\alpha_s}}\xspace}

\newcommand{\lqcd}{{\ensuremath{\Lambda_{\mathrm{QCD}}}}\xspace}

\def\CP                {{\ensuremath{C\!P}}\xspace}

\def\Vcs  {{\ensuremath{V_{\cquark\squark}^{\phantom{\ast}}}}\xspace}

\def\Vcb  {{\ensuremath{V_{\cquark\bquark}^{\phantom{\ast}}}}\xspace}

\def\Vuss  {{\ensuremath{V_{\uquark\squark}^\ast}}\xspace}

\def\Vubs  {{\ensuremath{V_{\uquark\bquark}^\ast}}\xspace}

\def\AT#1     {\ensuremath{A_{\mathrm{T}}^{#1}}\xspace}           

\def\C#1      {\ensuremath{\mathcal{C}_{#1}}\xspace}                       
\def\Cp#1     {\ensuremath{\mathcal{C}_{#1}^{'}}\xspace}                    
\def\Ceff#1   {\ensuremath{\mathcal{C}_{#1}^{\mathrm{(eff)}}}\xspace}        
\def\Cpeff#1  {\ensuremath{\mathcal{C}_{#1}^{'\mathrm{(eff)}}}\xspace}       
\def\Ope#1    {\ensuremath{\mathcal{O}_{#1}}\xspace}                       
\def\Opep#1   {\ensuremath{\mathcal{O}_{#1}^{'}}\xspace}                    

\newcommand{\ket}[1]{\ensuremath{|#1\rangle}}              

\newcommand{\nospaceunit}[1]{\ensuremath{\text{#1}}}       
\newcommand{\aunit}[1]{\ensuremath{\text{\,#1}}}       

\newcommand{\tev}{\aunit{Te\kern -0.1em V}\xspace}
\newcommand{\gev}{\aunit{Ge\kern -0.1em V}\xspace}
\newcommand{\mev}{\aunit{Me\kern -0.1em V}\xspace}
\newcommand{\kev}{\aunit{ke\kern -0.1em V}\xspace}
\newcommand{\ev}{\aunit{e\kern -0.1em V}\xspace}
 
\newcommand{\mevc}{\ensuremath{\aunit{Me\kern -0.1em V\!/}c}\xspace}
\newcommand{\gevc}{\ensuremath{\aunit{Ge\kern -0.1em V\!/}c}\xspace}
\newcommand{\mevcc}{\ensuremath{\aunit{Me\kern -0.1em V\!/}c^2}\xspace}
\newcommand{\gevcc}{\ensuremath{\aunit{Ge\kern -0.1em V\!/}c^2}\xspace}

\def\cm   {\aunit{cm}\xspace}

\def\mum  {\ensuremath{\,\upmu\nospaceunit{m}}\xspace}

\def\fb   {\ensuremath{\aunit{fb}}\xspace}
\def\invfb   {\ensuremath{\fb^{-1}}\xspace}

\def\sec  {\ensuremath{\aunit{s}}\xspace}

\def\ps   {\ensuremath{\aunit{ps}}\xspace}

\def\mhz  {\ensuremath{\aunit{MHz}}\xspace}

\def\order{{\ensuremath{\mathcal{O}}}\xspace}

\def\gsim{{~\raise.15em\hbox{$>$}\kern-.85em
          \lower.35em\hbox{$\sim$}~}\xspace}
\def\lsim{{~\raise.15em\hbox{$<$}\kern-.85em
          \lower.35em\hbox{$\sim$}~}\xspace}

\newcommand{\Imag}{\ensuremath{\mathcal{I}m}\xspace}

\def\pt         {\ensuremath{p_{\mathrm{T}}}\xspace}

\def\mrad{\aunit{mrad}\xspace}
\def\rad{\aunit{rad}\xspace}

\def\tell1  {TELL1\xspace}
\def\ukl1   {UKL1\xspace}

\def\runone {\textit{Run~1}\xspace}
\def\runtwo {\textit{Run~2}\xspace}
\def\runthree {\textit{Run~3}\xspace}
\def\runfour {\textit{Run~4}\xspace}
\def\runonetwo {\textit{Run 1} and \textit{2}\xspace}

\def\hp      {{\ensuremath{\hadron^+}}\xspace}
\def\hm      {{\ensuremath{\hadron^-}}\xspace}
\def\hzero   {{\ensuremath{\hadron^0}}\xspace}

\def\hpm     {{\ensuremath{\hadron^\pm}}\xspace}

\def\pisp    {{\ensuremath{\pion^{+}_{\mathrm{ \scriptscriptstyle tag}}}}\xspace}
\def\pism    {{\ensuremath{\pion^{-}_{\mathrm{ \scriptscriptstyle tag}}}}\xspace}
\def\DpOrDsp {{\ensuremath{\D^+_{(\squark)}}}\xspace}
\def\DzOrDzb {\kern 0.18em\optbar{\kern -0.18em \ensuremath{D}}{}{\ensuremath{^0}}\xspace}
\def\fOrfb   {\kern 0.18em\optbar{\kern -0.18em f}{}{}\xspace}

\def\KK {{\ensuremath{\Kp\Km  }}\xspace}
\def\PP {{\ensuremath{\pip\pim}}\xspace}
\def\RS {{\ensuremath{\Km\pip }}\xspace}
\def\WS {{\ensuremath{\Kp\pim }}\xspace}
\def\HH {{\ensuremath{\hp\hm }}\xspace}

\def\DzHH {\decay{\Dz}{\HH}}
\def\DzKK {\decay{\Dz}{\Kp\Km}}
\def\DzPP {\decay{\Dz}{\pip\pim}}
\def\DzRS {\decay{\Dz}{\Km\pip}}
\def\DzWS {\decay{\Dz}{\Kp\pim}}

\newcommand{\DY}[1]{{\ensuremath{\Delta Y_{#1}}}\xspace}
\newcommand{\agamma}[1]{{\ensuremath{A_{\Gamma}^{#1}}}\xspace}
\newcommand{\ycp}[1]{{\ensuremath{y_{\CP}^{#1}}}\xspace}

\def\DstM{{\ensuremath{m(\Dz\pisp)}}\xspace}

\def\af      {{\ensuremath{A_f}}\xspace}

\def\abf     {{\ensuremath{\bar{A}_f}}\xspace}
\def\abfb    {{\ensuremath{\bar{A}_{\bar{f}}}}\xspace}
\newcommand{\Acpdec}[1]{{\ensuremath{a^{d}_{#1}}}\xspace}
\newcommand{\Araw}[1]{{\ensuremath{A_\textnormal{raw}(#1)}}\xspace}
\newcommand{\Adet}[1]{{\ensuremath{A_\textnormal{det}(#1)}}\xspace}
\newcommand{\Aprod}[1]{{\ensuremath{A_\textnormal{prod}(#1)}}\xspace}
\def\deltakpi{{\ensuremath{\delta^{K\pi}_{\D}}}\xspace}

\def\SUF {{\ensuremath{SU(3)_\textnormal{F}}}\xspace}

\newcommand{\cgev}{\ensuremath{\,c/\nospaceunit{Ge\kern -0.1em V}}\xspace}
\newcommand{\kevcc}{\ensuremath{\aunit{ke\kern -0.1em V\!/}c^2}\xspace}

\graphicspath{{./figs/}}

\usepackage{cleveref}

\title{Recent advances in charm mixing and $C\!P$ violation \\ at LHCb}

\author[]{Tommaso Pajero}

\affiliation[]{Department of Physics, University of Oxford,\\Denys Wilkinson Building, Keble Road, Oxford OX1 3RH, United Kingdom}

\emailAdd{tommaso.pajero@physics.ox.ac.uk}

\abstract{
After playing a pivotal role in the birth of the Standard Model in the 70's, the study of charm physics has undergone a revival during the last decade, triggered by a wealth of precision measurements from the charm and \B factories, and from the \cdf and especially the \lhcb experiments.
In this article, we sum up how the unique phenomenology of charmed hadrons can be used to test the Standard Model and we review the latest measurements performed in this field by the \lhcb experiment.
These include the historic first observations of \CP violation and of a nonzero mass difference between the charmed neutral-meson mass eigenstates, the most precise determination of their decay-width difference to date, and a search for time-dependent \CP violation reaching the unprecedented precision of $10^{-4}$.
These results challenge our comprehension of nonperturbative strong interactions, and their interpretation calls for further studies on both the theoretical and experimental sides.
The upcoming upgrades of the \lhcb experiment will play a leading role in this quest.
}

\keywords{Charm physics; \CP violation; neutral-meson mixing; flavour-changing neutral currents.}

\arxivnumber{2208.05769}

\begin{document}
\emergencystretch 3em
\maketitle
\flushbottom

\section{Introduction}
Flavour physics constitutes a sensitive test bed of the standard model (SM), thanks to two peculiar properties of its Lagrangian.
On the one hand, flavour changing neutral currents (FCNC) are suppressed by the Glashow--Iliopoulos--Maiani (GIM) mechanism~\cite{Glashow:1970gm,Buras:2020_GIM}.
On the other, it encompasses only one observed source of violation of the \CP symmetry,\footnote{A second source, ascribable to the strong interaction, is experimentally negligible~\cite{Peccei:2006as}.}
that is, a single irreducible complex phase in the Cabibbo--Kobayashi--Maskawa (CKM) matrix governing the interaction of quarks with the \W boson~\cite{Cabibbo:1963yz,Kobayashi:1973fv,PDG2020_reviews}.
As a consequence, \CP-violation observables are overconstrained and follow a well defined pattern.
Precision measurements of FCNCs and of \CP violation are thus sensitive probes for new interactions beyond the SM (BSM), which could modify their size through diagrams including particles off the mass shell, even if their energy scales are larger than those available to the current particle colliders.
Historically, this line of research has turned out to be very fruitful.
Notable examples include the proposal of the GIM mechanism and the prediction of the existence of the charm quark in 1970~\cite{Glashow:1970gm}, based on the suppression of the branching fraction of \decay{\KL}{\mup\mun} decays~\cite{Bott-Bodenhausen:1967vka,Foeth:1969hi}; the observation of \CP violation in \Kz mesons, which suggested the existence of a third generation of quarks in 1973~\cite{Christenson:1964fg,Kobayashi:1973fv}; and the first evidence of \Bz mixing, which set a lower bound on the mass of the top quark in the 80s~\cite{Campbell:1981rg,ARGUS:1987xtv}.
Thus, it is not unlikely that future studies of rare flavour-changing processes will shed light on the structure of the BSM interactions that are needed to explain the shortcomings of the SM, such as the missing explanation of the nature of dark matter~\cite{PDG2020_reviews} and of the cosmological baryon asymmetry~\cite{Dine:2003ax}.

Charmed hadrons are the only hadrons where precision measurements of \mbox{FCNCs} and of \CP violation involving the decay of up-type quarks can be performed.
Therefore, they are sensitive to a different class of interactions with respect to \B and \kaon mesons, where the decaying quark is of type down.
Their phenomenology is also peculiar due to a stronger GIM suppression --- a consequence of the smaller mass of the beauty than the top quark involved in the respective loop diagrams, and of the smallness of the CKM matrix elements connecting the first two generations of quarks with the third.
In particular, \CP violation is proportional to the following combination of CKM matrix elements, \mbox{$\Imag(\Vcb\Vubs/\Vcs\Vuss)\approx -6\times 10^{-4}$}~\cite{CKMfitter2005}, leading to \CP asymmetries typically of the order of \mbox{$10^{-4}$} to \mbox{$10^{-3}$}~\cite{Grossman:2006jg}.
The smallness of FCNCs and of \CP violation in charm has frustrated their search for a long time.
It was only during the last decade that experimental progress with the \B factories, the \cdf experiment and, most prominently, the \lhcb experiment has eventually allowed these phenomena to be observed for the first time.

The entrance of charm physics into the era of precision measurements poses several challenges not only on the experimental, but also on the theoretical side.
In fact, while the enhanced GIM suppression potentially provides excellent sensitivity to BSM interactions, the theoretical interpretation of the measurements is complicated by the contributions from nonperturbative strong interactions involving strange and down quarks~\cite{Buccella:1994nf}.
Not only the size of these contributions is difficult to calculate, but they are subject to large cancellations, as they vanish in the \SUF limit, where the masses of the \squark, \dquark and \uquark quarks are neglected with respect to the typical hadronic scale of charm decays, \lqcd.
Therefore, a rigorous assessment of the agreement of the measurements with the SM will require a combination of advances of the available theoretical tools and an extensive experimental program of auxiliary measurements.

This review is structured as follows.
\Cref{sect:framework} introduces the theoretical framework to describe mixing and \CP violation in charmed hadrons and outlines the status of the theoretical predictions.
An introduction to the \lhcb experiment and to the typical analysis methods adopted in charm measurements is provided in \cref{sect:lhcb}, before describing the most important time-integrated and time-dependent measurements performed during the last few years in \cref{sect:cpv-decay,sect:time-dep-studies}, respectively.
Finally, we conclude by sketching the prospects for experimental progress in the coming years in \cref{sect:conclusions}.
Throughout this article, the first quoted uncertainties are statistical and the second are systematic.

\section{Theoretical framework}
\label{sect:framework}
The next sections provide a brief theoretical introduction to the phenomena of mixing and \CP violation in the decay of charmed hadrons.

\subsection{\Dz mixing}
\begin{figure}[bt]
  \begin{center}
    \includegraphics[height=0.165\linewidth]{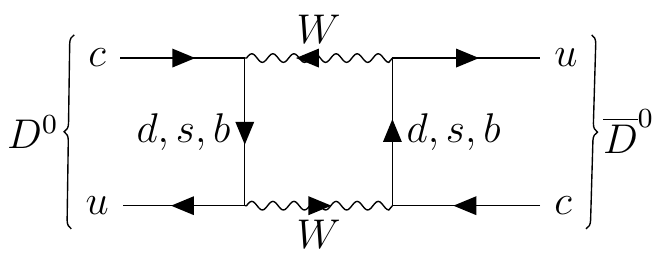}
    \includegraphics[height=0.165\linewidth]{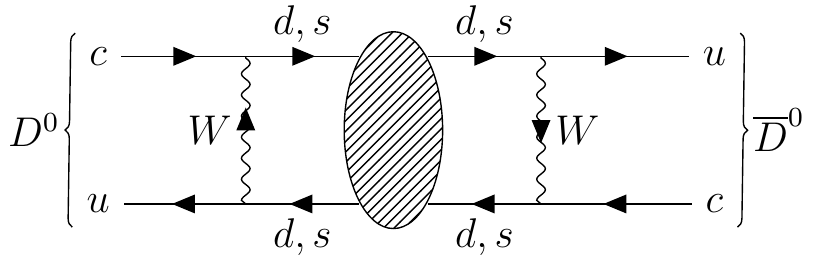}
    \vspace*{-0.2cm}
  \end{center}
  \caption{
    Examples of diagrams of (left) perturbative and (right) nonperturbative contributions to \Dz mixing.
    In the right diagram, the dashed area represents low-energy QCD interactions, possibly involving the exchange of hadrons on the mass shell.
    The nonperturbative contributions dominate, since they avoid the suppression from the loop factor, and can have a much milder \SUF-breaking (GIM) suppression~\cite{Grossman:2006jg}.
  }
  \label{fig:mixing}
\end{figure}

One of the most interesting phenomena involving FCNCs is mixing,\footnote{We refer the reader to ref.~\cite{Gisbert:2020vjx} for a recent review of leptonic and semileptonic decays of charmed hadrons, which are not discussed in this article, and to refs.~\cite{LHCb-PAPER-2021-035,LHCb-PAPER-2022-029} for updated experimental results.} that is, the quantum oscillation of a neutral flavoured meson such as the \Dz meson, made up of a $\cquark\uquarkbar$ quark pair, into its antiparticle, and vice versa; see \cref{fig:mixing}.
This process is parametrised through the dimensionless mixing parameters $x_{12}$ and $y_{12}$, defined as \mbox{$x_{12} \equiv 2 \lvert M_{12} / \Gamma \rvert$} and \mbox{$y_{12} \equiv \lvert \Gamma_{12} / \Gamma \rvert$}, where \mbox{$\bm{H}\equiv \bm{M} - \tfrac{i}{2}\bm{\Gamma}$} is the effective Hamiltonian of the subspace spanned by \ket{\Dz} and \ket{\Dzb} and $\Gamma$ is the \Dz decay width~\cite{Grossman:2009mn,Kagan:2009gb,Kagan:2020vri}.\footnote{We employ natural units throughout this article.}
The mixing parameter $x_{12}$ ($y_{12}$) is proportional to the size of the transition amplitudes between \Dz and \Dzb mesons through off-shell (on-shell) intermediate states, and is equal to the magnitude of the normalised difference between the masses (decay widths) of the two mass eigenstates, conventionally denoted as
$x \equiv \Delta m / \Gamma$ ($y \equiv \Delta\Gamma / 2\Gamma$),
up to second order in the small \CP-violation parameter $\phi_{12} \equiv \arg(M_{12}/\Gamma_{12})$.
Experimentally, $x_{12}$ and $y_{12}$ are equal to $(4.07  \pm 0.48) \times 10^{-3}$ and $(6.45  \pm 0.24) \times 10^{-3}$, respectively~\cite{pajero:charm-fitter}.

The absorptive mixing amplitude can be written in the SM as~\cite{Kagan:2020vri}
\begin{equation}
    \label{eq:gamma12_SM}
    \Gamma_{12}^\textnormal{SM} =   \frac{(\lambda_s - \lambda_d)^2}{4}\Gamma_2
                                  + \frac{(\lambda_s - \lambda_d)\lambda_b}{2}\Gamma_1
                                  + \frac{(\lambda_b)^2}{4}\Gamma_0,
\end{equation}
where $\lambda_i \equiv V_{ci}V_{ui}^*$, unitarity of the CKM matrix is assumed ($\lambda_d + \lambda_s + \lambda_b = 0$), and $\Gamma_{2,1,0}$ are the $\Delta U_3 = 0$ matrix elements of the $\Delta U = 2,1,0$ transitions, respectively.\footnote{$U$ spin is the $SU(2)$ subgroup of \SUF relating the \dquark and \squark quarks.}
In terms of their flavour structure, they are equal to
\begin{equation}
\begin{aligned}
    \label{eq:gamma_uspin}
    \Gamma_2 &= \Gamma_{dd} - 2 \Gamma_{ds} + \Gamma_{ss} \sim (\dquark\dquarkbar - \squark\squarkbar)^2 = \mathcal{O}(\epsilon^2),\\
    \Gamma_1 &= \Gamma_{dd}                - \Gamma_{ss} \sim (\dquark\dquarkbar - \squark\squarkbar)(\dquark\dquarkbar + \squark\squarkbar) = \mathcal{O}(\epsilon),\\
    \Gamma_0 &= \Gamma_{dd} + 2 \Gamma_{ds} + \Gamma_{ss} \sim (\dquark\dquarkbar + \squark\squarkbar)^2 = \mathcal{O}(1),
\end{aligned}
\end{equation}
where $\Gamma_{ij}$ designates the absorptive part, proceeding through on-shell intermediate states, of the box diagrams with internal quarks $i$ and $j$, and the rightmost terms show the order of $\Gamma_{0,1,2}$ in terms of the $U$-spin breaking parameter $\epsilon \approx 0.3$, assuming that a perturbative expansion in this parameter is possible~\cite{Chala:2019fdb}.
Owing to the hierarchy of CKM elements,
$(\lambda_s - \lambda_d)/2 \approx 0.22 - i\,6.6 \times 10^{-5}$ and
$\lambda_b / 2 \approx 3.0 \times 10^{-5} + i\,6.6\times 10^{-5}$,
the first term in \cref{eq:gamma12_SM} is the dominant one, even if it arises only at second order in $U$-spin breaking~\cite{Falk:2001hx,Gronau:2012kq}.
On the other hand, \CP violation requires the contribution of a second amplitude with a different weak phase~\cite{PDG2020_reviews},\footnote{Weak phases are defined as the phases that change their sign under the \CP transformation, like those of the CKM matrix elements.
On the other hand, phases that do not change their sign under the \CP transformation, such as those arising from QCD, are called strong phases.}
namely, the second term in the right-hand side of \cref{eq:gamma12_SM} or an additional term from BSM interactions.
Conventionally, it is parametrised through the phase of $\Gamma_{12}$ with respect to its dominant $\Delta U = 2$ component, $\phi^\Gamma_2 \equiv \arg\big(\Gamma_{12}/\big[(\lambda_s - \lambda_d)^2 \Gamma_2\big]\big)$, where the subscript denotes the chosen convention~\cite{Kagan:2020vri}.

An analogous discussion and weak phase, $\phi_2^M$, can be defined also for the dispersive matrix element, $M_{12}$, describing off-shell intermediate states.
The only difference is that $M_1$ and $M_0$ receive additional contributions, $2(M_{sb} - M_{db})$ and $4(M_{bb} - M_{sb} - M_{db})$, which could compensate in part for the CKM suppression.

Providing predictions for the mixing parameters is notoriously a formidable task.
Heavy quark expansion has been successfully used to predict the lifetimes of \Dz, \Dp and \Dsp mesons, suggesting that an inclusive calculation from a perturbative expansion in terms of $\lqcd / m_c \sim 0.3$ and of $\as(m_c) \sim 0.33$ might be viable~\cite{King:2021xqp}.
The individual contributions to $y_{12}$ from single $\Gamma_{ij}$ amplitudes are five times larger than the experimental value prior to GIM cancellations, but the size of such cancellations is not controlled yet; see ref.~\cite{Lenz:2020awd} for a review.
In the very long term, the size of the mixing parameters may be predicted through lattice calculations, by building on the methods described in ref.~\cite{Hansen:2012tf}.
On the contrary, exclusive approaches to estimate $y_{12}$, which sum over the contributions from all the final states shared by \Dz and \Dzb mesons~\cite{Cheng:2010rv,Jiang:2017zwr}, are unlikely to provide precise predictions.
In fact, the contributions from different final states within the same $U$-spin multiplet tend to cancel; hence, the precision with which the branching fractions and the strong phases of the decay amplitudes are known --- in particular for multibody decays --- significantly limits the achievable precision~\cite{Falk:2001hx}.
Moreover, since the exclusive methods are based on charm experimental data, they are unable to distinguish SM contributions from new interactions.
It is, however, interesting to note that the authors of ref.~\cite{Falk:2001hx} showed well before a nonzero value of $y_{12}$ was measured that even only $U$-spin breaking from the different phase space of multibody final states, neglecting all dynamical effects, can account for a value of $y_{12}$ as large as 1\%, which is consistent with the experimental value~\cite{LHCb-PAPER-2021-033}.
Finally, a dispersion relation between the mixing parameters has been derived in ref.~\cite{Falk:2004wg} in the heavy-quark limit, predicting values of $x_{12}$ between $10^{-3}$ and $10^{-2}$ if $y_{12}$ is of the order of 1\%, in keeping with the experimental data.

As far as \CP violation is concerned, the phases $\phi_2^M$ and $\phi_2^\Gamma$ can be estimated from \cref{eq:gamma12_SM} to be of the order of $2 \mrad$, though enhancements of up to one order of magnitude cannot be excluded~\cite{Bigi:2011re,Bobrowski:2010xg,Kagan:2020vri,Li:2020xrz}.
An upper bound of $5\mrad$ has recently been argued for $\phi^\Gamma_2$~\cite{Kagan:2020vri}.
The experimental limits are less precise by one order of magnitude, as the weak phases are currently measured to be \mbox{$\phi_2^M = (0.030 \pm 0.021)\rad$} and \mbox{$\phi_2^\Gamma = (0.044 \pm 0.027)\rad$}~\cite{pajero:charm-fitter}.

Even in absence of precise SM predictions, the small size of mixing and \CP violation parameters can be employed to set stringent limits on the scale of new BSM interactions, up to more than $10^4 \tev$, by assuming that contributions BSM saturate the measured values of $x_{12}$ and $\phi_2^M$~\cite{Golowich:2007ka,Carrasco:2014uya,Carrasco:2015pra,Bazavov:2017weg,Kirk:2017juj,Alpigiani:2017lpj}.

\subsection{\CP violation in the decay}
\CP violation can arise also in the decay amplitudes, where it is conventionally quantified through the parameter
\begin{equation}
    \label{eq:acp-def}
    \Acpdec{f} \equiv \frac{\lvert \af \rvert^{2} - \lvert \abfb \rvert^{2}}
                           {\lvert \af \rvert^{2} + \lvert \abfb \rvert^{2}},
\end{equation}
where \af and \abfb denote the decay amplitudes of a charmed hadron and of the corresponding anti-hadron into the \CP-conjugate final states $f$ and $\bar{f}$, respectively.
This phenomenon is expected to be observable only in singly Cabibbo-suppressed (SCS) \decay{\cquark}{\uquark\squark\squarkbar} and \decay{\cquark}{\uquark\dquark\dquarkbar} transitions, which receive contributions from QCD penguin and chromomagnetic-dipole operators.
In contrast, Cabibbo-favoured (CF) \decay{\cquark}{\uquark\squark\dquarkbar} and doubly Cabibbo-suppressed (DCS) \decay{\cquark}{\uquark\dquark\squarkbar} transitions cannot be influenced by these operators, as they involve quarks of four different flavours.
Therefore, any signs of \CP asymmetries larger than $10^{-5}$ in CF or DCS decays would be an unambiguous evidence of new BSM interactions~\cite{Bergmann:1999pm,Grossman:2006jg}.
The only exception is that of CF decays that contain an odd number of \KS kaons in their final state, since these decays also receive a contribution from a DCS amplitude, whose interference with the CF amplitude may enhance the \CP asymmetry up to $10^{-4}$ level~\cite{Bigi:1994aw}.

To predict the size of \Acpdec{f}, it is useful to decompose the decay amplitude in terms of CKM factors as
\begin{equation}
    \label{eq:decay-ampl-parametrisation}
    \af \equiv A_{sd} \,\frac{\lambda_s - \lambda_d}{2} - A_b \,\frac{\lambda_b}{2},
\end{equation}
where only two terms appear thanks to CKM unitarity, and the choice of using $(\lambda_s - \lambda_d)/2$ for the first is dictated by $U$-spin symmetry arguments.
While $A_{sd}$ is in some cases dominated by tree diagrams, it can receive contributions also from exchange, annihilation and broken-penguin (annihilation) diagrams, where the broken penguin is the $U$-spin breaking difference of the penguin diagrams with internal quarks \squark and \dquark.
Analogously, also $A_b$ contains all the categories listed above, with the exception that broken penguins are substituted by penguin diagrams with internal quark \bquark  and by the $U$-spin conserving average of penguin diagrams with internal quarks \squark and \dquark~\cite{Brod:2012ud,Muller:2015lua}.

\begin{figure}[bt]
  \begin{center}
    \includegraphics[width=0.3\linewidth]{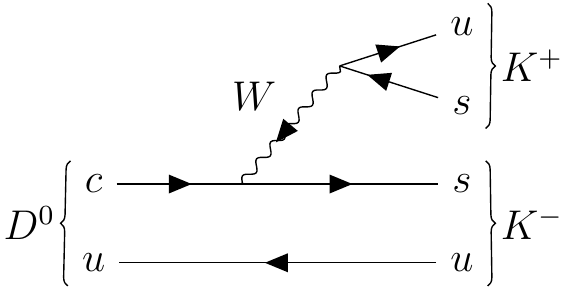}
    \includegraphics[width=0.3\linewidth]{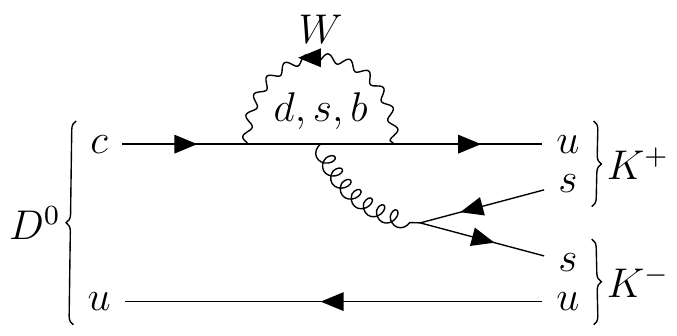}
    \includegraphics[width=0.35\linewidth]{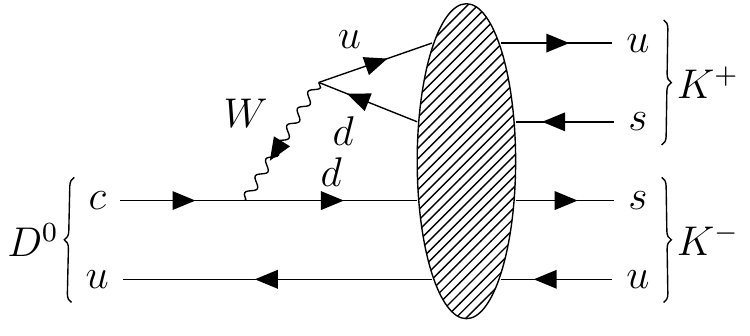}
    \vspace*{-0.2cm}
  \end{center}
  \caption{
        Examples of diagrams contributing to the \DzKK decay.
        The tree diagram (left) is proportional to $\lambda_s$, while the penguin diagram (centre) is proportional to $\lambda_q$, where $q$ is the internal quark of the loop.
        The right diagram represents rescattering of an on-shell state including a \dquark\dquarkbar quark pair such as \PP, represented by the dashed area, into the \KK final state.
        This diagram is proportional to $\lambda_d$ and thus contributes to $A_{sd}$ ($A_b$) with a relative minus sign (with the same sign) as the tree diagram.
        Therefore, it gives rise to \CP violation if it has a different strong phase with respect to the tree diagram, as expected from rescattering.
  }
  \label{fig:feynman-kk}
\end{figure}

The action of the \CP transformation corresponds to the complex conjugation for the CKM matrix elements, while it has no effects on the strong matrix elements $A_{sd}$ and $A_b$.
Therefore, the \CP violation in the decay is equal to
\begin{equation}
    \label{eq:cp-asymm-parametrisation}
    \Acpdec{f} \approx \Imag \left(\frac{2\lambda_b}{\lambda_s - \lambda_d}\right) \Imag \left(\frac{A_b}{A_{sd}}\right)
               = (-5.8 \pm 0.2) \times 10^{-4} \left\lvert\frac{A_b}{A_{sd}}\right\rvert \sin\delta,
\end{equation}
where terms of order higher than one in $\lambda_b$ are neglected, the CKM factor is taken from ref.~\cite{CKMfitter2005}, and the strong-phase difference $\delta$ is defined as $\delta \equiv \arg(A_b/A_{sd})$.
Since the second term in \cref{eq:decay-ampl-parametrisation} is heavily CKM suppressed, the decay width into a given final state is described to excellent approximation by the first term only, and the magnitude of $A_{sd}$ can be calculated from the branching fraction of the decay.
Thus, to provide predictions for \Acpdec{f}, one then needs to estimate the size of $A_b$ and of $\delta$.
Penguin diagrams are naively suppressed by a factor of $\as/\pi$ with respect to the tree diagrams~\cite{Grossman:2006jg}, leading to an additional suppression of \Acpdec{f} by approximately a factor of 5 on top of the already small CKM factor.
However, final-state rescattering from nonperturbative strong interactions is topologically equivalent to a penguin diagram, see \cref{fig:feynman-kk}, and could enhance this prediction by up to one order of magnitude, leading to asymmetries of the order of $10^{-3}$~\cite{Buccella:1994nf,Franco:2012ck}.
Even larger enhancements may take place in decays into two neutral kaons, where \CP violation is expected to arise from the interference of the tree-level transitions \decay{\cquark\uquarkbar}{\squark\squarkbar} and \decay{\cquark\uquarkbar}{\dquark\dquarkbar}~\cite{Brod:2011re,Nierste:2015zra,Nierste:2017cua}.
If confirmed, large rescattering effects could also explain the large \SUF breaking, order of 30\%, that is observed in some branching fractions --- for example, in the ratio $\BF(\DzKK) / \BF(\DzPP)$ and in the relatively large value of $\BF(\decay{\Dz}{\KS\KS})$~\cite{Brod:2012ud}.
Improving our understanding of these nonperturbative effects, for example, through studies of the resonances that might contribute to rescattering~\cite{Schacht:2021jaz}, or through precise measurements of rescattering close to the mass scale of charmed hadrons~\cite{Franco:2012ck}, is crucial to provide precise predictions of \CP asymmetries.

Nevertheless, it is already possible to relate the size of \CP asymmetries of different decay channels through \SUF or isospin symmetry arguments, which may be violated by BSM interactions~\cite{Grossman:2012eb,Grossman:2012ry,Muller:2015rna}.
For example, \cref{eq:cp-asymm-parametrisation} implies that in the limit of \SUF symmetry the \CP asymmetries for two decay channels related by a $U$-spin transformation, such as \DzKK and \DzPP, have equal magnitude but opposite sign.
Further examples will be provided in \cref{sect:hh0}.

\section{The \lhcb detector and charm physics}
\label{sect:lhcb}
As both \Dz oscillations and \CP violation in charmed hadrons are effects of the order of $10^{-3}$ or smaller, they require the collection of ten million decays or more to be observed.
While samples of this size have first been collected at the \B factories and by the \cdf experiment, the last decade has witnessed a further leap forward in precision thanks to the start of operations of the \lhcb experiment, which benefits from the high luminosity and huge production cross-section of charmed hadrons at the Large Hadron Collider~\cite{LHCb-PAPER-2013-004,LHCb-PAPER-2016-042} as well as from a detector design dedicated to the study of heavy-quark hadrons.

The \lhcb detector~\cite{LHCb-DP-2008-001,LHCb-DP-2014-002} is a single-arm forward spectrometer covering the \mbox{pseudorapidity} range $2 < \eta < 5$, where the production of heavy quarks is concentrated~\cite{bbangles}.
It has recently been upgraded for \runthree of the LHC~\cite{LHCb-TDR-012,LHCb-TDR-016}.
The following description is of the experiment that operated during \runone and 2.
The tracking system consists of a silicon-strip vertex detector surrounding the proton-proton ($pp$) interaction region, a large-area silicon-strip detector located upstream of a vertical dipole magnet with a bending power of about $4{\mathrm{\,T\,m}}$, and three stations of silicon-strip detectors and straw drift tubes placed downstream of the magnet.
The minimum distance of a track to a primary $pp$ collision vertex (PV), the impact parameter (IP), is measured with a resolution of $(15+29/\pt)\mum$, where \pt is the component of the momentum transverse to the beam, in \gevc.
This exquisite performance is crucial for triggering on the displaced decay vertices of heavy-quark hadrons, rejecting the huge background of tracks originating in the PV.
The momentum of charged particles is measured with a relative uncertainty which varies from 0.5\% at low momentum to 1.0\% at $200 \gevc$, guaranteeing a mass resolution of the order of $8\mevcc$ for charmed-hadron decays into charged hadrons.
Together with the particle identification capabilities ensured by a system of two ring-imaging Cherenkov detectors, this resolution ensures an excellent signal-to-background ratio for most of the decays of interest.
Finally, the particle identification capabilities are complemented by a scintillating-pad and preshower detectors to distinguish photons and electrons, an electromagnetic and a hadronic calorimeter, and a muon detector composed of alternating layers of iron and multiwire proportional chambers.

The data sample collected to date corresponds to $1(2)\invfb$ of integrated luminosity of \proton\proton collisions at a centre-of-mass energy of $7(8)\tev$ accumulated during the \lhc \runone (2011--2012), and to $6\invfb$ at $13\tev$ accumulated during \runtwo (2015--2018).
The online event selection is performed by a hardware trigger, based on information from the calorimeter and muon systems, followed by a software trigger, which applies a full event reconstruction in two stages.
Because of the large production rate of charm decays in the \lhcb acceptance, of the order of $1\mhz$~\cite{LHCb-PAPER-2012-041,LHCb-PAPER-2015-041}, the amount of data that can be written to permanent storage is one of the factors that limit the number of decays that can be collected.
To mitigate this problem, since 2015 the alignment and calibration of the detector is performed in near real-time after the first stage of the software trigger~\cite{LHCb-PROC-2015-011}.
The results are then used in the second stage, ensuring offline data quality already at the trigger level.
This opens up the possibility to perform physics analyses directly using candidates reconstructed in the trigger, allowing only the triggered candidates to be stored to disk~\cite{LHCb-DP-2012-004,LHCb-DP-2016-001}.
The consequent reduction in the event size by one order of magnitude allows the rate at which data are collected to be increased, by loosening the trigger requirements.
This accounts for a large part of the increased yield of charmed hadrons at equal luminosity achieved during \runtwo compared to \runone.
Further gains are due to the higher charm production cross-section at larger centre-of-mass energy~\cite{LHCb-PAPER-2013-004,LHCb-PAPER-2016-042} and to the introduction of a new two-track line looking for displaced vertices at the first stage of the software trigger, which complements the single-track line looking for tracks with high momentum and IP used during \runone.

It should be emphasised that the collected yield of charmed hadrons, which are hereafter denoted as \D, depends critically on their decay topology and on the trigger requirements.
Contrary to charm and \B factories, tight selections on the momentum and IP of the final-state particles as well as on the \D flight distance are needed to distinguish the signal from the background of random combinations of light hadrons produced in the \proton\proton collision.
Moreover, the first-stage software trigger is optimised for collecting \B rather than \D hadrons, and \B hadrons are characterised by larger momentum and longer flight distance on average.
This often constitutes the bottleneck for the charm collection efficiency, which can be much smaller than unity especially at low decay times and for multibody final states, where the momentum and IP of the final-state particles are smaller on average than those of two-body decays.
For multibody decays, requirements on these variables can also provoke undesirable variations of the efficiency across the final-state phase space.
The efficiency is even smaller for final states including \KS mesons and hyperons, which often decay outside of the vertex tracker or even after the tracker upstream of the magnet.
Therefore, they are not reconstructed in the first-stage software trigger, which only relies on tracks producing a signal in the vertex tracker, or in the second case they are excluded also from the offline reconstruction as their momentum cannot be accurately measured.
Even lower efficiencies are achieved for neutral pions since, even when the two photons in which they decay can be distinguished in the calorimeter, the \piz mass resolution is limited to approximately 9\mevcc; besides, the neutral pion cannot be associated to a single decay vertex.
Tight momentum requirements are thus needed to minimise the combinatorial background~\cite{LHCb-DP-2014-002,LHCb-PAPER-2014-054}.
Finally, decay channels with a single \KL meson or neutrino could be reconstructed with approximate methods~\cite{Johns:1995jc}, but the unknown initial state of the \D hadron poses additional challenges to the separation from background with respect to charm and \B factories.
No measurements of these decays have been published to date.

\subsection{Measuring \CP asymmetries}
The most common observable employed to search for \CP violation is the time-dependent asymmetry between the decay widths of \decay{\D}{f} and \decay{\Db}{\bar{f}} decays,
\begin{equation}
    \label{eq:acp}
    A_{\CP}(f,t) \equiv \frac{\Gamma(\decay{\D}{f},t) - \Gamma(\decay{\Db}{\bar{f}},t)}
                             {\Gamma(\decay{\D}{f},t) + \Gamma(\decay{\Db}{\bar{f}},t)},
\end{equation}
where \D denotes a charmed hadron, and $t$ denotes its proper decay time.
In the case of charged mesons, \Dp or \Dsp, of baryons, or when the time-dependent contribution is irrelevant to the final result, all the quantities in \cref{eq:acp} are integrated over decay time.

This asymmetry is measured starting from the raw asymmetry between the yields of \D and \Db decays at decay time $t$.
However, this encompasses additional contributions, which will globally be referred to as ``nuisance asymmetries'', as follows,
\begin{equation}
    \label{eq:araw}
    \Araw{f,t} \approx A_{\CP}(f,t) + A_\textnormal{det}(f) + \Aprod{\D},
\end{equation}
where $A_\textnormal{det}(f)$ is the detection asymmetry of the final state, $A_\textnormal{prod}(\D)$ is the production asymmetry of \D hadrons, and terms of third order in the asymmetries are neglected.
Detection asymmetries arise from the interplay of several factors.
For a given magnet polarity, low-momentum particles of one charge at large or small emission angles in the horizontal plane may be deflected out of the detector or into the uninstrumented beam pipe, whereas particles with the opposite charge are more likely to remain within the acceptance.
This effect is cancelled to a large extent by periodically reversing the polarity of the magnet.
Smaller residual asymmetries remaining after the averaging are due to right-left misalignment of detector elements and to the small shift of the collision point with respect to the symmetry axis of the detector; to different beam-beam crossing angles for opposite magnet polarities; and to variations of the detection efficiency over time.
The different interaction cross-section of positively and negatively charged kaons and pions with matter plays a role as well, especially when the selection at the hardware-trigger level is based on the information from the hadronic calorimeter.
On the other hand, the production asymmetry is due to the asymmetric hadronisation of \cquark\cquarkbar pairs into the different species of charmed hadrons and anti-hadrons, since the \proton\proton initial state is not self-conjugate.

Nuisance asymmetries are not reproduced by simulation with the required level of precision, and must be corrected for through methods based on collected data.
This is usually done by measuring the difference between the raw asymmetry of the SCS of interest with one or more CF decays which share the same nuisance asymmetries, but the dynamical asymmetries of which are expected to be negligible; or with SCS decays whose \CP asymmetry is known with better precision.

Finally, when analysing decays into a final state that is shared by \Dz and \Dzb mesons, their initial flavour can be identified only through their production mechanism.
This is done by measuring the charge of the accompanying particle in strong \decay{\theDstarp}{\Dz\pip} decays or in inclusive \decay{\B}{\Dzb\mup X} decays, where \B stands for a \bquark hadron and $X$ for an arbitrary set of unreconstructed particles.
Hereafter the \theDstarp meson is referred to as \Dstarp, and the two tagging categories as \Dstarp-tagged and \mun-tagged decays.
Both categories introduce a further detection asymmetry due to the tagging particle, \pip or \mun, in \cref{eq:araw}.
This asymmetry is corrected for in the same way as the others.
The \Dstarp-tagged sample is larger by around a factor of three and is purer, thanks to the larger production cross-section of charm with respect to beauty hadrons and to the low $Q$ value\footnote{The $Q$ value of a decay is defined as the energy released in the decay, and is equal to the mass of the decaying particle minus the sum of the masses of its decay products.} of the \Dstarp decay.
The low $Q$ value ensures excellent mass resolution for the \Dstarp meson, and causes the pion momentum to form a very small angle with that of the \Dz meson, thus reducing the combinatorial background.
On the other hand, this small angle implies a poor resolution on the \Dstarp decay vertex, which is therefore constrained to originate in the PV to achieve the best possible resolutions on the \Dstarp mass, around $0.5\mevcc$, and on the \Dz decay time~\cite{Hulsbergen:2005pu}.
The resolution on decay time, approximately $40\ps$, corresponds to around 0.1 \Dz lifetimes and is a factor of three better than that of the \mun-tagged sample.
However, the constraint biases the measured decay time of \Dz from secondary \Dstarp mesons that are produced in \B decays to larger values.
Therefore, secondary mesons are treated as a background in most measurements.
This background can be troublesome, as the production asymmetry of secondary mesons differs from that of the prompt ones; moreover, their fraction increases with decay time.
On the other hand, \mun-tagged candidates benefit from looser trigger requirements, allowing a larger reconstruction efficiency at low \Dz decay times as well as a flatter efficiency across the final-state phase space in multibody decays to be obtained, although at the cost of increased background.
This background can be reduced by using doubly tagged \decay{\B}{\Dstarm(\decay{}{\Dzb\pim})\mup X} decays, but the resulting yield is considerably smaller.

\section{Searches for \CP violation in the decay}
\label{sect:cpv-decay}
The next sections review the most recent \lhcb searches for \CP violation in the decay, starting from the historic first observation of \CP violation achieved in 2019.

\subsection{Observation of \CP violation in \DzHH decays}
\label{sect:observation}

\CP violation was observed for the first time in charm decays by the \lhcb collaboration in 2019, through a measurement of the difference between the time-integrated \CP asymmetries of \DzKK and \DzPP decays, $\Delta A_{\CP}$~\cite{LHCb-PAPER-2019-006,Betti:2669175}.
This is a very convenient observable, since nuisance asymmetries cancel in the difference, whereas the \CP asymmetries in the decay are expected to have opposite signs and their magnitudes add up; see \cref{eq:cp-asymm-parametrisation}.
The contribution to $\Delta A_{\CP}$ from time-dependent \CP violation also tends to cancel in the difference, as the time-dependent asymmetry of the \Dz and \Dzb decay rates into a \CP-even final state $f$ is equal to
\begin{equation}
    \label{eq:acp-expansion}
    A_{\CP}(f,t) \equiv \frac{\Gamma(\decay{\Dz}{f},t) - \Gamma(\decay{\Dzb}{f},t)}
                             {\Gamma(\decay{\Dz}{f},t) + \Gamma(\decay{\Dzb}{f},t)}
    \approx \Acpdec{f} + \DY{f} \frac{t}{\tauDz},
\end{equation}
where terms of order higher than two in the mixing parameters are neglected, \Acpdec{f} is the \CP asymmetry in the decay, $\tauDz$ is the \Dz lifetime, and the expression of the parameter \DY{f} in terms of the theoretical mixing parameters is given in \cref{sect:dy}.\footnote{The parameter \DY{f} is equal to the negative of the parameter \agamma{f} sometimes used in the literature, defined as the asymmetry of the effective decay widths of \Dz and \Dzb mesons into the final state $f$, up to a multiplicative factor which differs from unity by less than 1\%~\cite{LHCb-PAPER-2020-045,Pajero:2021jev}.}
The time-integrated \CP asymmetry is therefore equal to
\begin{equation}
    \label{eq:acp-f}
    A_\CP(f) \approx \Acpdec{f} + \DY{f}\frac{\langle t \rangle_{f}}{\tauDz},
\end{equation}
where \mbox{${\langle t \rangle}_{f}$} is the average decay time of the analysed sample, and depends on the selection requirements.
Assuming that \DY{f} is independent of the final state, see \cref{sect:dy}, and denoting it with \DY{}, the following relation holds,
\begin{equation}
    \label{eq:dacp-correction}
    \Delta A_{\CP} \approx \Acpdec{\KK} - \Acpdec{\PP} + \DY{} \;\frac{\langle t \rangle_{\KK} - \langle t \rangle_{\PP}}{\tauDz}.
\end{equation}
The contribution to $\Delta A_{\CP}$ from time-dependent \CP violation is very small, since the average decay times of the selected samples of \DzKK and \DzPP decays differ by less than $0.15\,\tauDz$, and \DY{} is measured to be consistent with zero with a precision close to $10^{-4}$~\cite{LHCb-PAPER-2014-069,LHCb-PAPER-2016-063,LHCb-PAPER-2019-032,LHCb-PAPER-2020-045}.

\begin{figure}[bt]
  \begin{center}
    \includegraphics[width=0.45\textwidth]{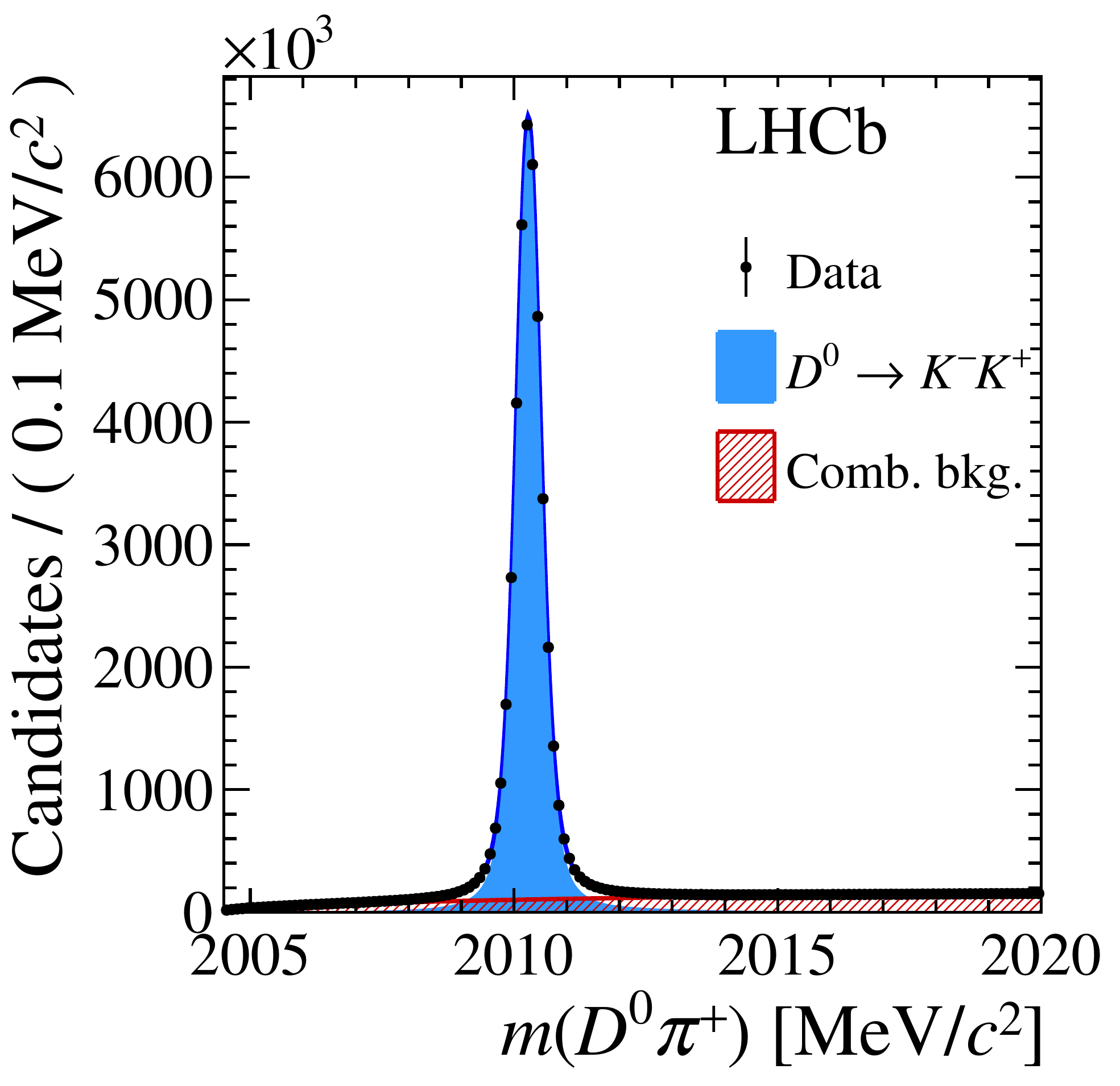}
    \includegraphics[width=0.45\textwidth]{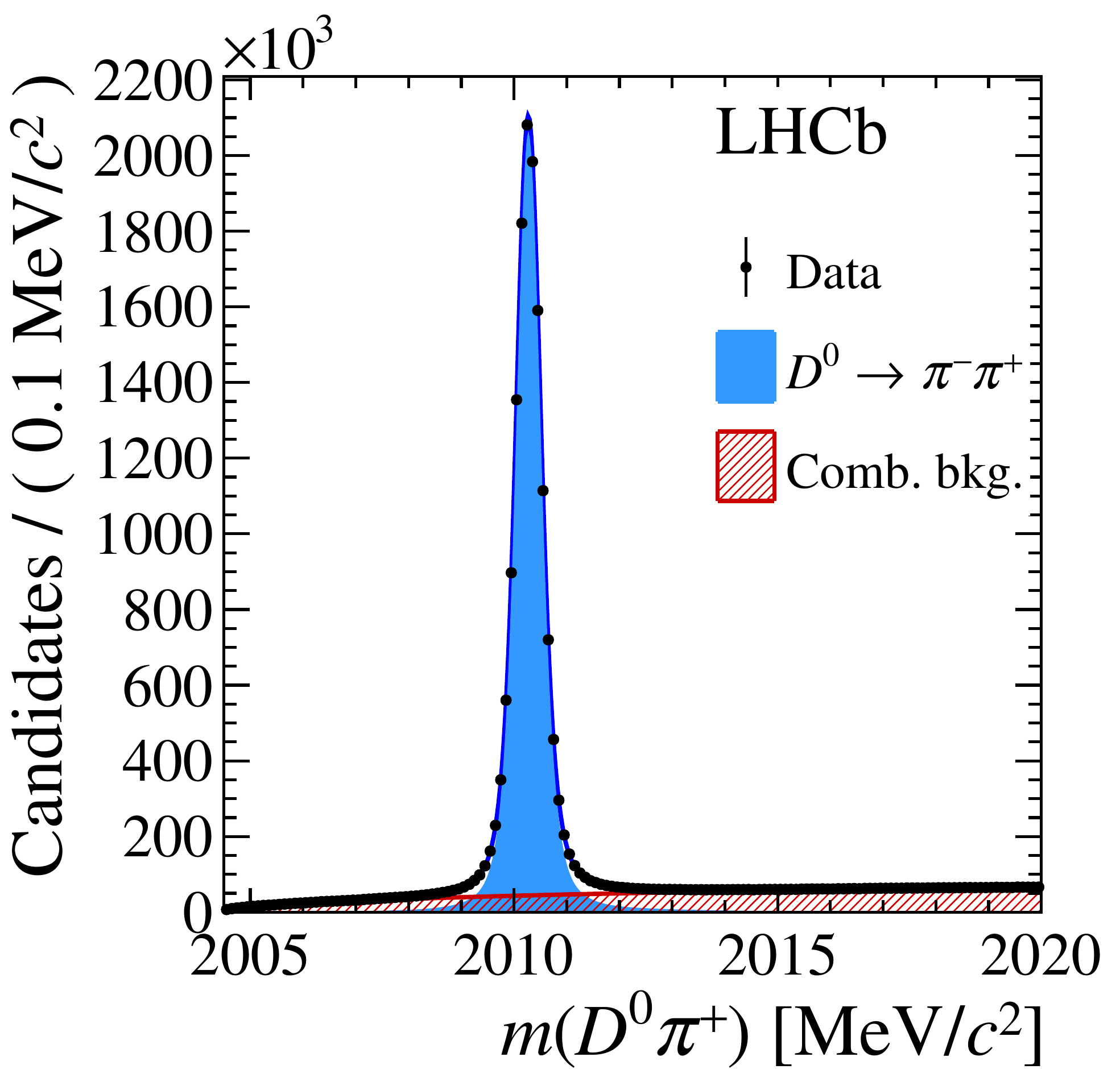}
  \end{center}
  \vspace*{-4mm}
  \caption{
    Invariant-mass distribution of the \Dstarp candidates yielding the first observation of \CP violation in charm decays in the (left) \DzKK and (right) \DzPP final states.
    The \mun-tagged sample is considerably smaller.
    Figures taken from ref.~\cite{LHCb-PAPER-2019-006}.
  }
  \label{fig:dacp}
\end{figure}

The observation of \CP violation is based on the \runtwo data sample, including both \Dstarp and \mun tagged candidates, which corresponds to around 53 million \DzKK and 17 million \DzPP decays.
The mass distributions of the \Dstarp-tagged candidates, which constitute the vast majority of the sample, are shown in \cref{fig:dacp}.
The result is
\[
  \Delta A_{\CP} = (-1.82 \pm 0.32 \pm 0.09) \times 10^{-3},
\]
where the systematic uncertainty is significantly smaller than the statistical one, thanks to the cancellation of most systematic effects in the difference, and is expected to be reducible when larger samples will become available.
A combination with the result of the \mun-tagged sample and with previous determinations~\cite{LHCb-PAPER-2014-013,LHCb-PAPER-2015-055}, including a small correction for a residual contribution from time-dependent \CP violation~\cite{LHCb-PAPER-2014-069,LHCb-PAPER-2016-063,LHCb-PAPER-2019-032}, see \cref{eq:dacp-correction}, yields
\[
  \Acpdec{\KK} - \Acpdec{\PP} = (-1.57 \pm 0.29) \times 10^{-3},
\]
which is inconsistent with the hypothesis of \CP symmetry at the level of 5.3 standard deviations.

This value is larger by a factor of five than perturbative estimates~\cite{Grossman:2006jg,Bigi:2011re,Grossman:2019xcj,Nierste:2020eqb} and predictions from light-cone sum rules~\cite{Khodjamirian:2017zdu,Chala:2019fdb}.
While its unexpectedly large size has prompted several BSM interpretations~\cite{Chala:2019fdb,Dery:2019ysp,Bause:2020obd}, other authors have claimed that it can be explained within the SM through a mild nonperturbative enhancement of $A_b$ due to final-state interactions~\cite{Franco:2012ck,Grossman:2019xcj,Cheng:2019ggx}.
Explicit models of such an enhancement, which may also explain the factor-of-three difference between the branching fractions of \DzKK and \DzPP decays without invoking large $U$-spin breaking effects~\cite{Brod:2012ud}, have been proposed in refs.~\cite{Schacht:2021jaz,Bediaga:2022sxw}.

Progress of \emph{ab initio} theoretical predictions is ultimately needed to rigorously assess the compatibility of the measurement with the SM.
Meanwhile, additional measurements of \CP asymmetries and of poorly known branching fractions of other decay channels, as well as of the scalar resonances that might be responsible for rescattering,\footnote{For example, it would be important to measure the branching fractions of the $f_0(1710)$ and $f_0(1790)$ resonances into the \PP and \KK final states~\cite{Schacht:2021jaz}.
In addition, if large rescattering through these resonances is responsible for the enhancement of the \CP asymmetries of \DzKK and \DzPP decays, the same mechanism could enhance also other observables like the branching fraction of $\decay{\Dz}{\gamma\gamma}$ decays~\cite{Fajfer:2001ad,Burdman:2001tf,Belle:2015pzk}.} would constitute an invaluable tool to pin down the nature of the nonperturbative effects at play, and to test available models~\cite{Li:2012cfa,Buccella:2019kpn,Cheng:2019ggx,Schacht:2021jaz}.\footnote{Note, however, that global fits to charm branching ratios and \CP asymmetries~\cite{Bhattacharya:2009ps,Pirtskhalava:2011va,Cheng:2012wr,Bhattacharya:2012ah,Li:2012cfa,Cheng:2012xb,Hiller:2012xm,Muller:2015lua,Buccella:2019kpn,Cheng:2019ggx}
are not necessarily able to distinguish the cases where \CP violation originates from the SM or requires the presence of BSM interactions, and the validity of the assumptions of some of these fits has recently been challenged~\cite{Bigi:2015cja,He:2018php,He:2018joe,Bhattacharya:2021ndt}.
}

In particular, an individual measurement of \Acpdec{\KK} or \Acpdec{\PP} would provide useful information on the size of $U$-spin breaking in $A_b$.
Since their difference is already measured through the $\Delta A_{\CP}$ observable, only $A_{\CP}(\KK)$ is directly measured and $A_{\CP}(\PP)$ is derived indirectly.
Measuring both asymmetries individually would not make an improvement, because their precision is limited by the correction for production and detection asymmetries, which would be the same for both decay channels (whereas such a correction is not needed for the measurement of $\Delta A_{\CP}$).

A very recent measurement determines $A_{\CP}(\KK)$ using the \runtwo \mbox{\Dstarp-tagged} data sample~\cite{LHCb-PAPER-2022-024,Maccolini:2824361}.
The correction for the nuisance asymmetries relies on multiple subtractions of raw asymmetries using the following CF decay channels,
\begin{equation}
    \label{eq:dp-corr}
    \begin{aligned}
  A_{\CP}(\KK) \approx &\, \Araw{\DzKK} - \Araw{\DzRS} \\
                       & + \Araw{\decay{\Dp}{\Km\pip\pip}} - \Araw{\decay{\Dp}{\KS\pip}} 
                         + \Adet{\Kz},
    \end{aligned}
\end{equation}
or
\begin{equation}
    \label{eq:dsp-corr}
    \begin{aligned}
  A_{\CP}(\KK) \approx &\,\Araw{\DzKK} - \Araw{\DzRS} \\
                      & + \Araw{\decay{\Dsp}{\Km\Kp\pip}} - \Araw{\decay{\Dsp}{\KS\Kp}} 
                      + \Adet{\Kz},
    \end{aligned}
\end{equation}
where the detection asymmetry of the neutral kaon includes a contribution from \CP violation in \Kz mixing and is calculated explicitly~\cite{Stahl:1997600,Yu:2017oky}.
The method based on \cref{eq:dsp-corr} had never been used before, and allows the statistical uncertainty to be improved by 37\% at equal integrated luminosity.
In fact, for both methods, before measuring the raw asymmetries, the momentum distributions of the \D mesons and of the final-state particles must be aligned by assigning per-candidate weights.
This ensures a proper cancellation of the nuisance asymmetries, which depend on kinematics.
However, due to the different topologies and number of final-state particles of these decays, the statistical uncertainty is significantly degraded by the weighting and is eventually limited by the \decay{\Dp}{\KS\pip} or \decay{\Dsp}{\KS\Kp} channels.
Adding the new decay chain in \cref{eq:dsp-corr} thus allows the statistical uncertainty to be significantly reduced even at equal \DzKK yield.

\begin{figure}[tb]
  \begin{center}
    \includegraphics[width=0.65\textwidth]{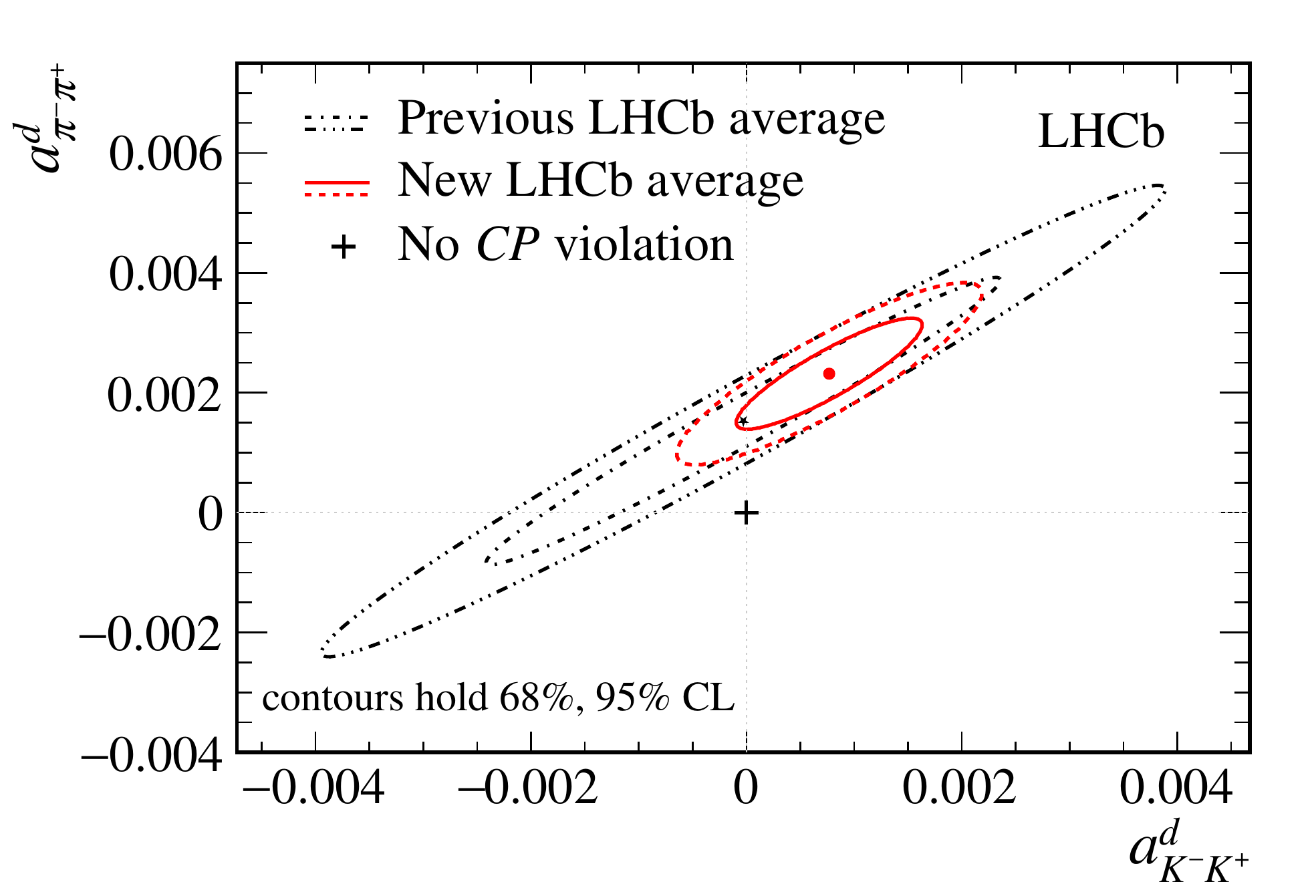}
  \end{center}
  \vspace*{-4mm}
  \caption{
    Best-fit values and two-dimensional confidence regions in the $(\Acpdec{\KK},\Acpdec{\PP})$ plane for the combinations of the \lhcb results obtained with the new $A_{\CP}(\KK)$ measurement and without.
    Figure taken from ref.~\cite{LHCb-PAPER-2022-024}.
  }
  \label{fig:acp-improvement}
\end{figure}

The average of the results obtained with the methods in \cref{eq:dp-corr,eq:dsp-corr} is
\begin{equation*}
    A_{\CP}(\DzKK) = (6.8 \pm 5.4 \pm 1.6) \times 10^{-4},
\end{equation*}
where the systematic uncertainty is, again, limited by the size of the calibration samples and could in principle be reduced as their size increases.
Contrary to $\Delta A_{\CP}$, in this case the contribution from time-dependent \CP violation in \cref{eq:acp-f} cannot be neglected, as the average decay time of the collected \DzKK decays is approximately equal to $1.7\,\tauDz$.
Therefore, the result is combined with previous determinations of $A_{\CP}(\KK)$~\cite{LHCb-PAPER-2014-013,LHCb-PAPER-2016-035}, $\Delta A_{\CP}$~\cite{LHCb-PAPER-2014-013,LHCb-PAPER-2015-055,LHCb-PAPER-2019-006} and \DY{}~\cite{LHCb-PAPER-2014-069,LHCb-PAPER-2016-063,LHCb-PAPER-2019-032,LHCb-PAPER-2020-045} to determine the \CP asymmetries in the decay \Acpdec{\KK} and \Acpdec{\PP}; see \cref{fig:acp-improvement}.
The numerical results are
\[
    \begin{aligned}
    \Acpdec{\KK} &= (\phantom{0}7.7 \pm 5.7) \times 10^{-4}, \\
    \Acpdec{\PP} &= (          23.2 \pm 6.1) \times 10^{-4},
    \end{aligned}
\]
where the uncertainties include systematic and statistical contributions and the correlation coefficient is equal to 0.88.
The results are consistent with the SM expectations for the size of $U$-spin breaking~\cite{Franco:2012ck,Cheng:2019ggx,Bediaga:2022sxw,Schacht:2022kuj}, and the second shows a departure from zero at the level of 3.8 standard deviations.

Larger data samples from \runthree and beyond are needed to establish a first observation of \CP violation in either of the two individual decay channels, and to assess the level of $U$-spin breaking in $A_b$.
In the mean time, measurements of other decay channels can provide complementary information on the origin of \CP violation in charm, as discussed in the next sections.

\subsection{Search for \CP violation in \decay{\Dz}{\KS\KS} decays}
\label{sect:ksks}
The size of \CP violation could be larger in \decay{\Dz}{\KS\KS} decays, where only exchange and penguin-annihilation diagrams that vanish in the $U$-spin limit contribute to $A_{sd}$, while the \CP-violating contributions to $A_b$ from the same diagrams do not cancel out.
Therefore, the \CP asymmetry in the decay might be as large as 1\%~\cite{Brod:2011re,Nierste:2015zra}, even if somewhat smaller values are favoured by most models~\cite{Li:2012cfa,Buccella:2019kpn,Cheng:2019ggx}.

The \lhcb collaboration has recently measured the time-integrated \CP asymmetry of this decay mode employing the \runtwo \Dstarp-tagged data sample~\cite{LHCb-PAPER-2020-047,Tuci:2765102}.
The \KS candidates are reconstructed in the $\pip\pim$ final state, either from tracks that generated signals in all the tracking stations including the vertex detector (if the \KS meson decayed early enough), or otherwise from the two trackers immediately upstream and downstream of the magnet only.
The two categories are named \emph{long} and \emph{downstream} and are labelled ``L'' and ``D'', respectively.
The \Dz candidates are classified accordingly into three categories: LL, LD and DD.
Downstream \KS candidates have a geometrical acceptance larger than that of long candidates by a factor of two, but they are not selected by the first-stage software trigger.
As a consequence, DD candidates are selected only if the first-stage trigger has been activated by unrelated tracks in the event, and they are fewer than LL and LD candidates.
Moreover, the \Dz-mass and decay-vertex resolutions are degraded for candidates containing downstream kaons.
This reduces the capability to distinguish prompt mesons from secondary mesons.
Therefore, contrary to most charm measurements, no requirements are applied to reject the latter category.

\begin{figure}[tb]
    \begin{center}
        \includegraphics[width=0.48\textwidth]{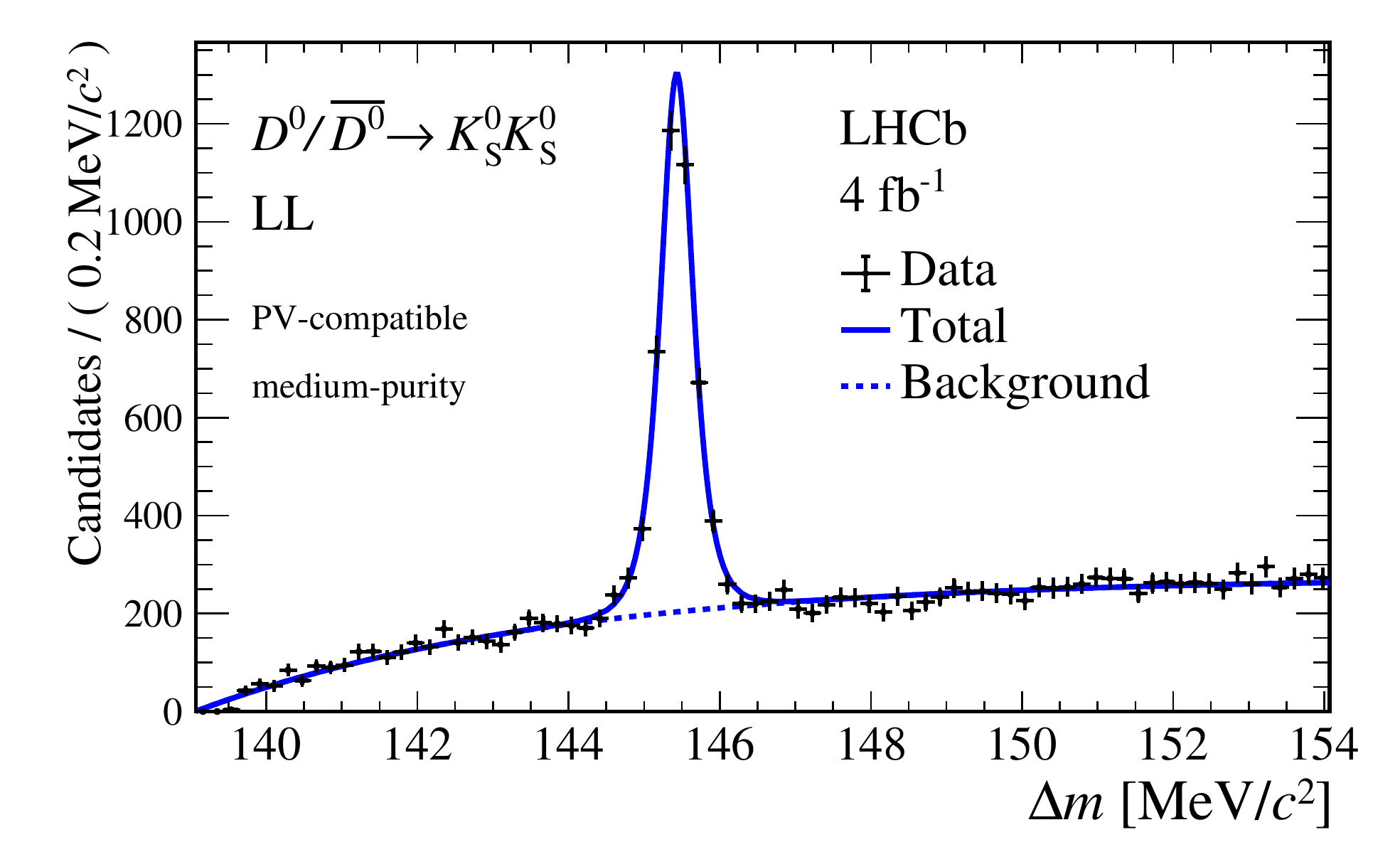}
        \includegraphics[width=0.48\textwidth]{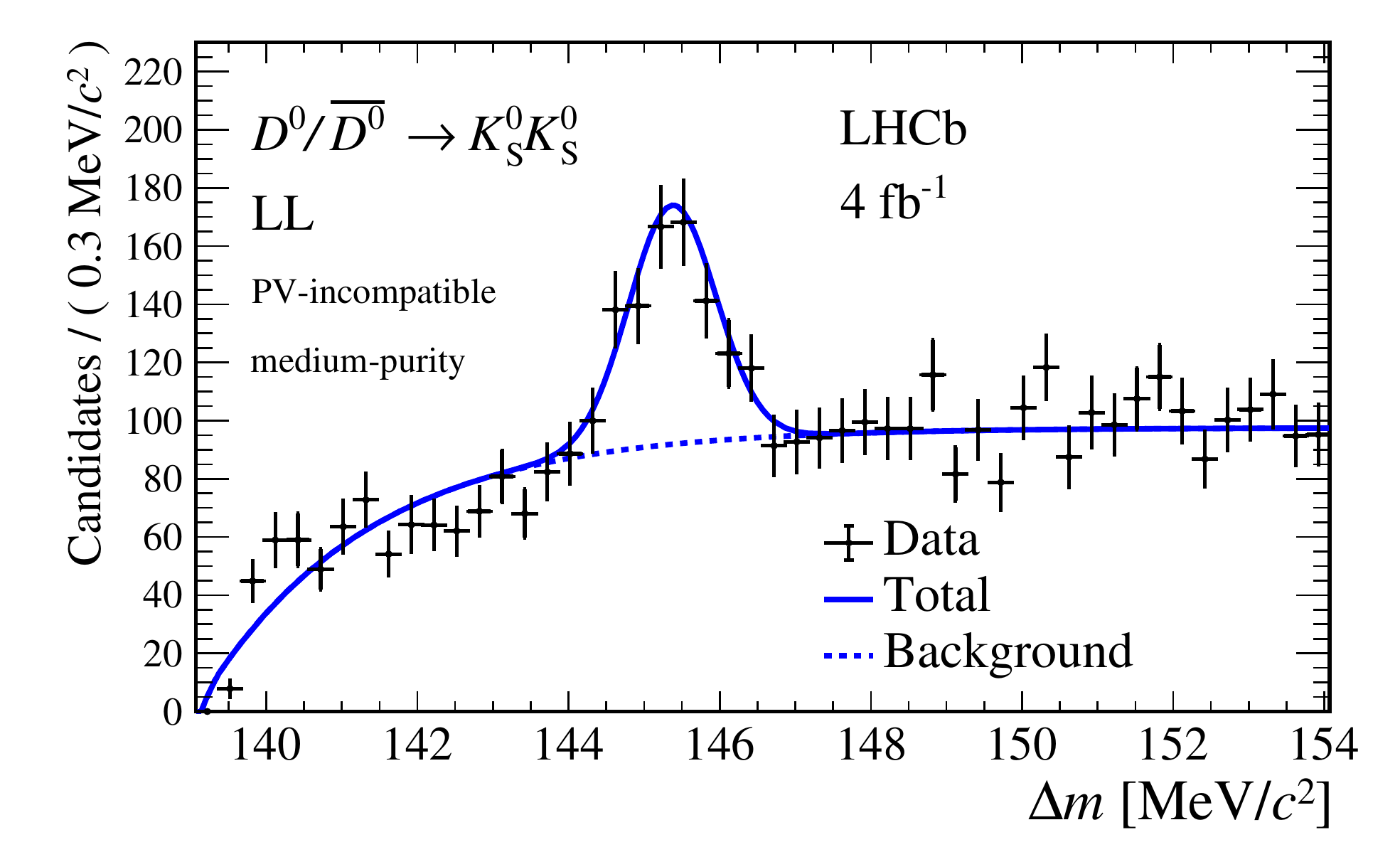} \\
        \includegraphics[width=0.48\textwidth]{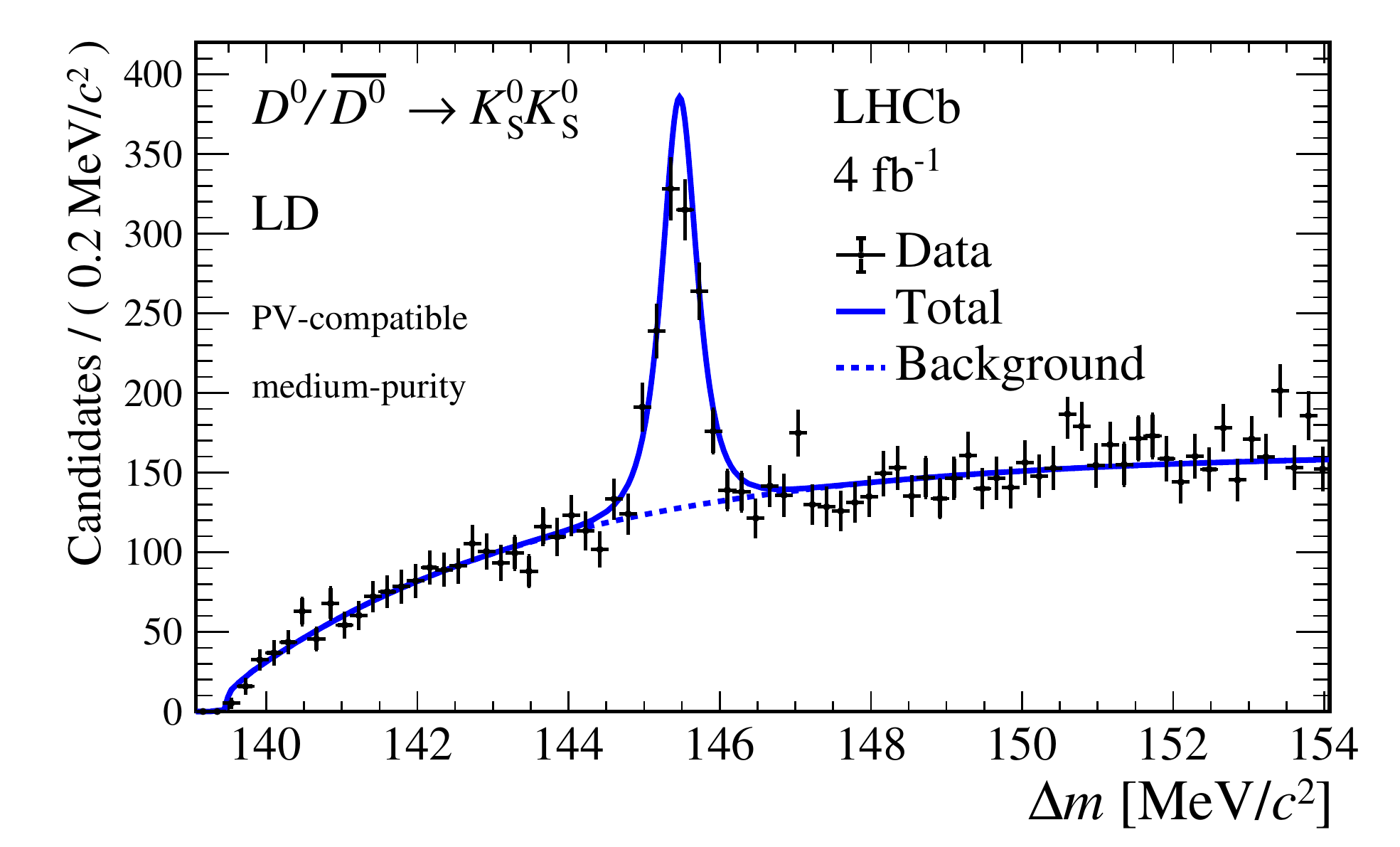}
        \includegraphics[width=0.48\textwidth]{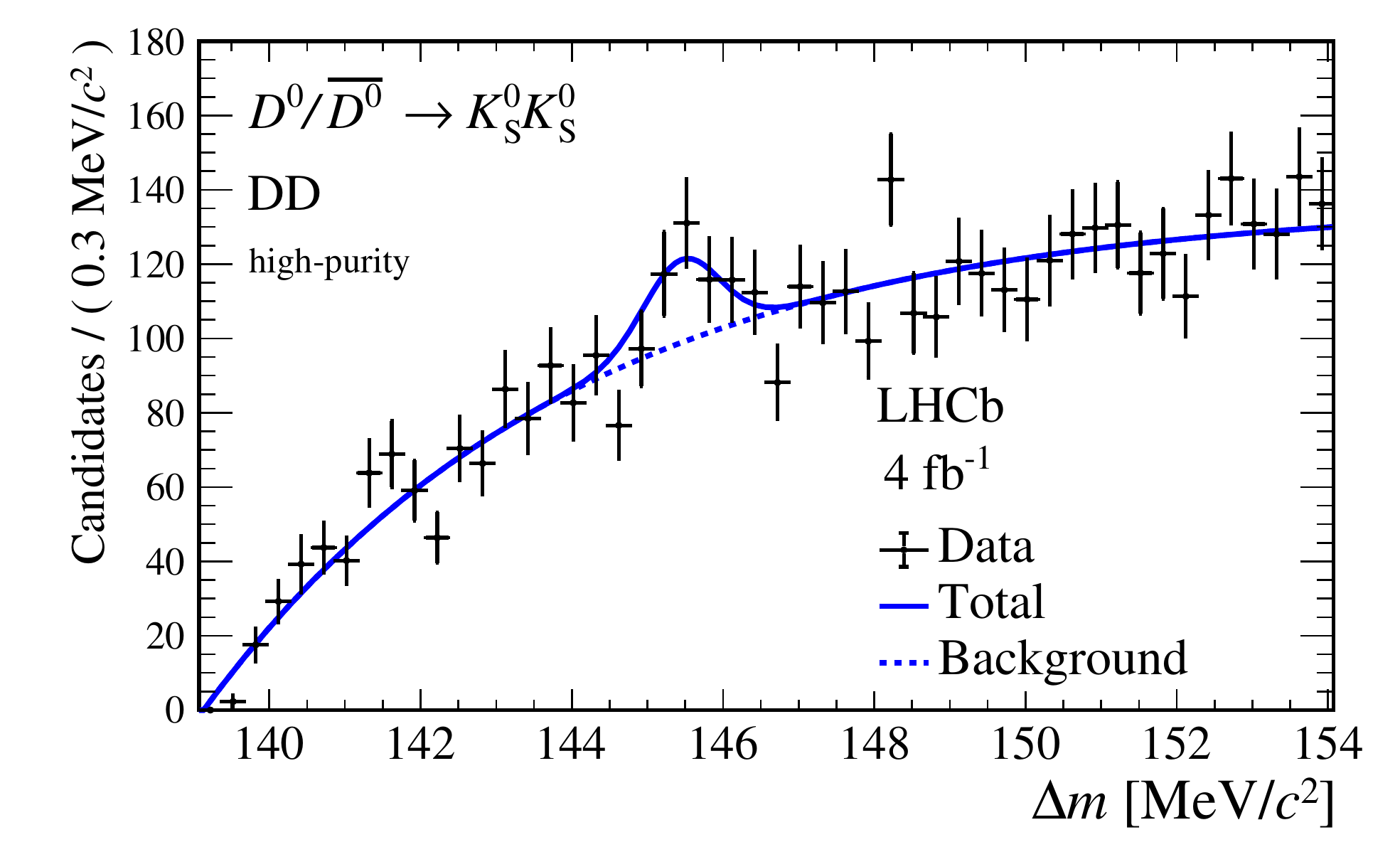}
    \end{center}
    \vspace*{-4mm}
    \caption{
        Distribution of the difference of the \Dstarp and \Dz invariant masses, for the most significant \decay{\Dz}{\KS\KS} categories.
        Fit projections are overlaid.
        Figure taken from ref.~\cite{LHCb-PAPER-2020-047}.
    }
    \label{fig:ksks}
\end{figure}

To maximise the statistical precision, the total sample is further subdivided into nine categories, based on the compatibility of the \Dz candidate with originating from the PV, and on the output of a multivariate classifier trained to reject combinatorial background.
This allows some categories to benefit from a better signal-to-background ratio; see \cref{fig:ksks}.
The nuisance asymmetries are removed by subtraction with the raw asymmetry of \DzKK decays, weighting each \Dz candidate with momentum $\vec{p}$ by
\begin{equation}
    w^\pm(\vec{p}) = \frac{n^+_{K^+ K^-}(\vec{p}) + n^-_{K^+ K^-}(\vec{p})}{2 n^\pm_{K^+ K^-}(\vec{p})}
                     \big[1 \pm A_{CP}(D^0 \to K^+ K^-)\big],
\end{equation}
where the plus (minus) sign applies to \Dz (\Dzb) candidates and $n^\pm_{K^+ K^-}(\vec{p})$ is the local density of \Dz (\Dzb) mesons with momentum $\vec{p}$ decaying into the \KK final state, as estimated with a multivariate classifier.
The selection requirements of the signal and control samples are aligned to ensure an effective cancellation of the detection asymmetry of the tagging pion and of the production asymmetries of prompt and secondary \Dstarp mesons.

Finally, the asymmetry is measured through a simultaneous fit to the weighted distributions of the invariant masses of the two \KS mesons and to the difference of the invariant masses of the \Dstarp and \Dz mesons.
The results for the nine categories are compatible with each other, and are combined to yield
\[
    A_{\CP}(\decay{\Dz}{\KS\KS}) = (-3.1 \pm 1.2 \pm 0.4 \pm 0.2)\%,
\]
where the third uncertainty is due to the precision with which the \CP asymmetry of the control channel is known~\cite{LHCb-PAPER-2016-035},
and the systematic uncertainty is dominated by the uncertainty on the shape of the mass distributions and by the statistical uncertainty on the weighting, which are both expected to decrease with future larger data samples.
This result supersedes that of ref.~\cite{LHCb-PAPER-2018-012} and is compatible, but more precise, than previous determinations~\cite{CLEO:2000opx,LHCb-PAPER-2015-030,Dash:2017heu}.
The new world average, $A_{\CP}(\decay{\Dz}{\KS\KS}) = (-1.9 \pm 1.0)\%$, is compatible with the absence of \CP violation within 1.9 standard deviations, with a precision equal to the upper edge of the SM predictions.

\subsection{Search for \CP violation in \decay{\DpOrDsp}{\hzero\hp} decays}
\label{sect:hh0}
While predicting the absolute size of \CP violation in charm in the SM is a formidable challenge, one can easily derive \SUF-based sum rules relating its size in different decay channels~\cite{Grossman:2012eb,Grossman:2012ry,Muller:2015rna,Gavrilova:2022hbx}.
These rules might be violated by new interactions beyond the level expected in the SM, where the size of \SUF breaking is around 30\%~\cite{Brod:2012ud,Chala:2019fdb}.
Another testable feature of the SM is that QCD penguin diagrams contribute only to $\Delta I = 1/2$ transitions, whereas $\Delta I = 3/2$ transitions are allowed only at tree-level and therefore cannot give rise to \CP violation~\cite{Grossman:2012eb}.
By measuring the branching ratios and \CP asymmetries of decays sharing the same isospin amplitudes, such as \decay{\Dz}{\pip\pim} and \decay{\Dz}{\piz\piz}, one can determine whether the individual \CP asymmetries can be interpreted only in terms of $\Delta I = 1/2$ amplitudes, or if they require the presence of $\Delta I = 3/2$ contributions from BSM interactions.
The \decay{\Dp}{\pip\piz} decay is particularly interesting as its \CP asymmetry is expected to be smaller than $10^{-5}$, even if it is SCS, since it is a pure $\Delta I = 3/2$ transition.
Unfortunately, the precision of all these tests is limited by decay channels involving neutral particles, which are reconstructed with low efficiency at hadron colliders.

The \lhcb collaboration has recently measured the \CP asymmetries of \decay{\DpOrDsp}{\hp\hzero} decays, where \hp stands for a \pip or \Kp meson, and \hzero for a \piz, \etaz or \etapr meson, with the data sample collected in \runonetwo~\cite{LHCb-PAPER-2021-001,LHCb-PAPER-2021-051}.
For the first time, \piz and \etaz mesons are reconstructed through Dalitz decays, \decay{\hzero}{\ep\en\gamma}, or two-photon decays where one of the photons converts into an electron-positron pair within the vertex detector.
The latter sample is larger by a factor of six.
Both decay chains allow triggering on the displaced \D-meson decay vertex, which would be impossible to reconstruct using bare two-photon decays.
While these decay chains account for only a small fraction of the total decays, this suppression is counterbalanced by the large charmed-hadron production cross-section with respect to \B factories, which allows for results to be obtained that are equally or more precise.
On the other hand, the signal-to-background ratio is lower than in other hadronic charm decays, and the measurement requires a careful correction for electron and positron bremsstrahlung in the magnetic field.
For \hp equal to \pip, the \etaz mesons are additionally reconstructed in the $\gamma\pip\pim$ final state, achieving a similar precision.
The same final state is used for the \etapr in \decay{\DpOrDsp}{\etapr\pip} decays, too.

\begin{figure}[h!]
    \begin{center}
        \includegraphics[width=0.31\textwidth]{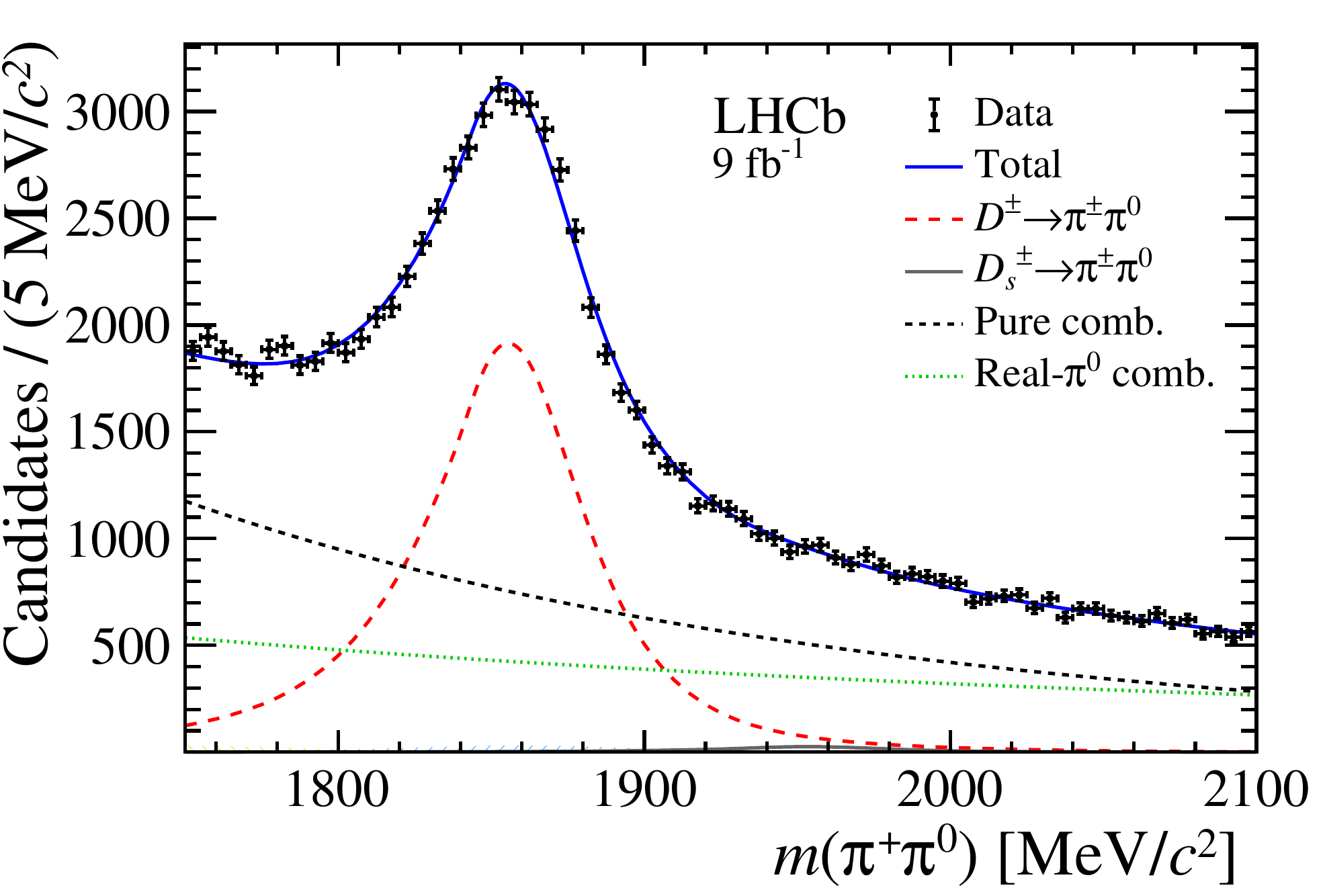}
        \includegraphics[width=0.31\textwidth]{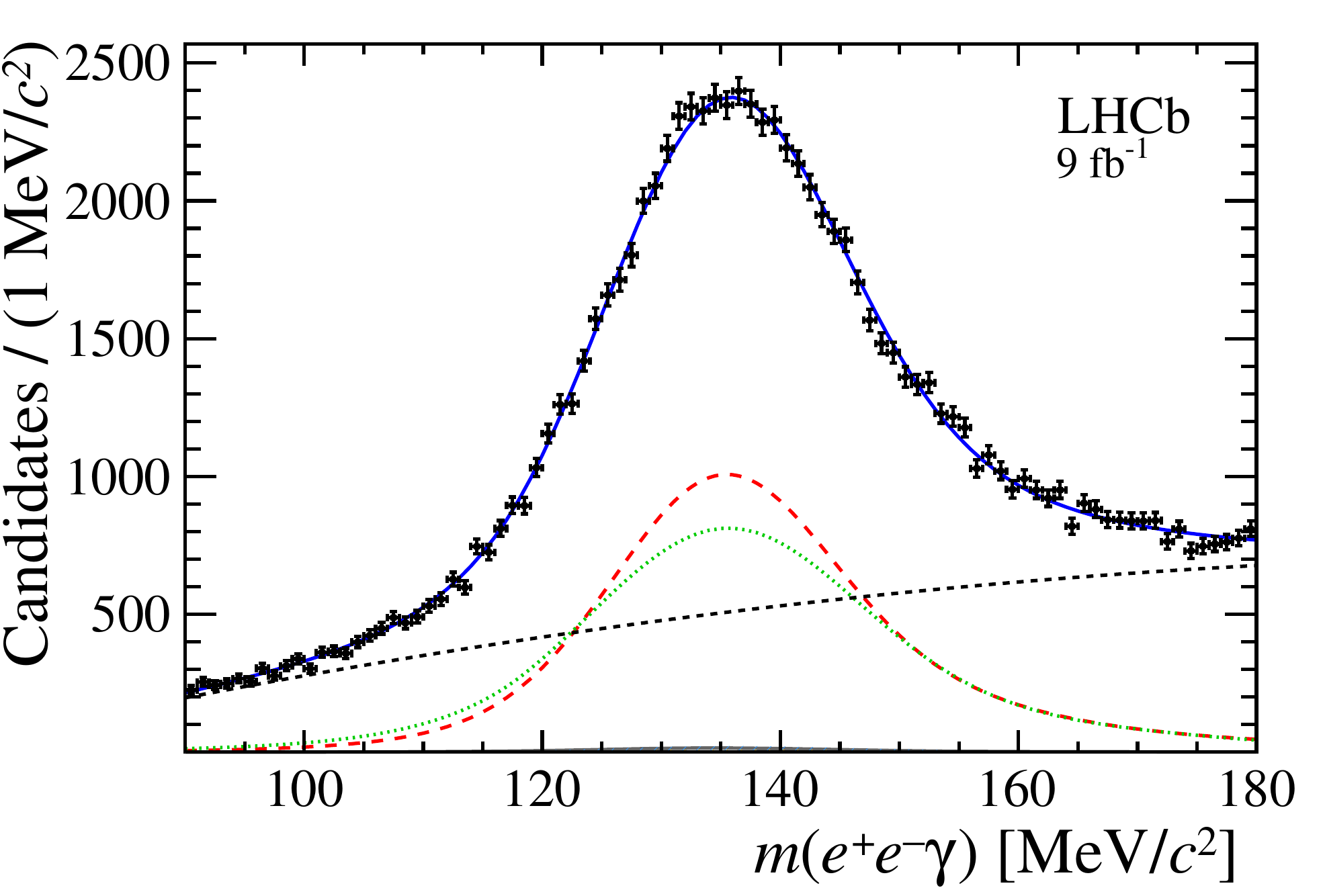}\\
        \includegraphics[width=0.31\textwidth]{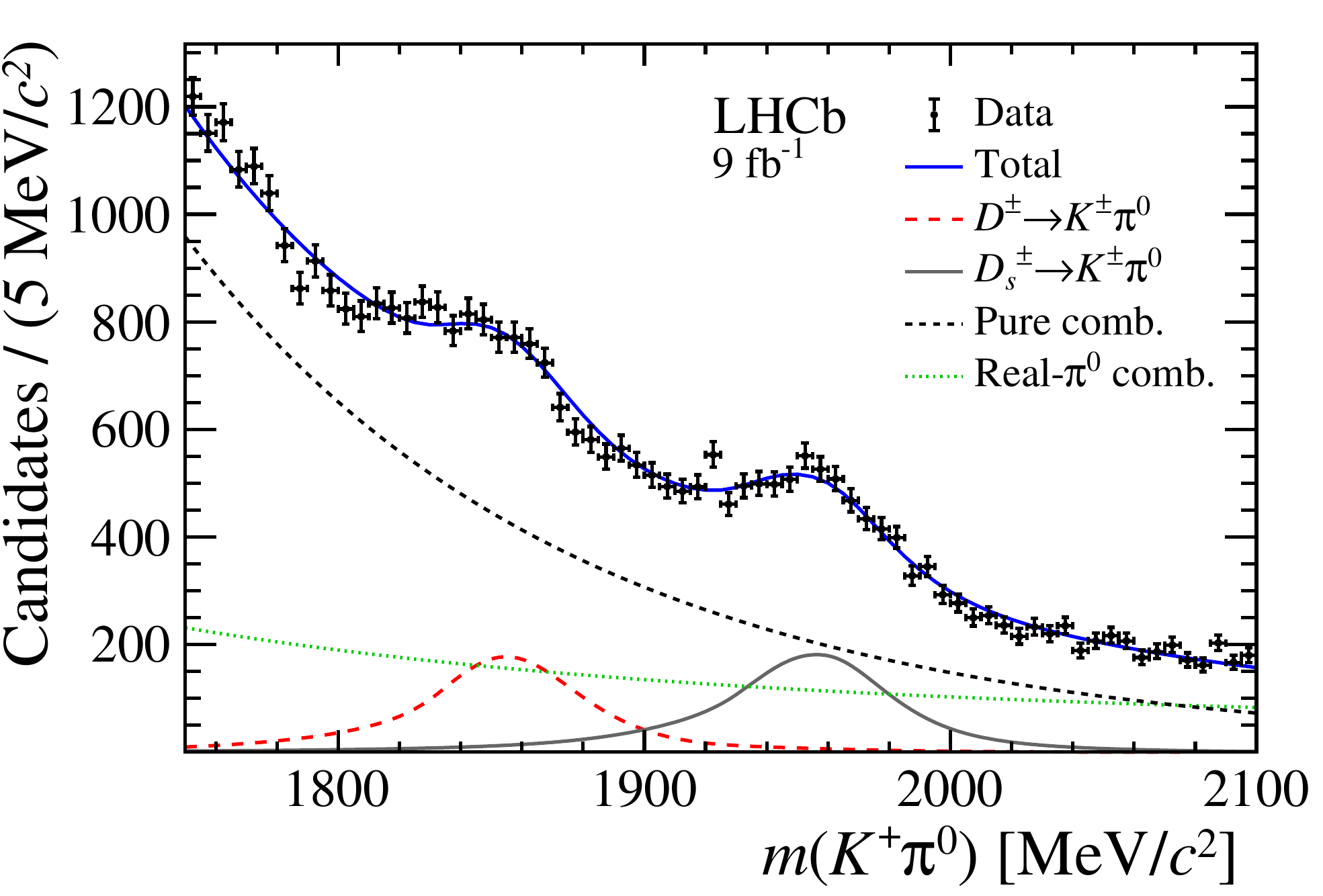}
        \includegraphics[width=0.31\textwidth]{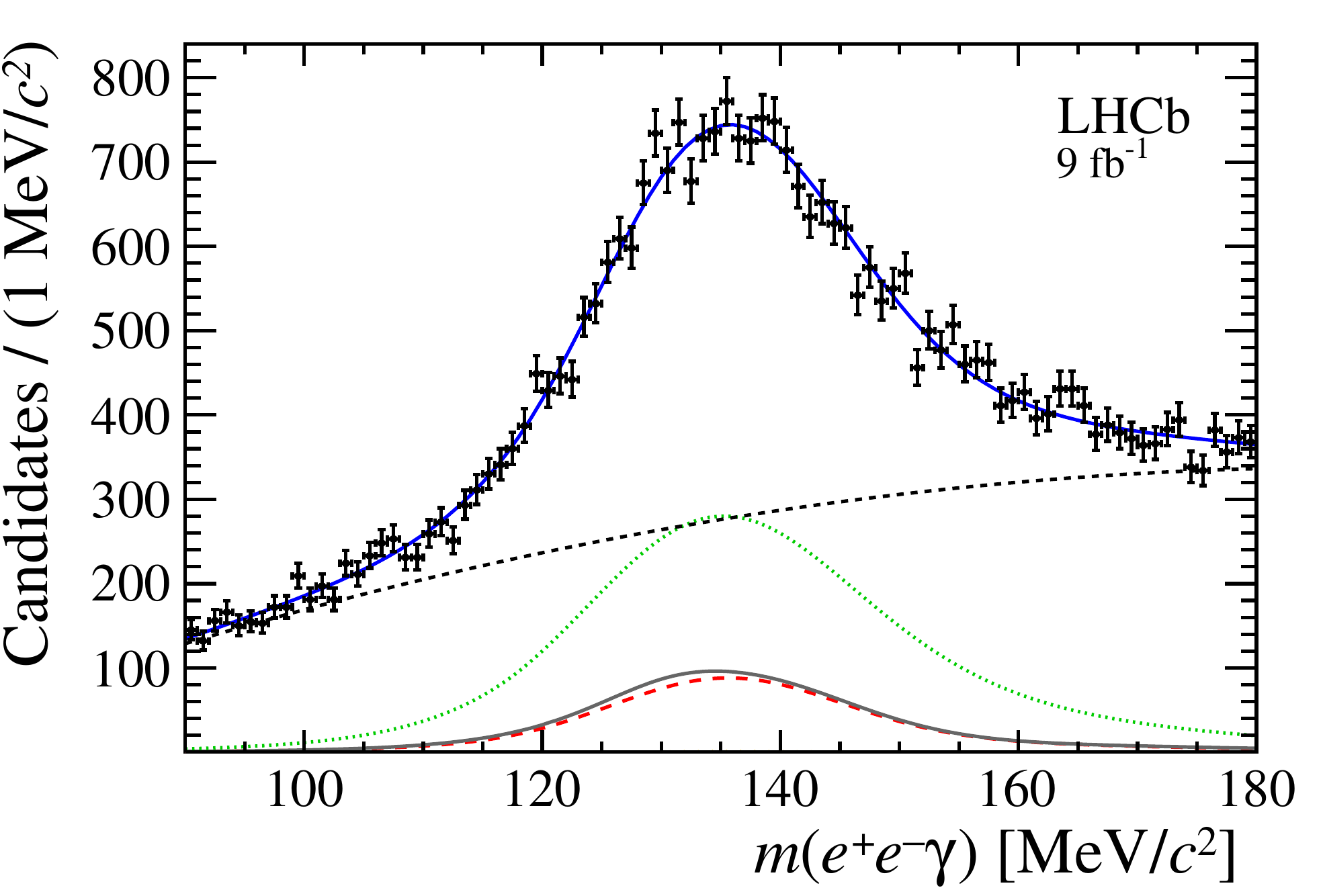}\\
        \includegraphics[width=0.31\textwidth]{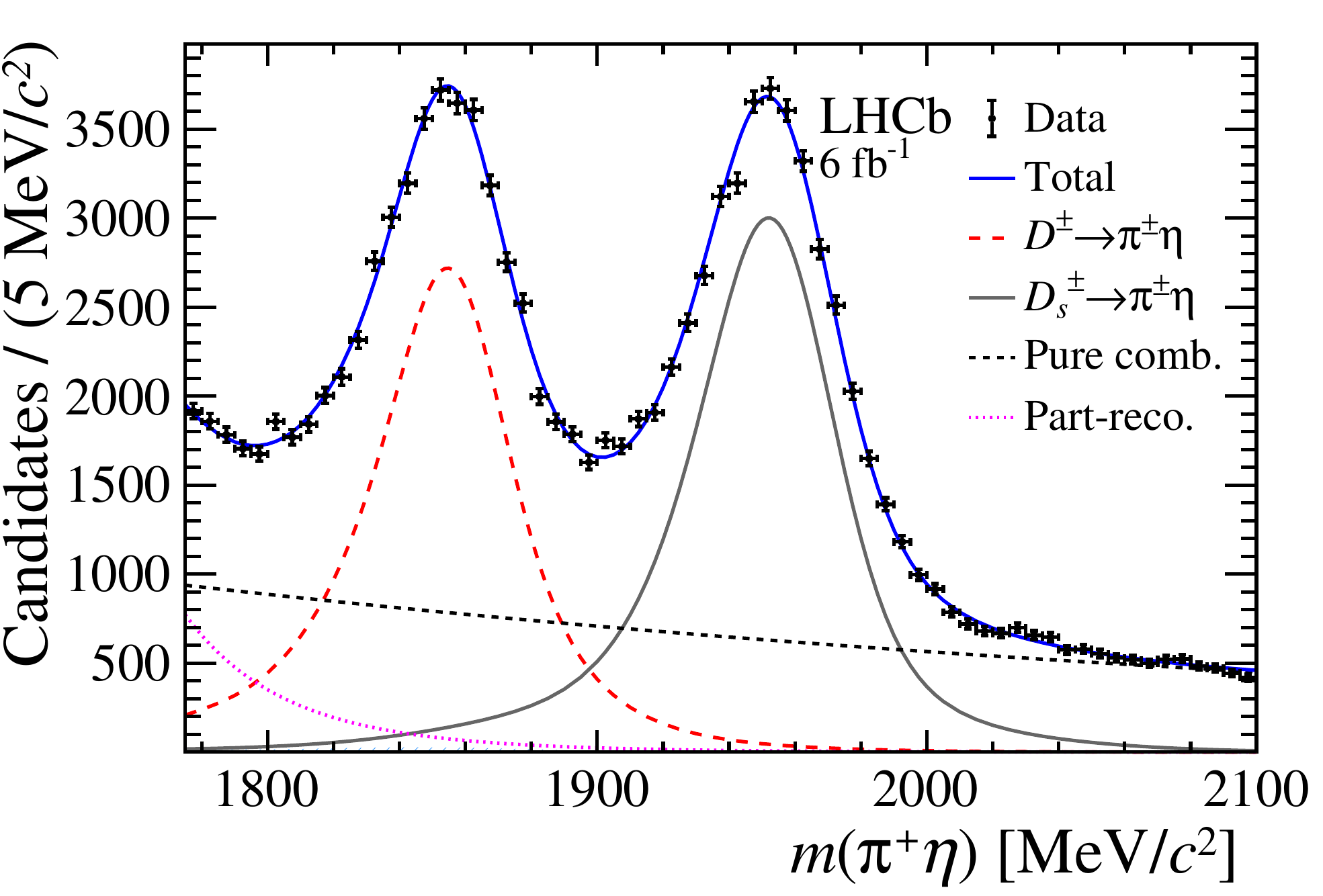}
        \includegraphics[width=0.31\textwidth]{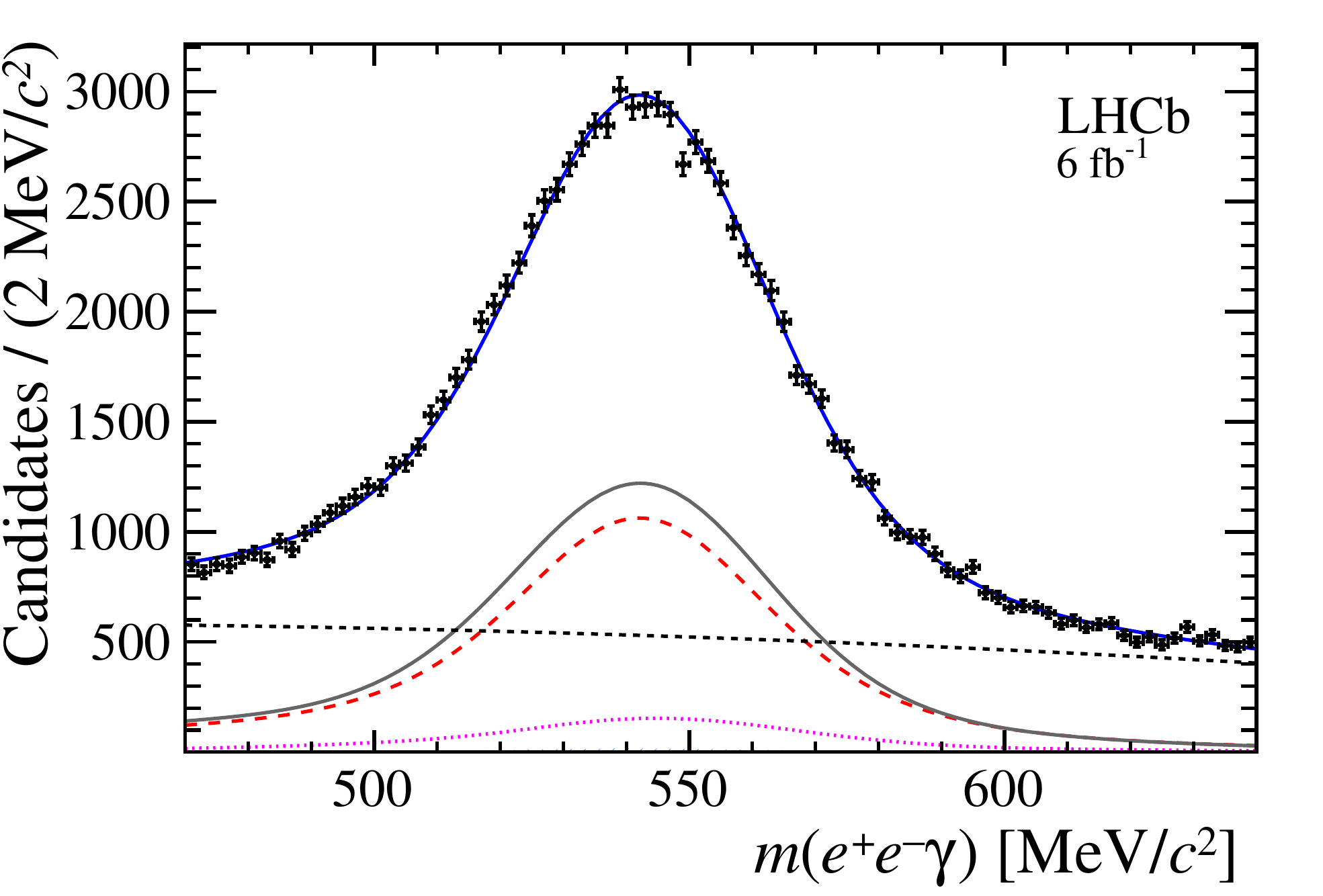}\\
        \includegraphics[width=0.31\textwidth]{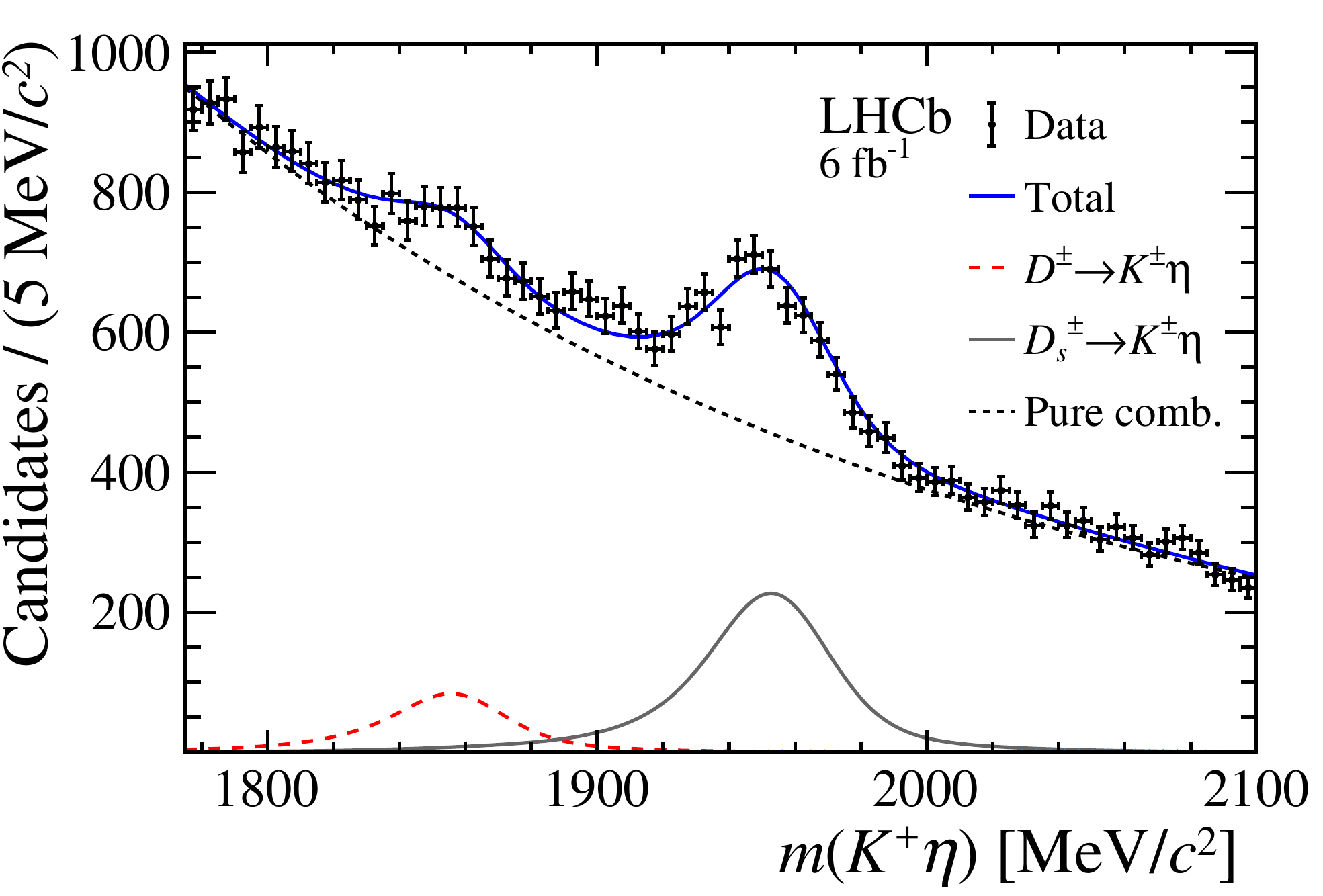}
        \includegraphics[width=0.31\textwidth]{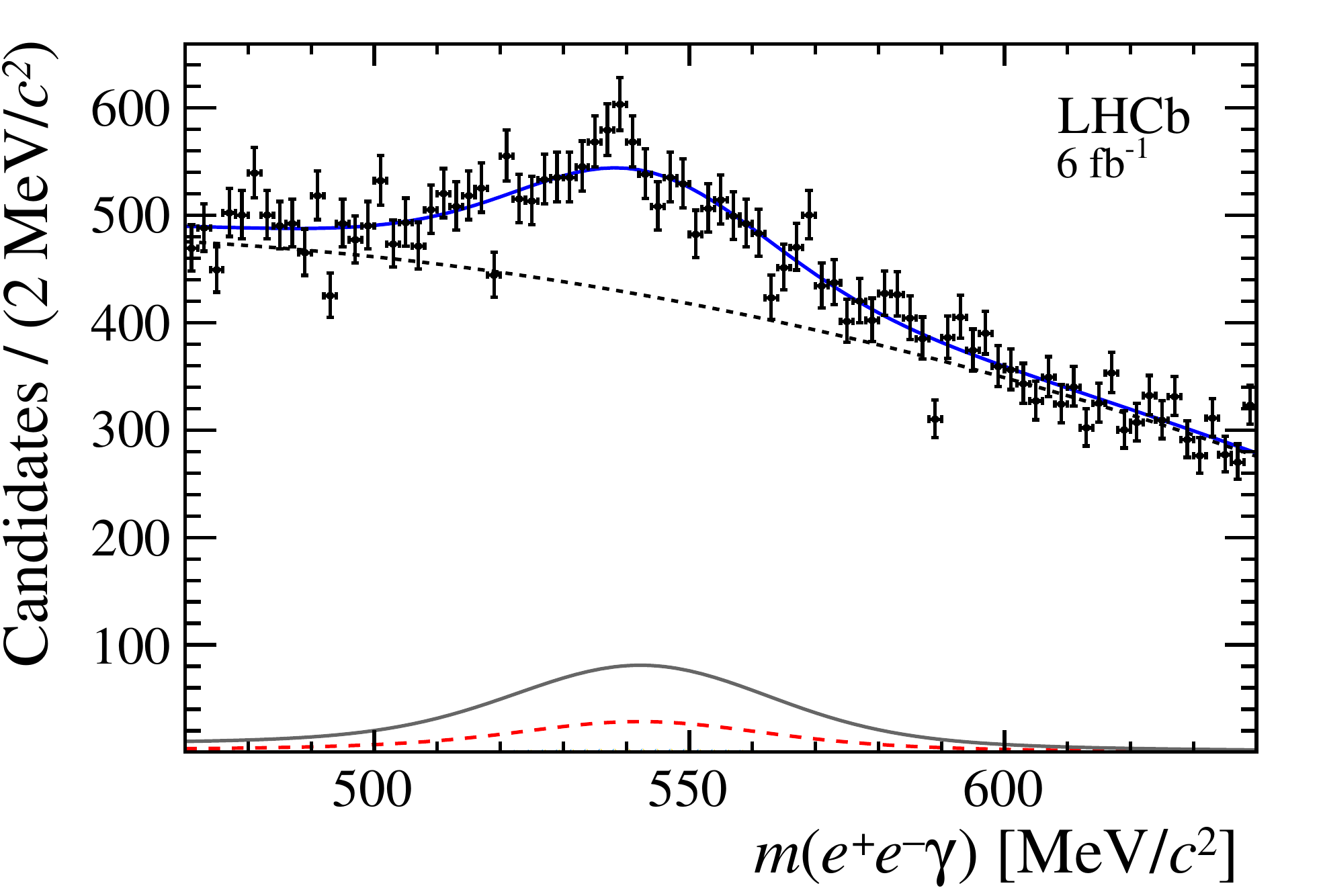} \\
        \includegraphics[width=0.31\textwidth]{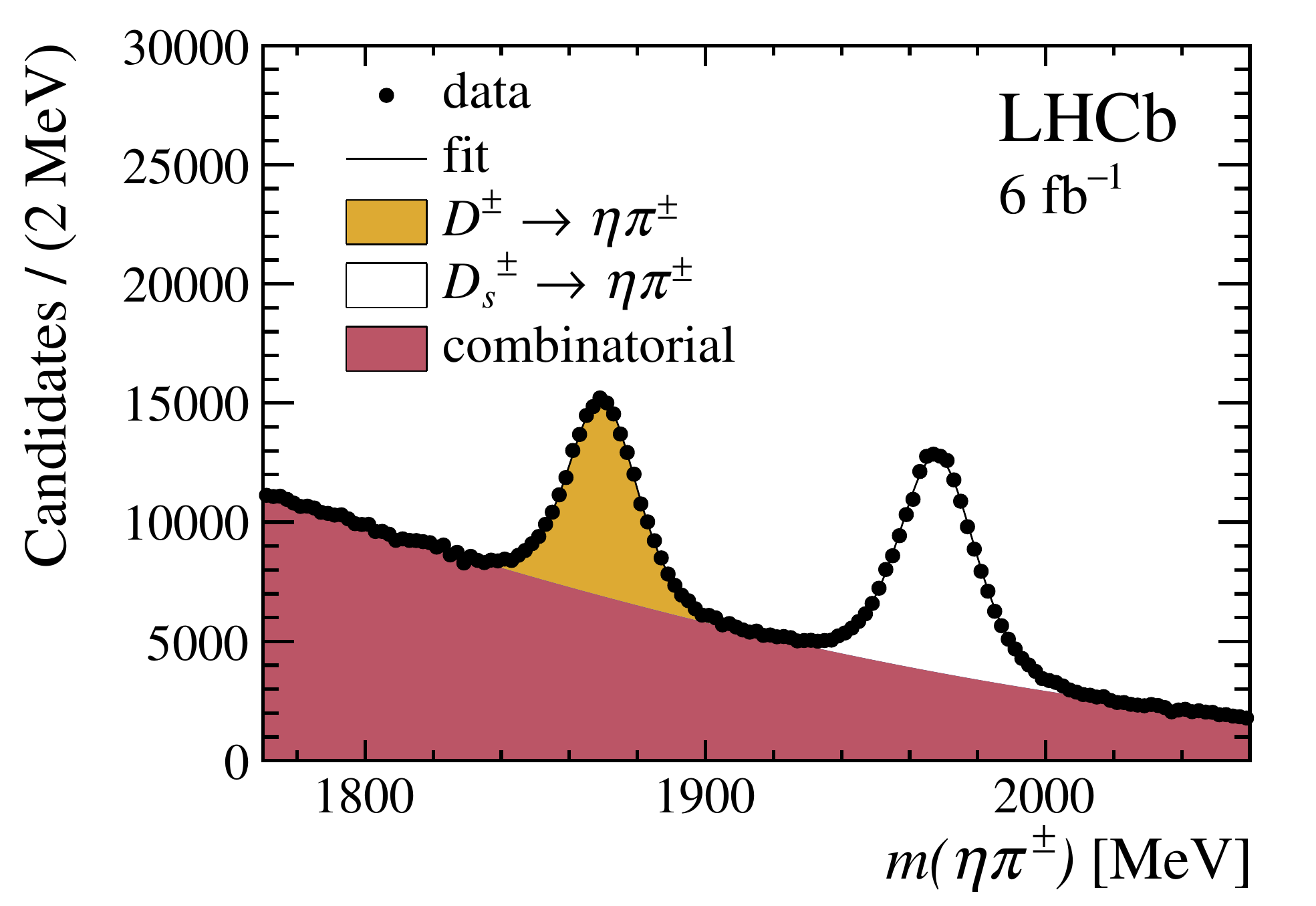}
        \includegraphics[width=0.31\textwidth]{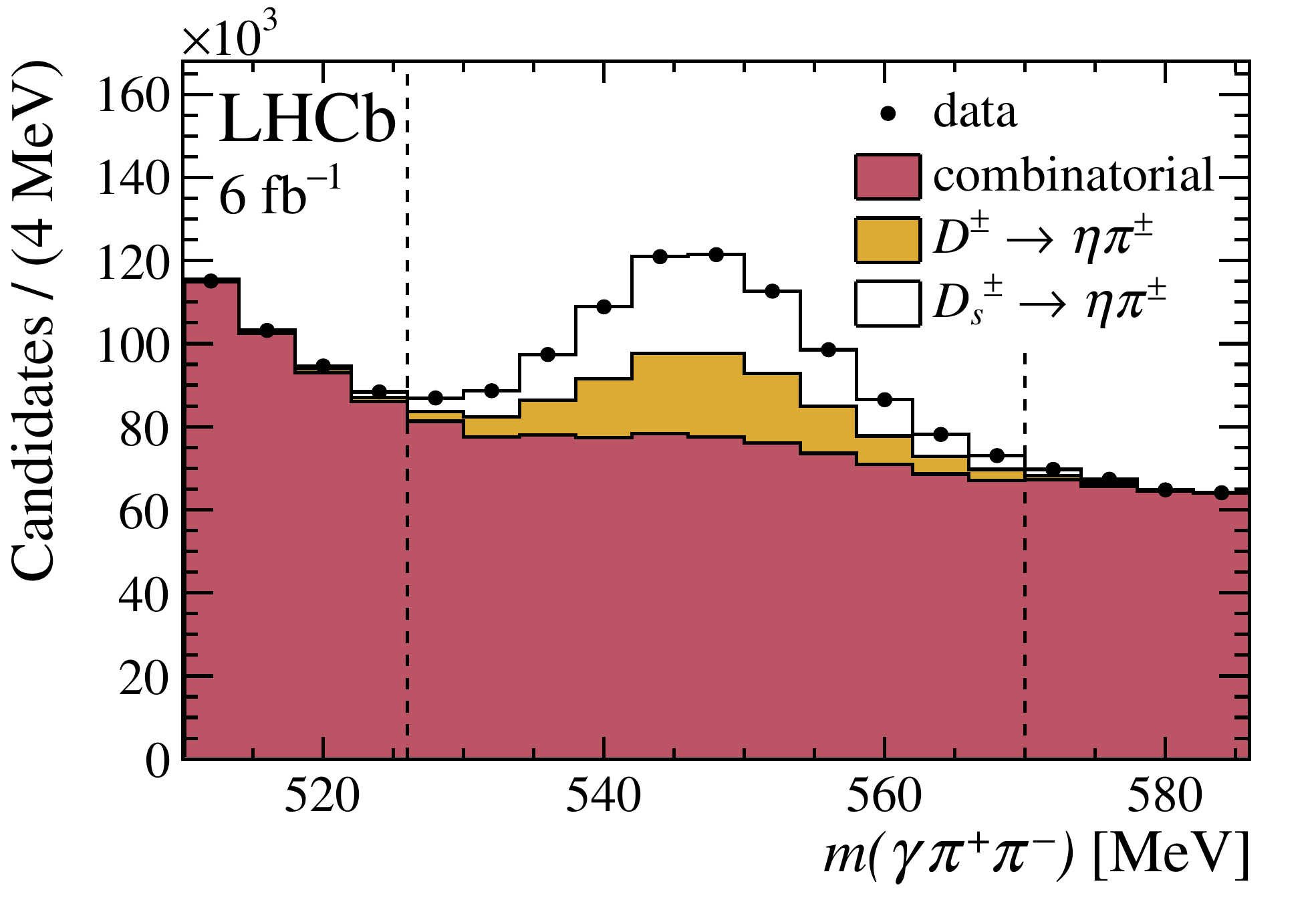} \\
        \includegraphics[width=0.31\textwidth]{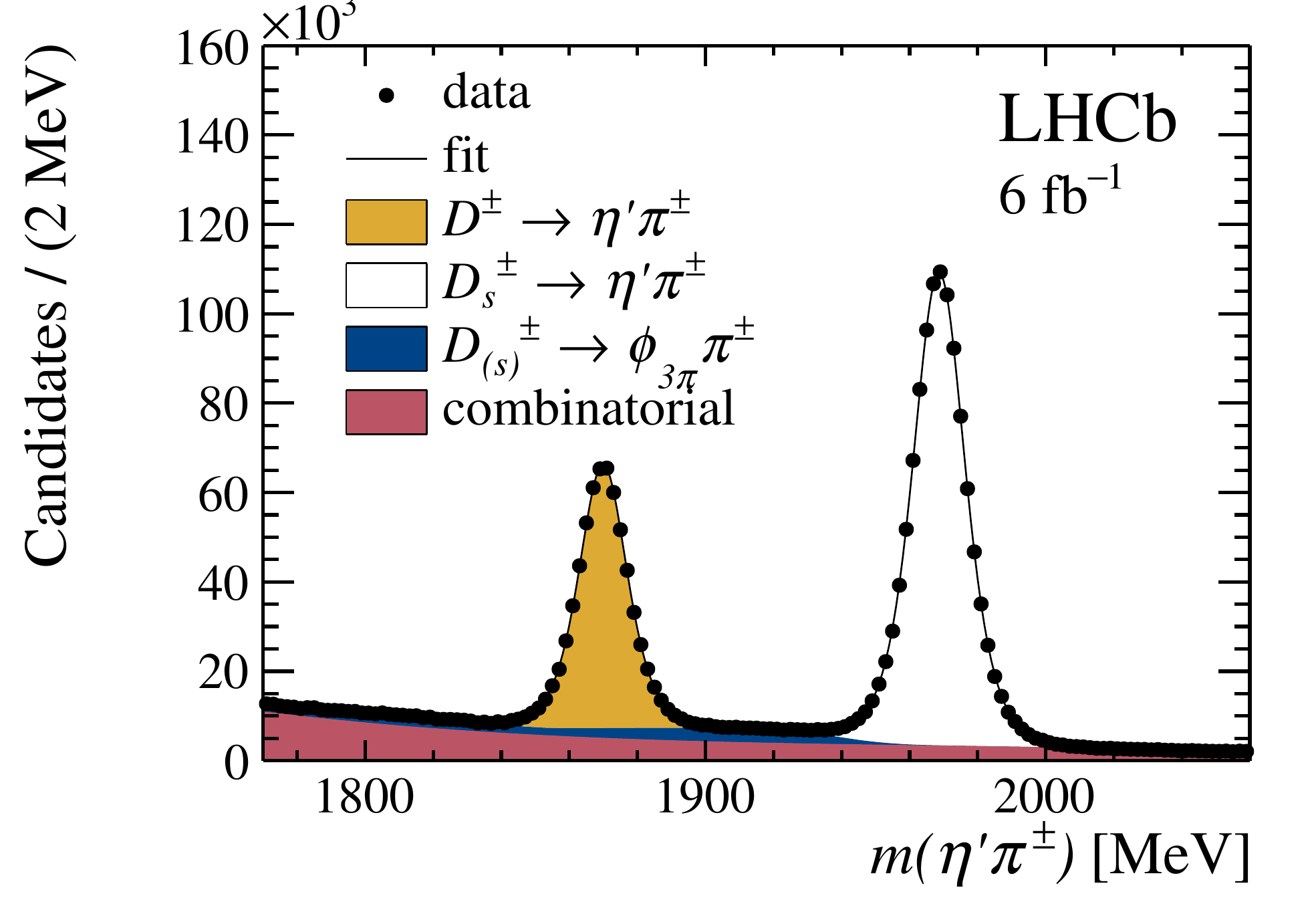}
        \includegraphics[width=0.31\textwidth]{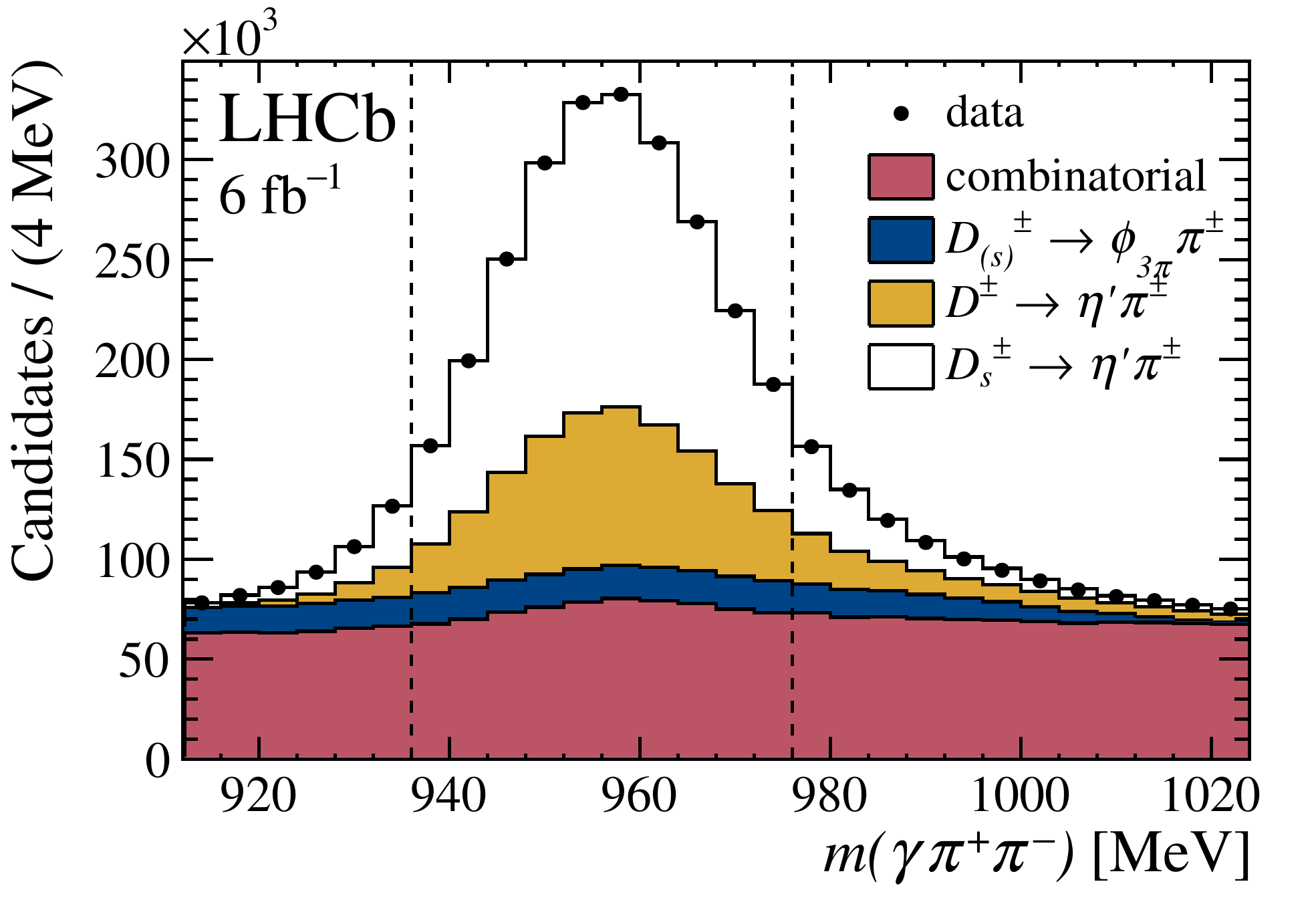}
    \end{center}
    \vspace*{-4mm}
    \caption{
        Distribution of (left) $m(\hp\hzero)$ and (right) the invariant mass of the corresponding neutral meson, for the various combinations of \DpOrDsp decays analysed in refs.~\cite{LHCb-PAPER-2021-001,LHCb-PAPER-2021-051}, from which the figures are taken.
        Projections of the fit results and individual fit components are overlaid.
        In the right figures of the two bottom rows, the $m(\gamma\pip\pim)$ mass range is enlarged with respect to the baseline fit, and the default mass range is indicated by the vertical dashed lines.
    }
    \label{fig:hh0}
\end{figure}

For all final states, the \CP asymmetries are measured from a two-dimensional fit to the invariant mass distributions of the \DpOrDsp and \hzero candidates, where the probability distributions are based on simulation and account for correlations, especially between the radiative tails.
The fit projections are shown in \cref{fig:hh0}.
Detection asymmetries are removed by subtraction with \decay{\DpOrDsp}{\hp\KS} or \decay{\DpOrDsp}{\hp\phi} calibration decays, the kinematics of which is weighted to match that of the signal decays; the \KS detection asymmetry is corrected for with an explicit calculation~\cite{Stahl:1997600}.
The results are
\setlength{\jot}{0pt}
\[
  \begin{aligned}
    A_{\CP}^{\decay{\Dp} {\Kp\piz}}    &= (-3.2\phantom{0} \pm 4.7\phantom{0} \pm 2.1\phantom{0})\%, \\
    A_{\CP}^{\decay{\Dp} {\pip\piz}}   &= (-1.3\phantom{0} \pm 0.9\phantom{0} \pm 0.6\phantom{0})\%, \\
    A_{\CP}^{\decay{\Dp} {\pip\etaz}}  &= (\phantom{1.}0.13 \pm 0.50 \pm 0.18)\%, \\
    A_{\CP}^{\decay{\Dp} {\pip\etapr}} &= (\phantom{1.}0.43 \pm 0.17 \pm 0.10)\%,
  \end{aligned}
  \hspace{10mm}
  \begin{aligned}
    A_{\CP}^{\decay{\Dp} {\Kp\etaz}} 	&= (-6\phantom{.1} \pm 10\phantom{.} \pm 4\phantom{.1})\%, \\
    A_{\CP}^{\decay{\Dsp}{\Kp\piz}} 	&= (-0.8 \pm 3.9 \pm 1.2)\%, \\
    A_{\CP}^{\decay{\Dsp}{\Kp\etaz}}  &= (\phantom{1.}0.9 \pm 3.7 \pm 1.1)\%, \\
    \phantom{A_{\CP}^{\decay{\Dsp}{\Kp\etaz}}}&
  \end{aligned}
\]
where the first and the subsequent rows correspond to DCS and SCS decays, respectively, and the systematic uncertainties are mainly due to the uncertainty on the fit models.
They are all consistent with the absence of \CP violation and they are in agreement with previous determinations at \B and charm factories~\cite{CLEO:2009fiz,Belle:2011tmj,Belle:2017tho,Belle:2021ygw}, while their precision is comparable or better.

\subsection{Multibody decays}
Even if the most precise searches for \CP violation to date concern two-body and quasi-two body decays~\cite{LHCb-PAPER-2019-006,LHCb-PAPER-2019-002,LHCb-PAPER-2022-024}, an extensive set of searches in \D multibody decays are also being pursued.
Their rich resonant structure and the variations of the strong phases across their final-state phase space may provide further handles to pin down the size of nonperturbative QCD effects and to understand the nature of the interactions responsible for \CP violation.
Some of these decays, such as \decay{\Dz}{\KS\Kpm\pimp}, could display enhanced \CP asymmetries, thanks to a mechanism analogous to that described in \cref{sect:ksks}~\cite{Nierste:2017cua}.

A whole variety of model-independent methods~\cite{Bediaga:2009tr,Bediaga:2012tm,Williams:2011cd,Gronau:2011cf,Durieux:2015zwa} have been developed and used for the study of multibody decays during \runone~\cite{LHCb-PAPER-2011-017,LHCb-PAPER-2013-041,LHCb-PAPER-2013-057,LHCb-PAPER-2014-046,LHCb-PAPER-2014-054,LHCb-PAPER-2016-044,LHCb-PAPER-2019-026} as an intermediate step towards amplitude analyses which would be crucial to theoretically interpret an observation of \CP violation.
Some efforts have been devoted also to amplitude analyses, either allowing for the presence of \CP violation~\cite{LHCb-PAPER-2015-026,LHCb-PAPER-2018-041} or not~\cite{LHCb-PAPER-2017-040,LHCb-PAPER-2022-016,LHCb-PAPER-2022-030}.
Many new measurements based on the \runtwo sample, which guarantees much larger signal yields thanks to the improvements in triggering~\cite{LHCb-DP-2016-001}, are underway.
Finally, a recent amplitude analysis of the \decay{\Lc}{\proton\Km\pip} decay and of the \Lc production polarisation in semimuonic \B decays~\cite{LHCb-PAPER-2022-002} paves the way to future measurements of its electric and magnetic dipole moments, which are sensitive to BSM interactions of the charm quark and to the QCD structure of the \Lc baryon, respectively~\cite{Botella:2016ksl,Fomin:2017ltw,Bagli:2017foe,Mirarchi:2019vqi,Aiola:2020yam}.

\section{Time-dependent measurements}
\label{sect:time-dep-studies}
Searches for time-dependent \CP violation are mostly sensitive to \CP violation in the mixing amplitudes~\cite{Kagan:2020vri}, and thus they are complementary to those of \CP violation in the decay.
However, they require even better experimental precision.
In fact, in the SM the contribution of \CP violation in the mixing to the \CP asymmetries is typically suppressed by one order in the $U$-spin breaking parameter $\epsilon$ with respect to that of \CP violation in the decay~\cite{Kagan:2020vri}.
While \CP violation in the mixing has not been observed yet, the past two years have witnessed a leap forward in the precision of its search and of the measurements of the mixing parameters, as described in the next sections.

\subsection{Search for time-dependent \CP violation in \decay{\Dz}{\HH} decays}
\label{sect:dy}
The \CP-even, SCS final states $f=\KK$ or \PP provide a clean way to measure the dispersive \CP-violating contributions to \Dz mixing.
The relevant observable is the slope, \DY{f}, of the time-dependent asymmetry of the \Dz and \Dzb decay widths defined in \cref{eq:acp-expansion}.
This observable is approximately equal to~\cite{Kagan:2020vri}
\begin{equation}
    \DY{f} \approx - x_{12} \sin\phi^{M}_{2} + y_{12}\Acpdec{f} \Big(1 + \frac{x_{12}}{y_{12}} \cot \delta_f\Big),
\end{equation}
where $\delta_f$ is the strong-phase difference between $A_b$ and $A_{sd}$.
In the very long term, the dependence of \DY{f} on the final state could be used to measure $\delta_f$, using external inputs for the mixing parameters, weak phase $\phi_2^M$ and \Acpdec{f}~\cite{Kagan:2020vri}.
However, assuming that $\delta_f$ is not fine-tuned to zero or $\pi$ (as expected from large rescattering at the charm mass scale), final-state dependent contributions are of the order of $10^{-5}$~\cite{LHCb-PAPER-2019-006,LHCb-PAPER-2021-033,LHCb-PAPER-2022-024} and can be neglected at the current level of experimental precision.
The \DY{f} parameter is consequently assumed to be independent of the final state and is denoted as $\DY{} \equiv -x_{12} \sin\phi_2^M$.
Its value sets a direct constraint on the dispersive weak phase $\phi^M_2$, thanks to independent determinations of $x_{12}$~\cite{LHCb-PAPER-2021-033}.
Reducing the uncertainty on \DY{f} is essential not only to constrain possible \CP-violating BSM interactions, but also to determine the parameter \Acpdec{\KK} from the measurements of the time-integrated asymmetry of \DzKK decays, as described in \cref{sect:observation}.

The \DY{} parameter has been recently measured using the \Dstarp-tagged data sample collected during \runtwo~\cite{LHCb-PAPER-2020-045,Pajero:2747731},
which comprises 58 million and 18 million \DzKK and \DzPP decays, respectively.
The main background, which comes from random associations of unrelated particles, is around 5\% of the signal and is removed through a sideband subtraction in the \DstM variable.

\begin{figure}[tb]
  \begin{center}
    \includegraphics[width=0.32\linewidth]{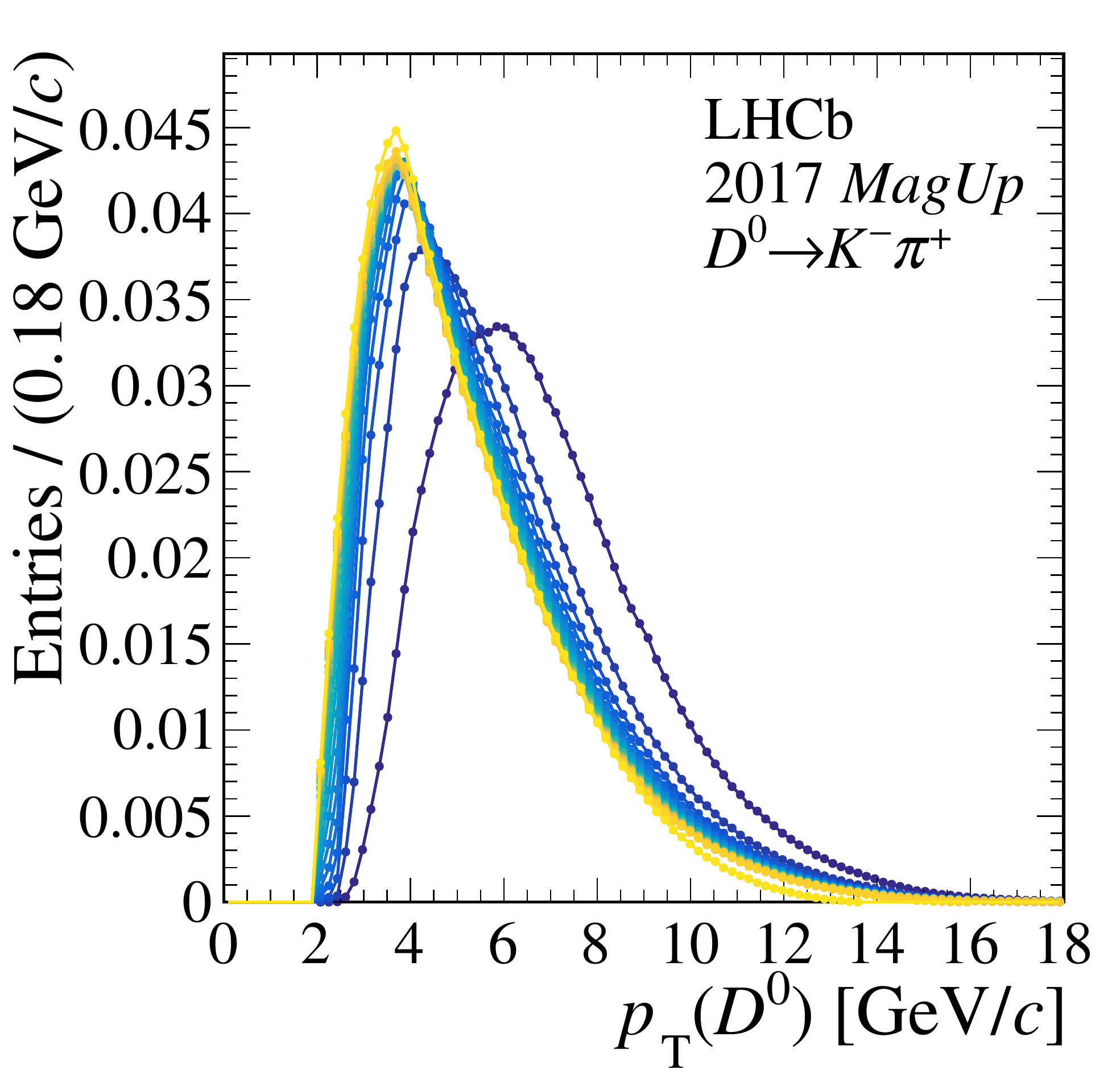}
    \includegraphics[width=0.32\linewidth]{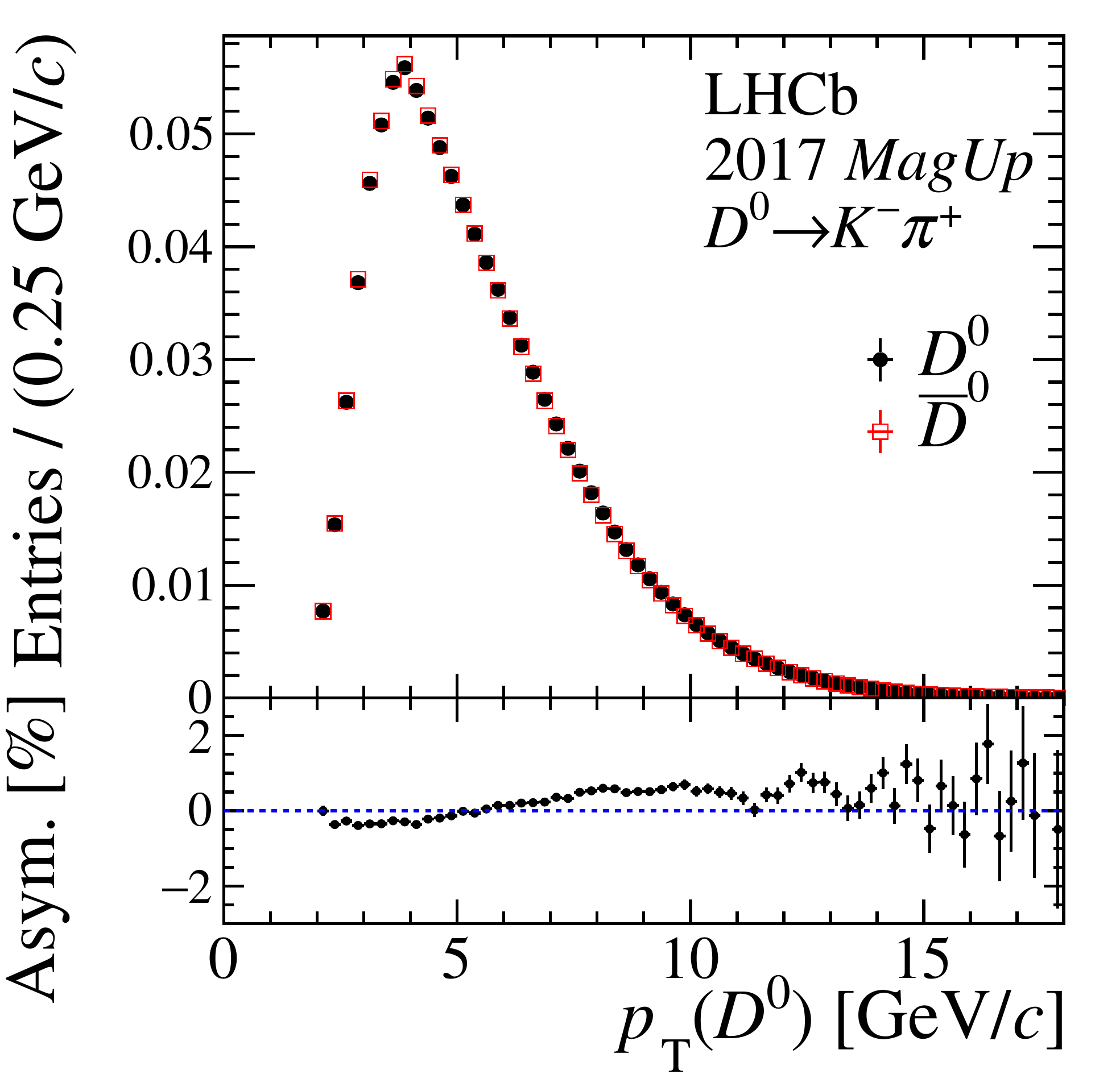}
    \includegraphics[width=0.32\linewidth]{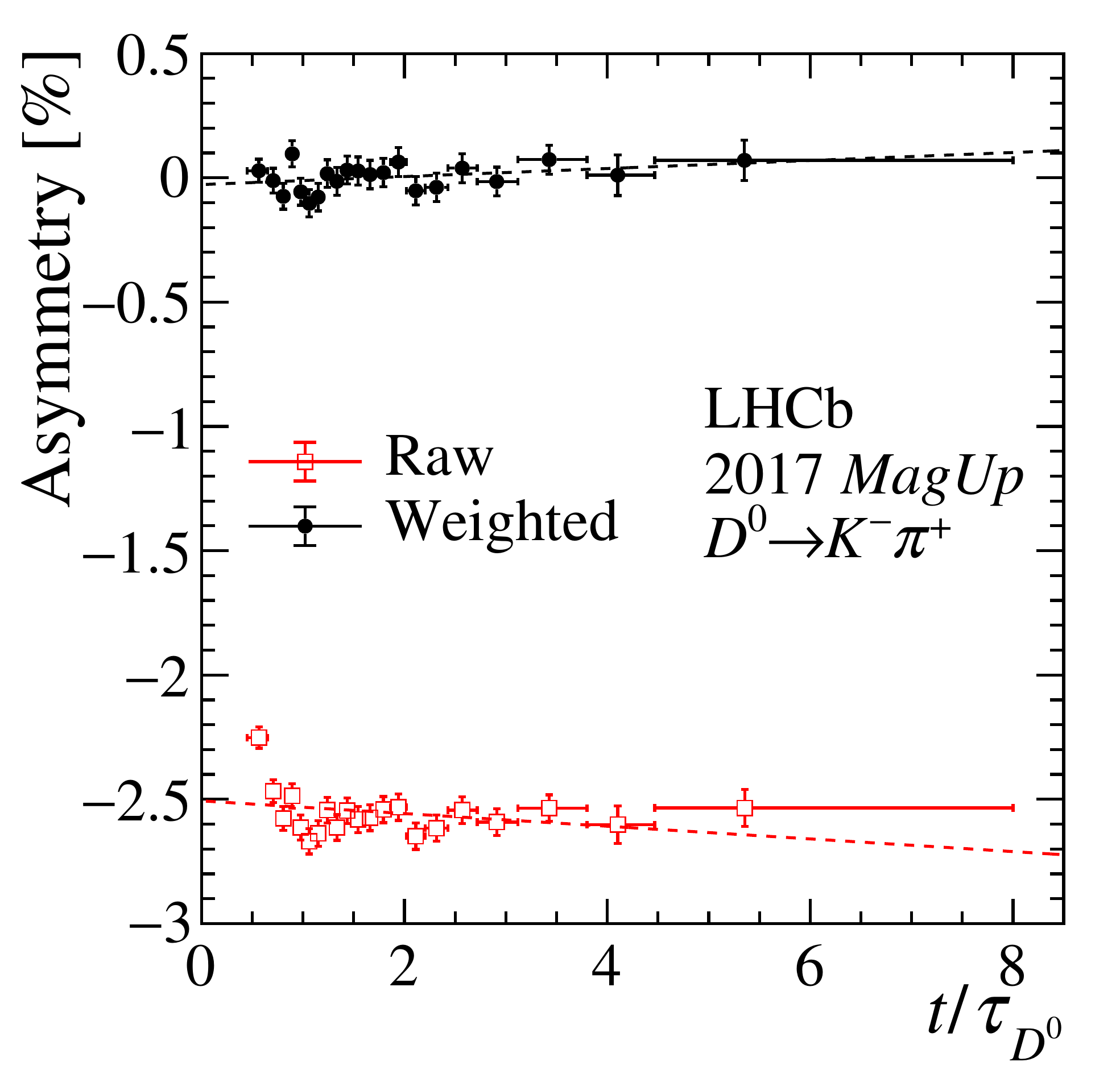}
  \end{center}
  \vspace*{-0.4cm}
  \caption{
    (Left) Normalised distributions of the \Dz transverse momentum, in different colours for each decay-time interval in \cref{fig:dy-results}.
    Decay time increases from blue to yellow colour.
    (Centre) Asymmetry between the normalised $\pt$ distributions of \Dz and \Dzb mesons.
    (Right)
    Linear fit to the time-dependent asymmetry (red) before and (black) after the kinematic equalisation, after which the time-integrated asymmetry is zero by construction.
    All plots correspond to \DzRS candidates collected in 2017 with the magnet polarity pointing upwards.
    Figures taken from ref.~\cite{LHCb-PAPER-2020-045}.
  }
  \label{fig:correlations}
\end{figure}

While the measurement is insensitive by construction to time-independent asymmetries, a time dependence of the detection and production asymmetries is indirectly introduced by the trigger requirements, even if these nuisance asymmetries depend explicitly only on the particles momenta.
In fact, requirements on displacement-related variables like the \Dz flight distance and the IP of its daughter particles with respect to the PV select \Dz mesons with low decay times only if their momentum is large enough.
As a result, low decay times correspond on average to larger momenta, and momentum-dependent asymmetries give rise to artificial time-dependent asymmetries; see \cref{fig:correlations}.
These effects are studied in a control sample of 518 million \DzRS decays, for which the dynamical time-dependent asymmetry is known to be smaller than the experimental precision~\cite{LHCb-PAPER-2020-045,Pajero:2021jev}.
The nuisance asymmetries are up to six times larger than the statistical uncertainty of the measurement, and are removed by equalising the vector-momentum distributions of the \pisp/\pism and \Dz/\Dzb mesons through per-event weights.
This removes by construction the dependence of the nuisance asymmetries on momentum and consequently on time; see \cref{fig:correlations} right.
While the equalisation marginally biases also the dynamical asymmetry, this effect is measured to be small and is corrected for.
On the other hand, the equalisation allows to avoid the loss of statistical precision due to the weighting of multiple calibration channels that would otherwise be needed to correct for the nuisance asymmetries, as done in time-integrated measurements; see \cref{sect:observation}.
Naturally this procedure allows to measure only the time-dependent contribution of the asymmetry, and not the time-independent parameter \Acpdec{f}.

The production asymmetry of secondary mesons differs from that of prompt mesons, and their fraction increases with decay time.
This background would thus bias the measurement even after the kinematic equalisation.
Both the asymmetry and the fraction of secondary mesons are measured through a fit to the two-dimensional IP distribution of the \Dz meson with respect to its PV, as a function of its decay time.
Most of the secondary mesons are rejected by requiring the IP of the \Dz meson to be greater than $60 \mum$, and a correction of $0.3 \times 10^{-4}$ is applied to the measurement of \DY{} to account for their residual 4\% contamination.

\begin{figure}[tb]
    \begin{center}
        \includegraphics[width=0.48\textwidth]{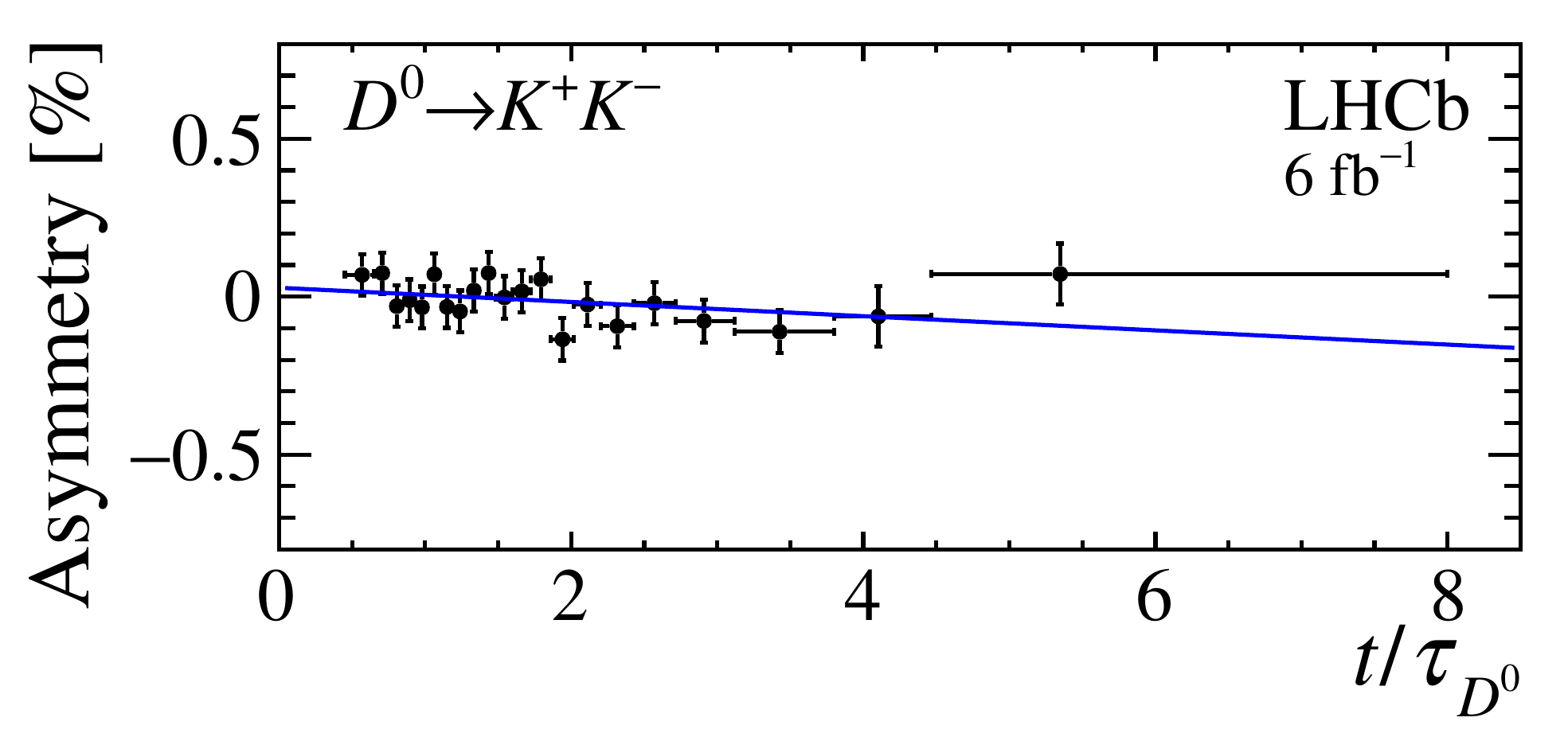}
        \includegraphics[width=0.48\textwidth]{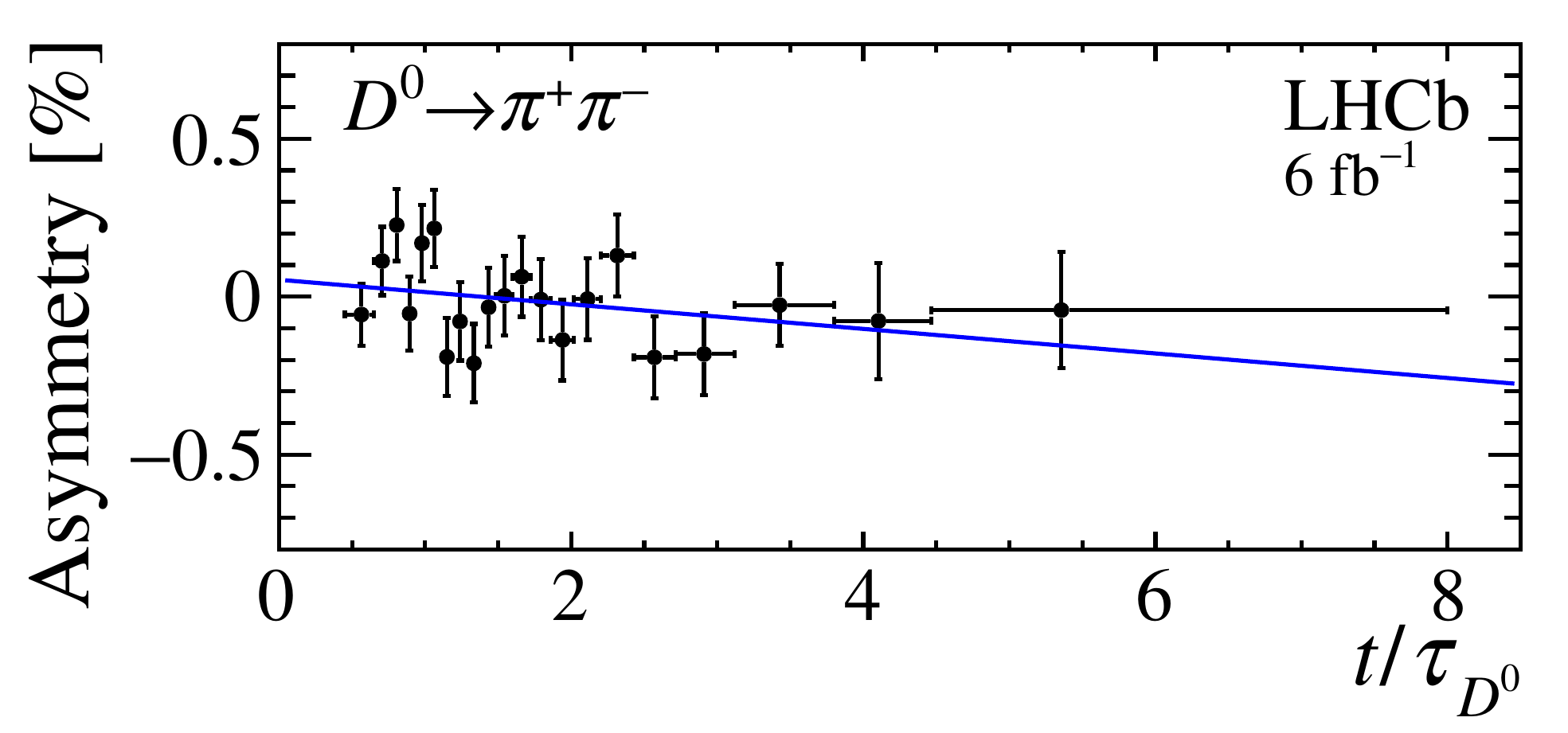}
    \end{center}
    \vspace*{-4mm}
    \caption{
        Asymmetry as a function of decay time for (left) \DzKK and (right) \DzPP candidates.
        A linear fit is superimposed.
        Figures taken from ref.~\cite{LHCb-PAPER-2020-045}.
    }
    \label{fig:dy-results}
\end{figure}

The analogue of \DY{} for the control sample is measured from a binned linear fit to the time-dependent asymmetry of the \Dz and \Dzb weighted yields.
The result, after the aforementioned correction for the bias from secondary mesons, is $\DY{\RS} = (-0.4 \pm 0.5 \pm 0.2) \times 10^{-4}$, and is compatible with zero (as expected) within an uncertainty smaller than that of the final measurement by more than a factor of two.
The linear fits to the time-dependent asymmetry for the signal samples are shown in \cref{fig:dy-results}.
The results,
\[
    \begin{aligned}
    \DY{\KK} &= (-2.3 \pm 1.5 \pm 0.3) \times 10^{-4},\\
    \DY{\PP} &= (-4.0 \pm 2.8 \pm 0.4) \times 10^{-4},
    \end{aligned}
\]
are compatible with the absence of \CP violation at the level of two standard deviations and are in keeping with previous determinations~\cite{Lees:2012qh,Aaltonen:2014efa,Staric:2015sta,LHCb-PAPER-2014-069,LHCb-PAPER-2016-063,LHCb-PAPER-2019-032}.
They constitute the most precise search for \CP violation performed at a hadron collider to date and, neglecting possible differences between the \KK and \PP final states, they improve the precision of the previous world average by nearly a factor of two~\cite{HFLAV18}, yielding
\[
    \DY{} = (-1.0 \pm 1.1 \pm 0.3) \times 10^{-4}.
\]
The reduction of the systematic uncertainty by a factor of three with respect to the previous most precise determination~\cite{LHCb-PAPER-2016-063}, as well as the fact that it is dominated by the statistical precision with which the asymmetry of the combinatorial background is known and thus is expected to decrease as the size of the analysed sample increases, pave the way for even more precise future measurements.

\subsection{Observation of a nonzero mass difference between the neutral charmed-meson mass eigenstates with \decay{\Dz}{\KS\pip\pim} decays}
\label{sect:kspipi}
Contrary to two-body decays~\cite{Pajero:2021jev,LHCb-PAPER-2017-046,LHCb-PAPER-2020-045,LHCb-PAPER-2021-041}, multibody \Dz decays allow to measure simultaneously all mixing parameters and \CP-violation phases, rather than a combination of these parameters.
This is made possible by the variation of the strong phase across their final-state phase space, caused by their rich resonance structure.
However, they imply the additional complication of a multi-dimensional analysis of the final-state phase space.
Especially at hadron colliders, the selection efficiency varies significantly across the phase space due to tight trigger requirements, and this effect must be accounted for.

\begin{figure}[tb]
    \begin{center}
        \includegraphics[width=0.5\textwidth]{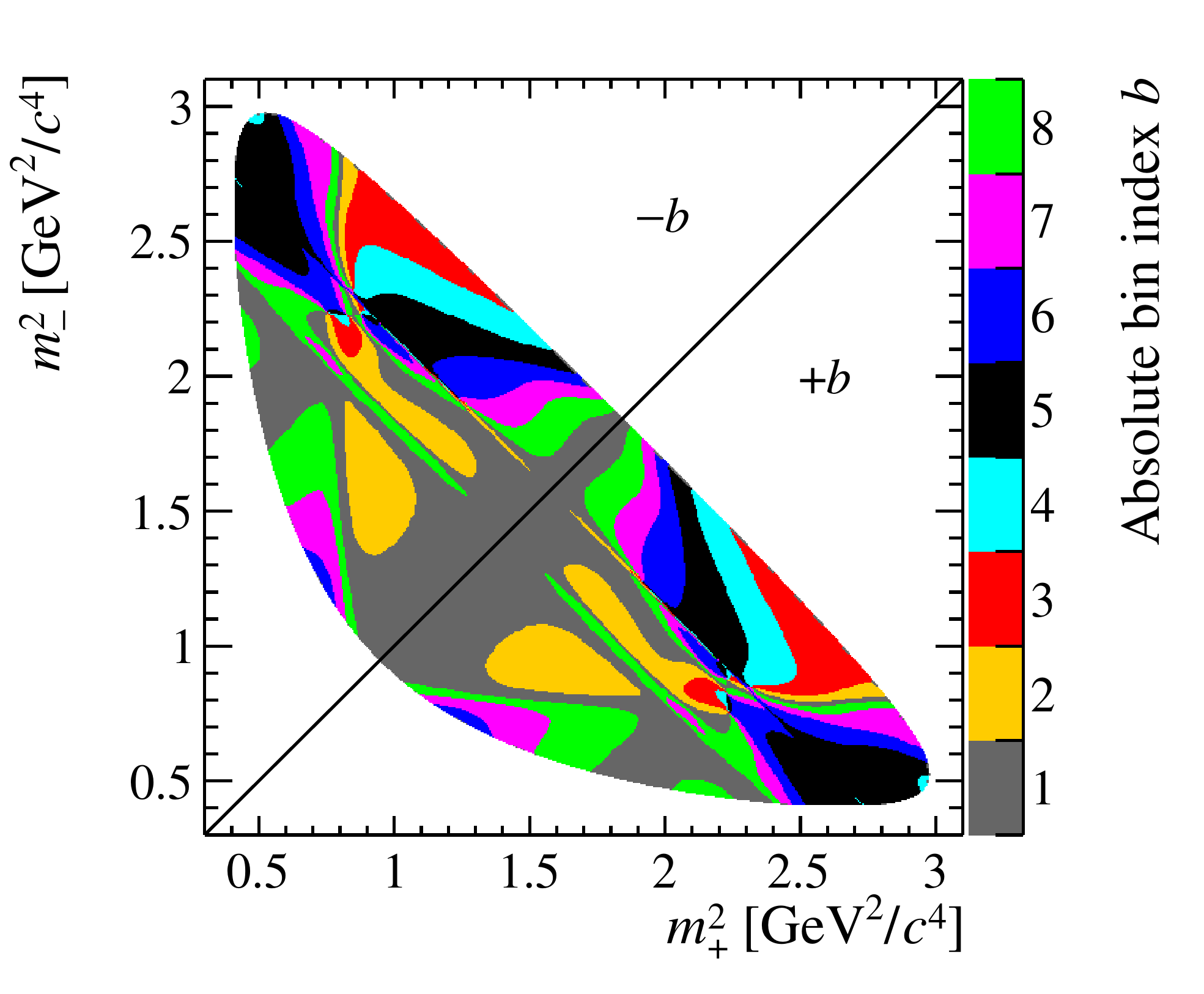}
        \includegraphics[width=0.48\textwidth]{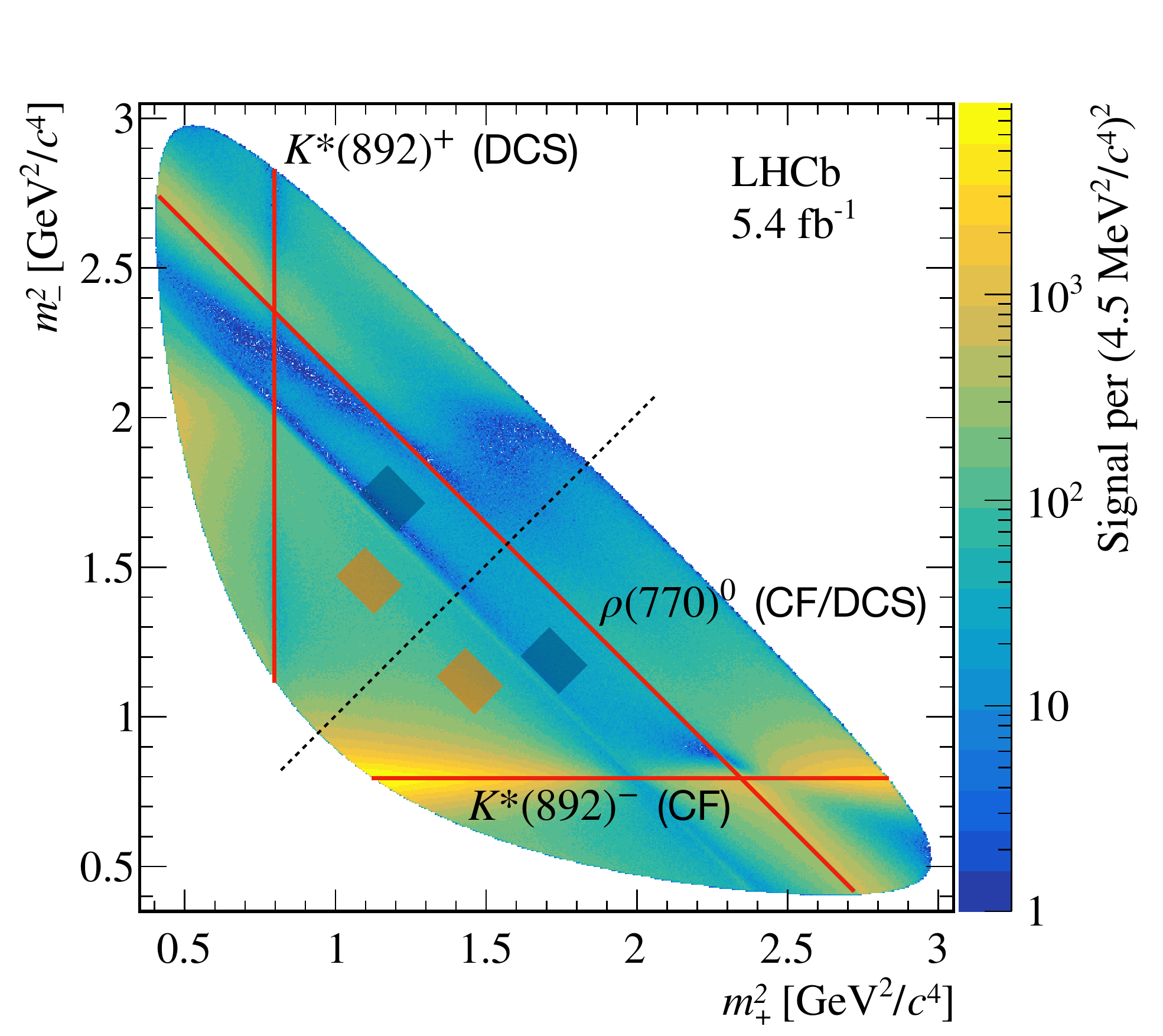}
    \end{center}
    \vspace*{-4mm}
    \caption{
        (Left)
        Iso-$\Delta\delta$ division of the Dalitz plot of \decay{\Dz}{\KS\pip\pim} decays.
        The variable $m_\pm$ is defined as $m(\KS\pipm)$ for \Dz decays, and as $m(\KS\pimp)$ for \Dzb decays.
        (Right)
        Dalitz plot of background-subtracted \decay{\Dz}{\KS\pip\pim} candidates.
        The masses of the most relevant resonances are highlighted by the red lines.
        The coloured rectangles are an example of the regions used to decorrelate the \Dz decay time and $m(\pip\pim)$ --- note that the size and position of the rectangles used in the actual measurement are different.
        Figures adapted from refs.~\cite{CLEO:2010iul,LHCb-PAPER-2021-009}.
    }
    \label{fig:binflip-dalitz}
\end{figure}

This problem can be mitigated by using model-independent analysis methods such as that proposed in ref.~\cite{DiCanto:2018tsd} for \decay{\Dz}{\KS\pip\pim} decays.
There, the Dalitz plot is divided into 8 regions symmetric with respect to its bisector (see \cref{fig:binflip-dalitz}), each having an approximately constant strong-phase difference, $\Delta\delta$, between the decay amplitudes in the two halves of the plot.
The ratios of the yields of the upper to the lower half are measured for each of the 8 regions, labelled ``$b$'', and in 13 intervals of decay time, labelled ``$j$'', separately for \Dz and \Dzb decays.
Biases due to efficiency variations across the plot mostly cancel in the ratio, since the efficiency is approximately symmetric with respect to the bisector.
A similar cancellation holds also for the production asymmetry and for the detection asymmetry of the tagging particle.
On the other hand, the ratios provide nearly optimal sensitivity to the mixing parameters.
In fact, the decays in the denominator mostly correspond to CF amplitudes and have nearly constant decay rates as a function of decay time, whereas the decay amplitudes in the numerator are mostly DCS, so that the fraction of CF decays following mixing is comparable in size and the decay rate significantly increases as a function of decay time.
In particular, the ratios are equal to 
\begin{equation}
    \begin{aligned}
    R_{bj}^\pm &\approx \frac{r_b
                              + \sqrt{r_b}\,{\rm Re}\big[X_b^*(z_{CP} \pm \Delta z)\big]\langle t \rangle_j
                              + \tfrac{1}{4}\big[\lvert z_{CP} \pm \Delta z\rvert^2 + r_b{\rm Re}(z^2_{CP} - \Delta z^2)\big] \langle t^2 \rangle_j}
                             {1_{\phantom{b}}
                              + \sqrt{r_b}\,{\rm Re}\big[X_b(z_{CP} \pm \Delta z)\big]\langle t \rangle_j
                              + \tfrac{1}{4}\big[{\rm Re}(z^2_{CP} - \Delta z^2) + r_b\lvert z_{CP} \pm \Delta z\rvert^2 \big] \langle t^2 \rangle_j}\\
        & \approx r_b + \sqrt{r_b}\big[s_b (1+r_b)(x_{CP} \pm \Delta x)-c_b (1-r_b)(y_{CP} \pm \Delta y)\big]\langle t \rangle_j,
    \end{aligned}
\end{equation}
where the plus (minus) sign applies to the ratio of \Dz (\Dzb) decays; the first and second line have been expanded up to second and to first order in the mixing parameters, respectively; $r_b$ is the ratio at zero decay time; $X_b \equiv  c_b + i s_b$ is the average of $e^{i\Delta\delta}$ in the region ``$b$'', as measured at charm factories~\cite{CLEO:2010iul,BESIII:2020khq}; and the two complex parameters $z_{CP} \equiv - (y_{CP} + i\,x_{CP})$ and $\Delta z \equiv - (\Delta y + i\,\Delta x)$ are defined as
\begin{equation}
  \label{eq:z-params}
  \begin{aligned}
    x_{CP} &\equiv x_{12}\cos\phi^M_2,
    &&\Delta x \equiv -y_{12}\sin\phi^\Gamma_2,\\
    y_{CP} &\equiv y_{12}\cos\phi^\Gamma_2,
    &&\Delta y \equiv x_{12}\sin\phi^M_2 = - \DY{}.
  \end{aligned}
\end{equation}

A new measurement~\cite{LHCb-PAPER-2021-009} based on the \Dstarp-tagged data sample collected in \runtwo increases the signal yield tenfold with respect to the previous determination based on the sample collected in \runone~\cite{LHCb-PAPER-2019-001}, thanks to improved triggering~\cite{LHCb-DP-2016-001}.
The selection requirements introduce correlations between the decay time and the Dalitz coordinates, in particular $m^2(\pip\pim)$.
Regions of constant $m^2(\pip\pim)$ correspond to bands orthogonal to the bisector; see for example the $\rho(770)^0$ lineshape in \cref{fig:binflip-dalitz} right.
Therefore, a decorrelation procedure is applied by weighting the time distribution of the sum of the candidates in pairs of rectangles symmetric with respect to the bisector (see \cref{fig:binflip-dalitz} right) to the time distribution of the total sample.
Since a nonzero value of $x_{\CP}$ has the effect of moving candidates into the position symmetric with respect to the bisector, the procedure does not bias this variable, whereas second-order biases to $y_{\CP}$ are corrected for.
Momentum-dependent detection asymmetries of the \pip and \pim mesons also do not cancel in the ratios.
They are corrected for based on a measurement of kinematically weighted \decay{\Dsp}{\pip\pim\pip} CF decays, where the additional \Dsp production asymmetry and the detection asymmetry of the third pion are removed by subtraction with \decay{\Dsp}{\phiz(1020)\pip} CF decays followed by $\decay{\phiz(1020)}{\Kp\Km}$.
This takes advantage of the fact that the detection asymmetry of the kaon pair is zero, as the $\phiz(1020)$ decay is self-conjugate and its decay width is small, while the same is not true for the resonant structure of the pion pair of \decay{\Dsp}{\pip\pim\pip} decays.

\begin{figure}[tb]
    \begin{center}
        \includegraphics[width=0.49\textwidth]{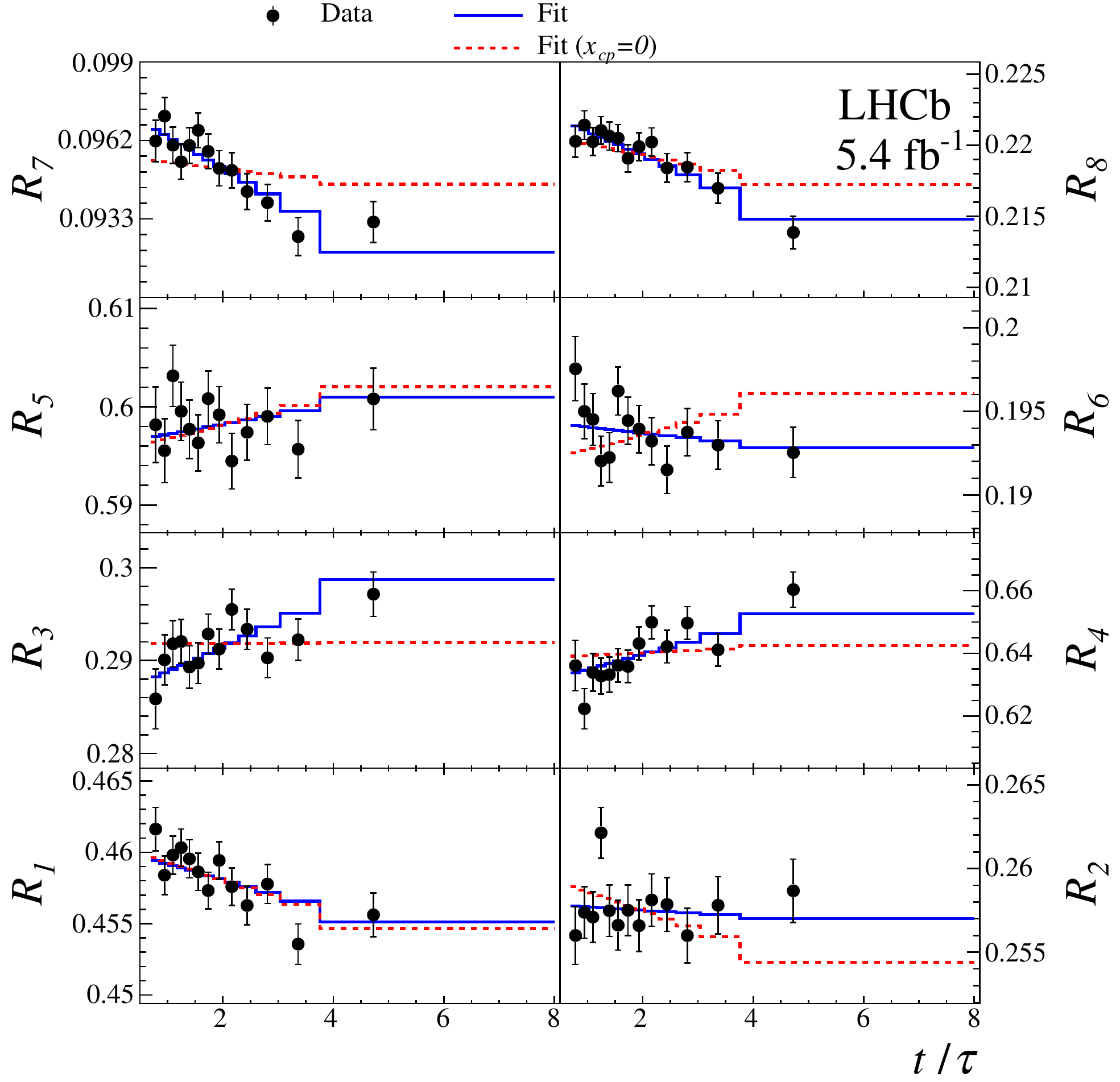}
        \includegraphics[width=0.49\textwidth]{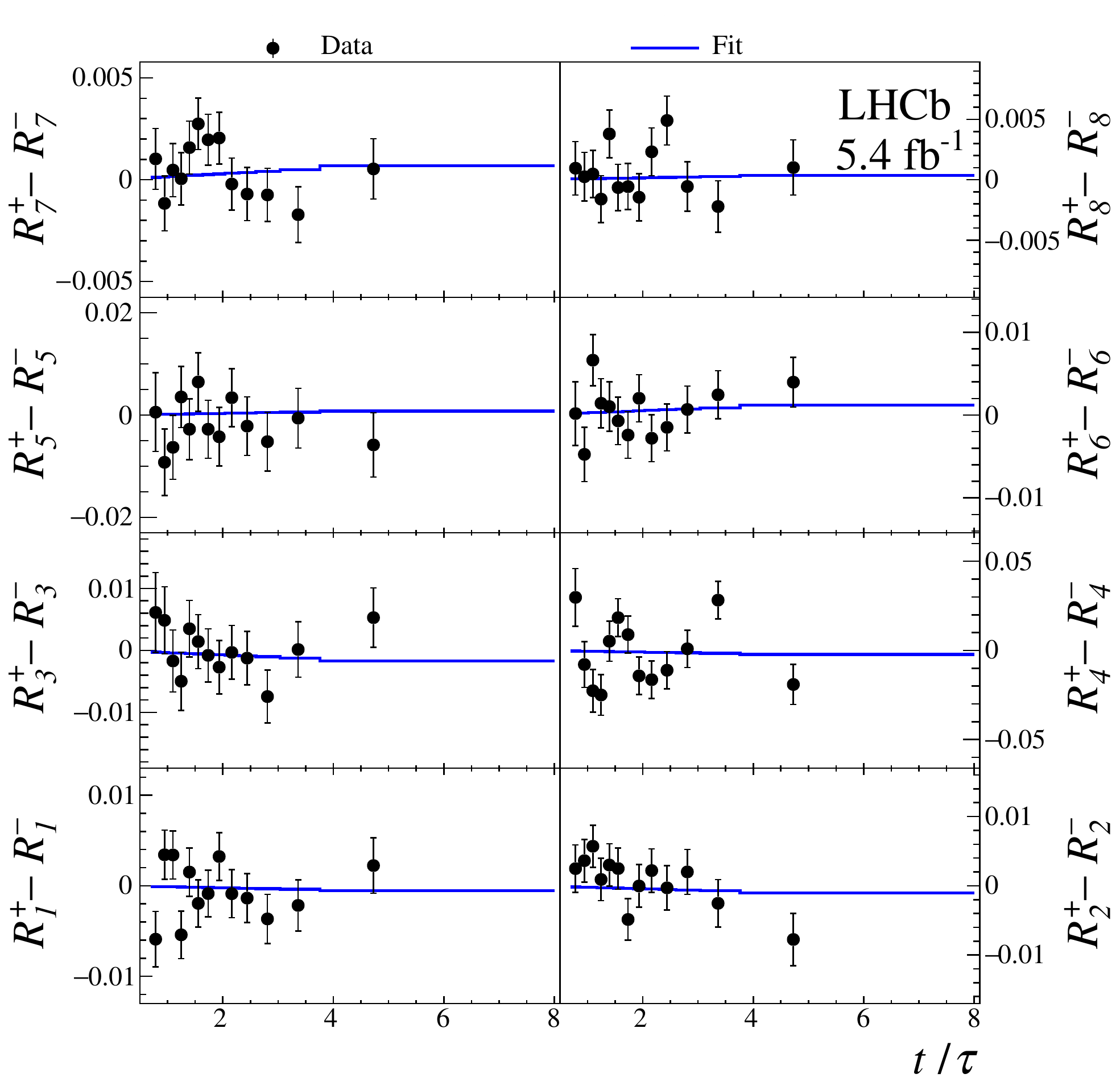}
    \end{center}
    \vspace*{-4mm}
    \caption{
        (Left) \CP-averaged yield ratios and (right) ratio differences of \Dz and \Dzb decays into the $\KS\pip\pim$ final state as a function of decay time, for each Dalitz-plot bin.
        Any deviations from constant functions would indicate the presence of (left) mixing or (right) \CP violation.
        Fit projections are overlaid.
        The dashed red line in the left plot corresponds to the projection of a fit where the parameter $x_{\CP}$ is fixed to zero.
        Figures taken from ref.~\cite{LHCb-PAPER-2021-009}.
    }
    \label{fig:binflip-fits}
\end{figure}

The projections of the fits to the corrected time-dependent ratios are shown in \cref{fig:binflip-fits}.
The results for the mixing and \CP violation parameters are, after averaging with a more recent but threefold less precise measurement based on the \runtwo \mun-tagged sample~\cite{LHCb-PAPER-2022-020},
\[
    \begin{aligned}
    x_{CP}    &= (           4.00 \pm 0.45 \pm 0.20) \times 10^{-3},
    &&\Delta x = (          -0.29 \pm 0.18 \pm 0.01) \times 10^{-3},\\
    y_{CP}    &= (           5.51 \pm 1.16 \pm 0.59) \times 10^{-3},
    &&\Delta y = (\phantom{-}0.31 \pm 0.35 \pm 0.13) \times 10^{-3},
    \end{aligned}
\]
where the statistical uncertainties include a component from the external knowledge of the strong-phase differences~\cite{CLEO:2010iul,BESIII:2020khq}, which accounts for approximately 50\% of the uncertainties of $x_{\CP}$ and $y_{\CP}$.
The systematic uncertainties on $x_{\CP}$ and $y_{\CP}$ are mainly due to neglecting the finite resolution of the Dalitz variables, to the assumption that the selection efficiency is constant across each Dalitz bin, and to the decorrelation procedure between decay time and the Dalitz coordinates.
The parameters $\Delta x$ and $\Delta y$ are compatible with zero within 1.4 standard deviations, in agreement with the hypothesis of no \CP violation.
The determinations of $x_{CP}$ and $\Delta x$ improve the precision of their world average by a factor of 3, and the former constitutes the first observation of a nonzero mass difference between the neutral charmed-meson mass eigenstates, with a significance greater than 7 standard deviations.
Furthermore, it confirms that the phase $\phi^M_2$ is approximately equal to zero rather than $\pi$, implying that the shorter-lived and nearly \CP-even eigenstate is also heavier.

\subsection{Measurement of the decay-width difference of the neutral charmed-mesons mass eigenstates with \DzKK and \DzPP decays}
\label{sect:ycp}

The \ycp{} parameter introduced in \cref{eq:z-params}, given the current constraints on the size of $\phi^\Gamma_2$~\cite{HFLAV18}, is nearly indistinguishable from the mixing parameter $y_{12}$, and provides the cleanest access to this observable.
It can be also measured with other decay channels~\cite{Belle:2009xzl,BaBar:2012bho,Belle:2015etc,LHCb-PAPER-2018-038,Belle:2019xha}; currently, the best precision is achieved using \DzKK and \DzPP decays.
Due to mixing, the time distribution of \Dz decays into these \CP eigenstates differs from an exponential and receives first-order corrections proportional to \ycp{}.
To measure this tiny deviation, it is essential to calibrate the measurement on a reference channel.
This would ideally be a flavour-specific decay such as a leptonic decay, \decay{\Dz}{\Km\ellp\neul}.
Owing to the poor mass resolution due to the missing neutrino, and to the desire to keep the selection as close as possible to that of the signal channel, \DzRS decays are usually preferred.
Here the contribution of mixing is not negligible; however, it is suppressed with respect to the decays into the \CP eigenstates, since the decay amplitude following mixing is DCS rather than CF (whereas for the \CP eigenstates both the decay amplitudes without and following mixing are CS).
The time-dependent ratio of the decay rates of the signal and calibration channels is equal to~\cite{Pajero:2021jev}
\begin{equation}
    \label{eq:ycp_from_ratio}
    \begin{aligned}
    &\frac{\Gamma(\decay{\Dz}{f,t}) + \Gamma(\decay{\Dzb}{f,t})}
         {\Gamma(\decay{\Dz}{\Km\pip,t}) + \Gamma(\decay{\Dzb}{\Kp\pim},t)}
         \approx \textnormal{const.} \times \Big\{1 - (\ycp{} - \ycp{\RS})\frac{t}{\tauDz} \\
    &\hspace{3.6cm} + \Big[\frac{1}{4}(x_{12}^2 + y_{12}^2)
    + \ycp{\RS}(\ycp{\RS} - \ycp{})\Big]\Big(\frac{t}{\tauDz}\Big)^2\Big\},
    \end{aligned}
\end{equation}
where $f$ stands for \KK or \PP, the size of \ycp{\RS} is around 6\% of that of \ycp{} (this value is approximately equal to the ratio of the magnitudes of the DCS to CF decay amplitudes) and its sign is the opposite of that of \ycp{}, and the selection efficiencies cancel to a large extent in the ratio.\footnote{The observable \ycp{} in \cref{eq:ycp_from_ratio} should actually be \ycp{f}, defined analogously to \ycp{} in \cref{eq:z-params}, but with the substitution $\phi^\Gamma_2 \to \phi^\Gamma_f \equiv \arg(\abf \Gamma_{12}/\af)$.
Since $\phi^\Gamma_f / \phi^\Gamma_2 = 1 + \order(\epsilon)$, where the parameter $\epsilon \approx 0.3$ quantifies $U$-spin breaking~\cite{Kagan:2020vri}, and the dependence of \ycp{f} on $\phi^\Gamma_f$ is at second order, this difference is negligible.}

This ratio has been recently measured by the \lhcb collaboration employing the \Dstarp-tagged data sample collected in \runtwo~\cite{LHCb-PAPER-2021-041,Pietrzyk:2803301}.
The measurement requires the time-dependent ratio of the selection efficiencies at numerator and denominator to be controlled with a precision better than the absolute uncertainty on \ycp{}, that is close to $10^{-4}$ level.
This is achieved by employing trigger lines designed to minimise the variation of the efficiency as a function of decay time, in particular by avoiding requirements on displacement-related variables~\cite{Kenzie:2110638}.
The main differences between the selection efficiencies of the signal and calibration channels are a consequence of the different masses of the final-state particles, which result in different momenta and opening angles of the two hadrons at equal \Dz momentum.
This effect is corrected for as follows: for each \DzKK decay, the momentum and direction of each kaon are recomputed under the hypothesis that the mass of the positively charged kaon be equal to that of a pion, keeping constant the decay angle in the \Dz rest frame.
The kinematics of the \DzKK decay is thus transformed into the one that it would have had if its final state had been \RS.
After this ``kinematic matching'', the same requirements on the momentum and IP of the particles are applied to both the \DzRS decay and the transformed \DzKK decay.
The requirements are tighter than the trigger requirements across the whole available phase-space, to ensure equal selection efficiencies for the two decay channels.
An analogous procedure is employed to transform the \DzRS kinematics into that of \DzPP decays.
Second order differences due to different particle identification efficiencies and to detection asymmetries that are momentum-dependent are removed by equalising the momentum distributions of the signal and calibration samples, assigning per-candidate weights.
The method is validated in simulation and by applying it also to the time-dependent ratio of \DzKK to \DzPP decays, for which the relative kinematic differences are larger.
This ratio is verified to be compatible with a constant, as expected, within an uncertainty of $0.5 \times 10^{-3}$.
Finally, the contribution of secondary mesons is explicitly accounted for, with a procedure similar to that described in \cref{sect:dy}.

\begin{figure}[tb]
    \begin{center}
        \includegraphics[width=0.48\textwidth]{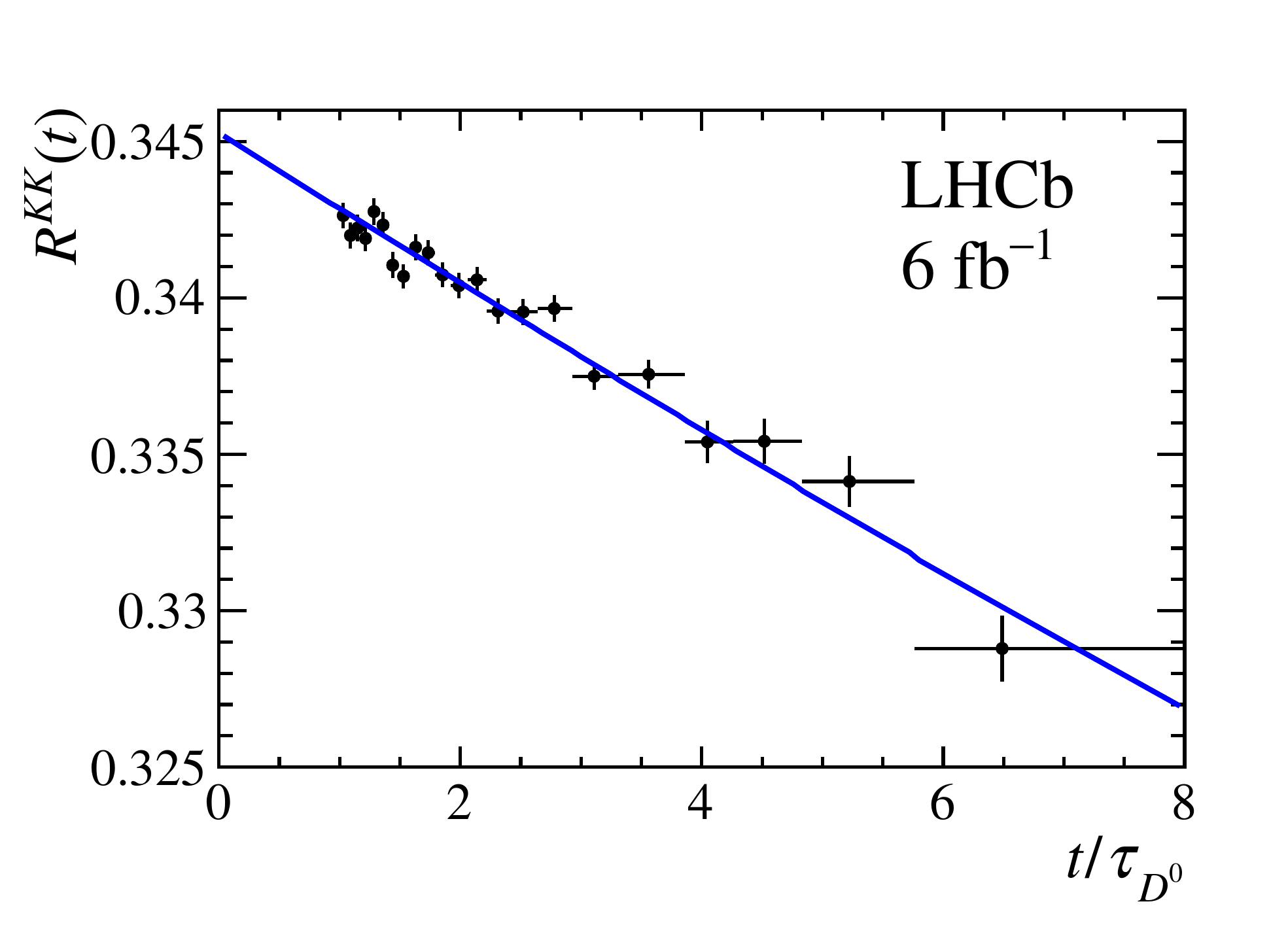}
        \includegraphics[width=0.48\textwidth]{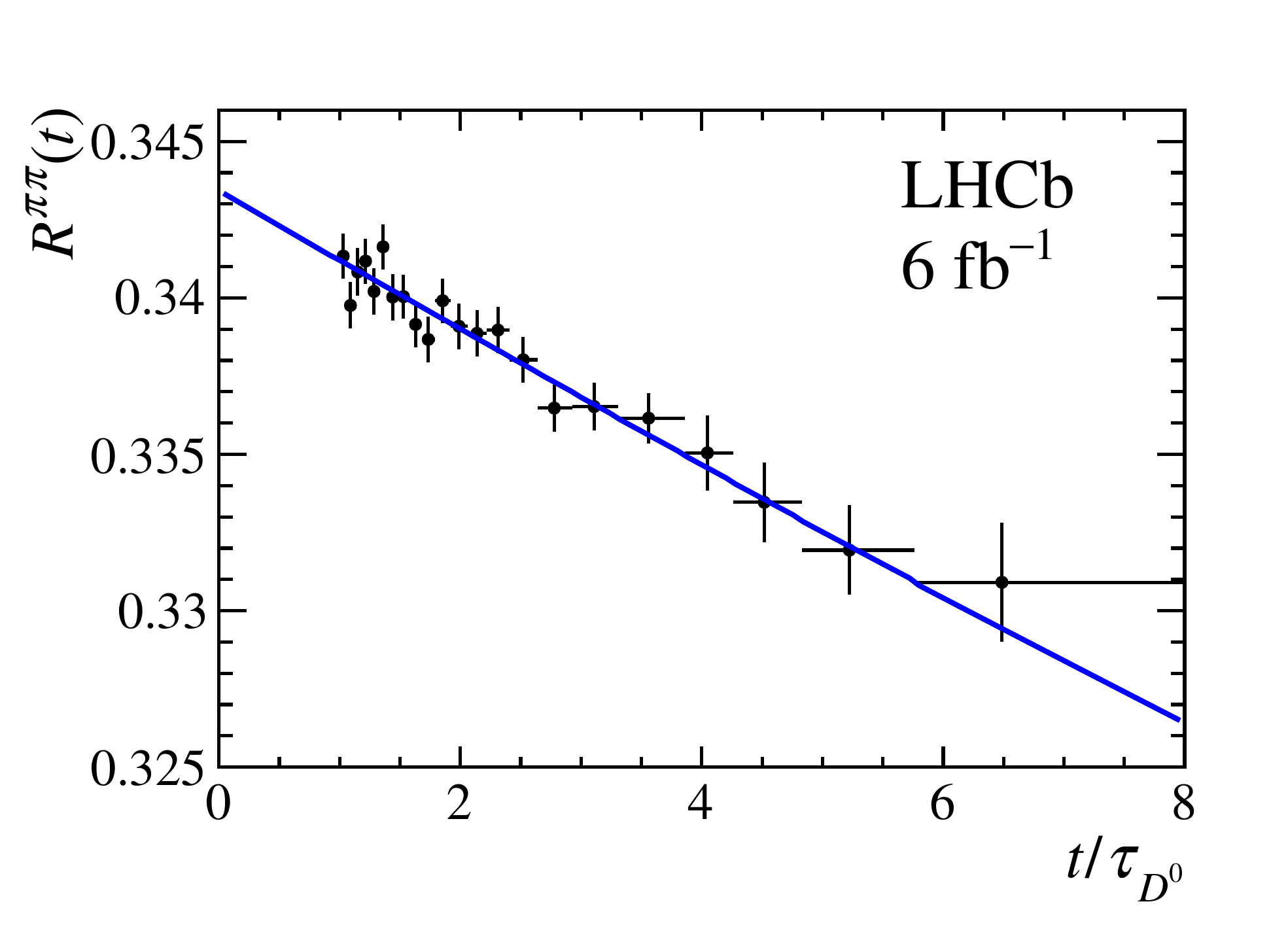}
    \end{center}
    \vspace*{-4mm}
    \caption{
        Ratio of the yields of (left) \DzKK and (right) \DzPP decays to that of \DzRS decays, after the kinematic matching and equalisation.
        An exponential fit is superimposed, where the deviation from a constant indicates the presence of nonzero $\ycp{} - \ycp{\RS}$.
        Differences between the quadratic term of \cref{eq:ycp_from_ratio} and that of the Taylor series of the exponential are accounted for among the systematic uncertainties.
        Figure taken from ref.~\cite{LHCb-PAPER-2021-041}.
    }
    \label{fig:ycp}
\end{figure}

The exponential fits to the yields ratios of the signal and calibration channels after the corrections above are shown in \cref{fig:ycp}.
The results of the two measurements are in agreement with each other.
Neglecting final-state dependent effects, their average is
\[
    \ycp{} - \ycp{\RS} = (6.96 \pm 0.26 \pm 0.13) \times 10^{-3},
\]
where the largest systematic uncertainties are due to the subtraction of the combinatorial background from random \Dz--\pip associations, and to the background of misreconstructed multibody \Dz decays.
This result is more precise than the previous world average~\cite{HFLAV18,BaBar:2012bho,Belle:2015etc,LHCb-PAPER-2018-038} by a factor of four,  
and improves significantly our knowledge of $y_{12}$, as shown in the next sections.

\subsection{Improvement in the knowledge of the mixing parameters}
\begin{figure}[bt]
    \begin{center}
        \includegraphics[width=0.48\textwidth]{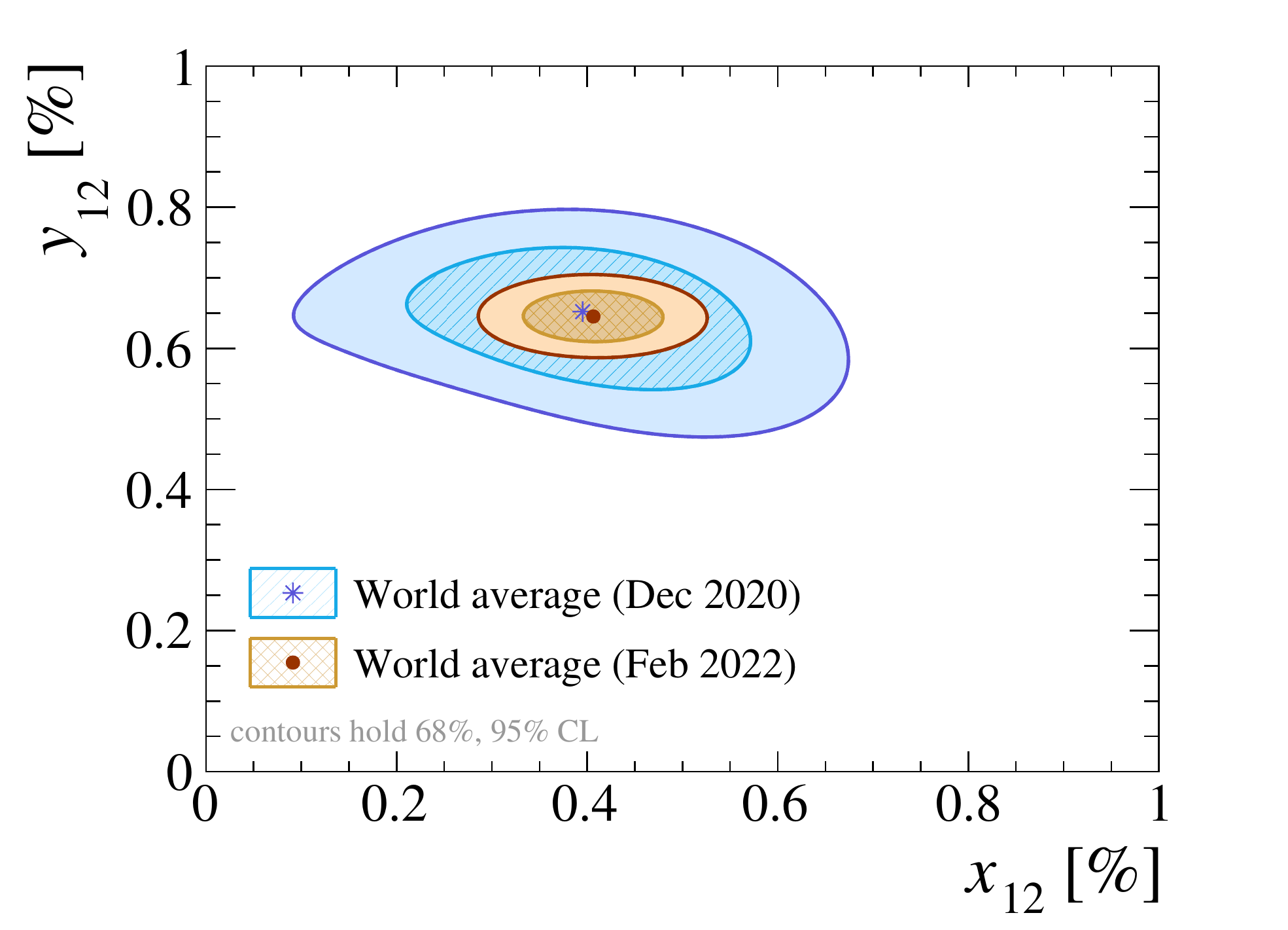}
        \includegraphics[width=0.48\textwidth]{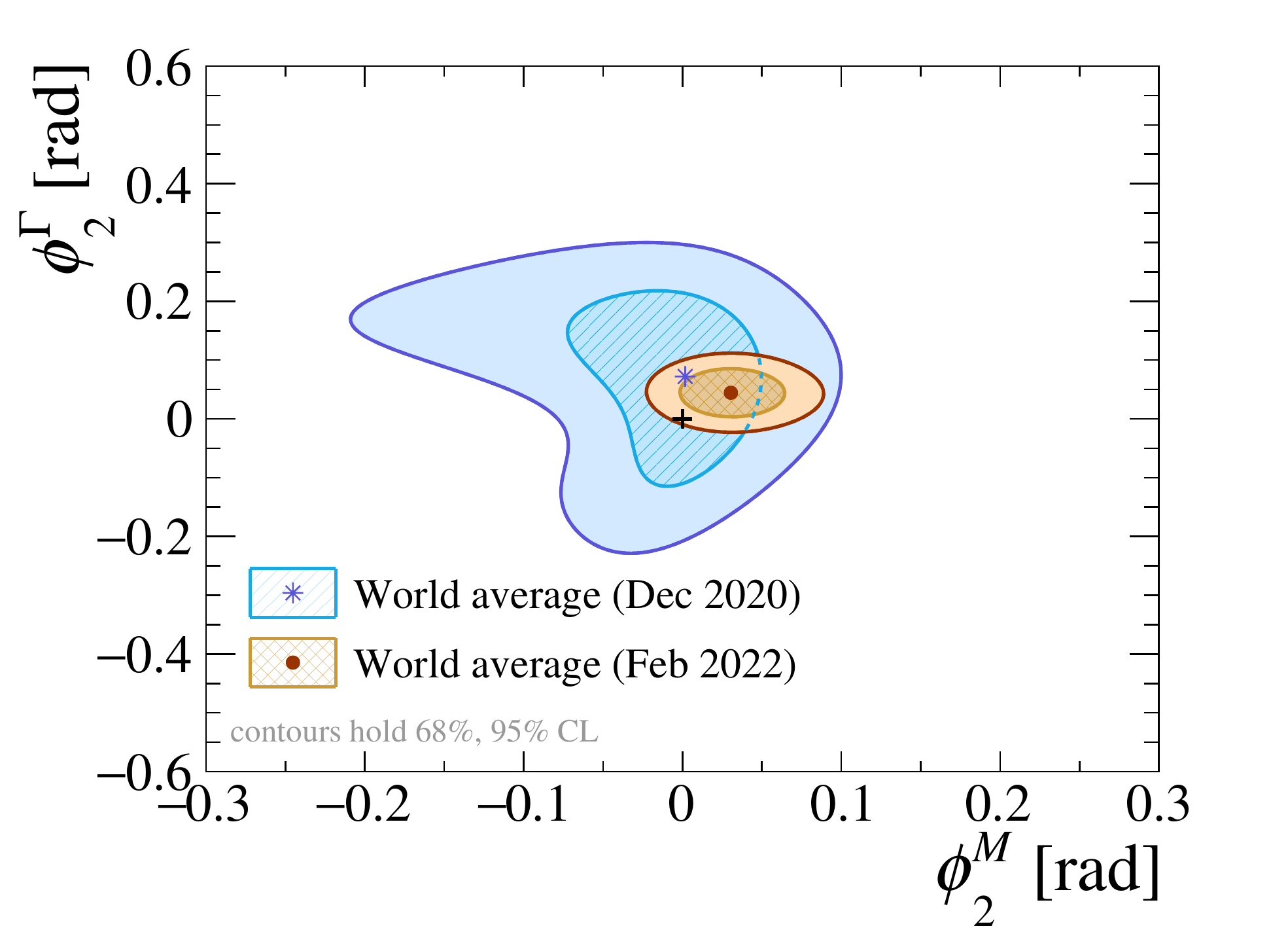}
    \end{center}
    \vspace*{-4mm}
    \caption{
        Improvement in the knowledge of the parameters of (left) mixing and (right) \CP-violation in the mixing of \Dz mesons, following the measurements presented in \cref{sect:dy,sect:kspipi,sect:ycp}.
        The black cross in the right plot corresponds to the case of no \CP violation.
        The improvements in $x_{12}$ and $\phi_2^\Gamma$ are driven by the measurement of \decay{\Dz}{\KS\pip\pim} decays, while the improvement in $y_{12}$ is driven by the measurement of \ycp{}.
        The precision on $\phi_2^M$ improves thanks to the interplay of the new measurement of \DY{} and to the improved precision on $x_{12}$ and $y_{12}$, which enhances the sensitivity to $\phi_2^M$ from measurements of both \DY{} and \decay{\Dz}{\Kp\pim} decays~\cite{LHCb-PAPER-2017-046}.
        Both plots have been produced using the public code in ref.~\cite{pajero:charm-fitter}.
    }
    \label{fig:charm}
\end{figure}

The improvement in the knowledge of the parameters of charm mixing and of \CP violation in the mixing following the measurements described in the previous sections is shown in \cref{fig:charm}.
The new world averages are
\[
    \begin{aligned}
    x_{12}          &= (4.06  \pm 0.44) \times 10^{-3},
    \hspace{9mm}\phi_2^M       = (0.031 \pm 0.021)\;{\rm rad}, \\
    y_{12}          &= (6.47  \pm 0.24) \times 10^{-3},
    \hspace{10.2mm}\phi_2^\Gamma  = (0.047 \pm 0.027)\;{\rm rad},
    \end{aligned}
\]
and improve by more than a factor of two with respect to the previous determinations.
The precision on the \CP-violation mixing phases is still around one order of magnitude larger than the SM predictions, and their best-point estimates agree with the hypothesis of no \CP violation within slightly less than two standard deviations.
Further investigations with additional decay channels and with the data from the upcoming \lhcb upgrade are needed to clarify whether this is a statistical fluctuation or not.

\subsection{First simultaneous combination of charm and beauty measurements}
\label{sect:gammacombo}
\begin{figure}[tb]
    \begin{center}
        \includegraphics[width=0.48\textwidth]{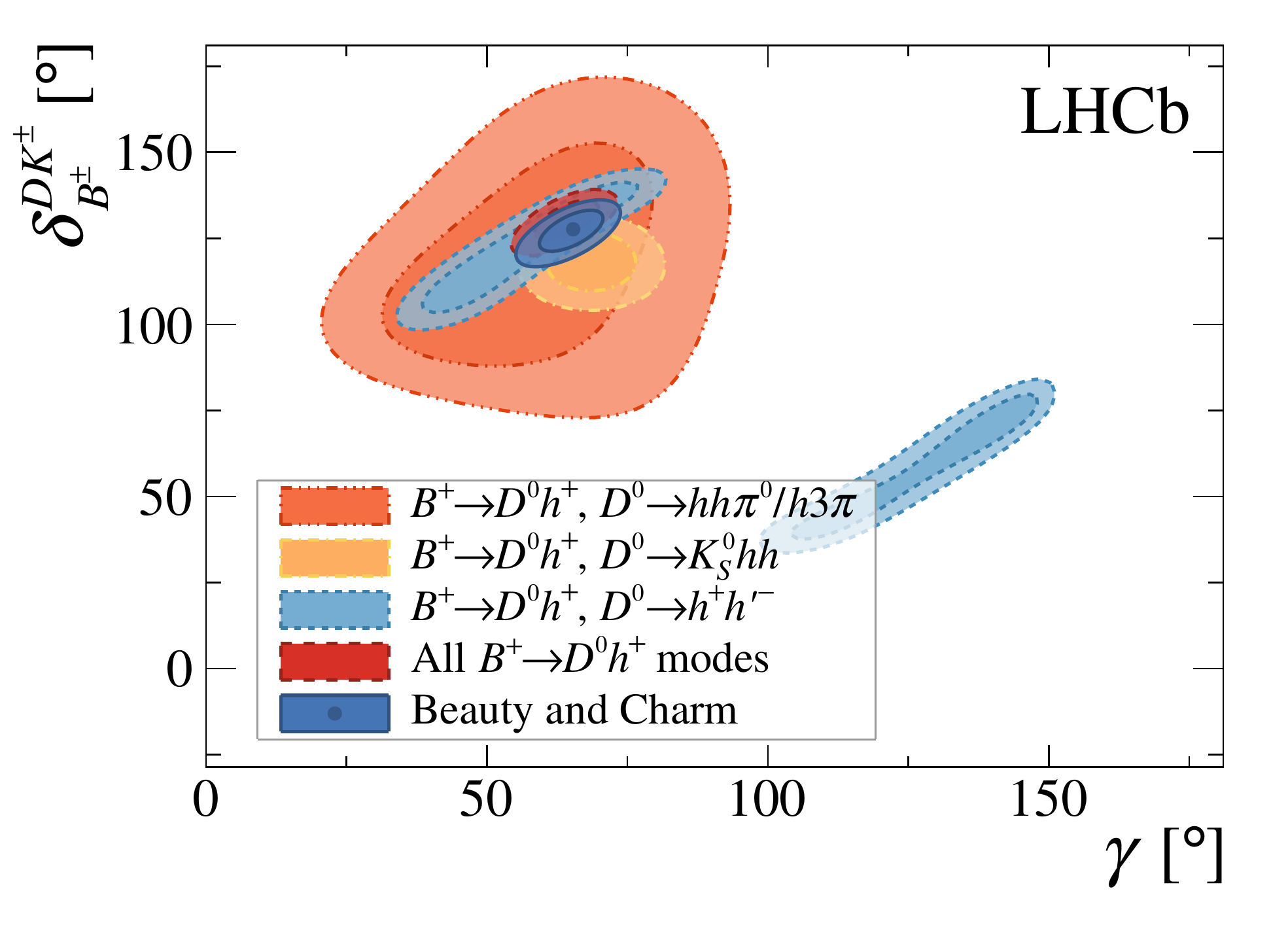}
        \includegraphics[width=0.48\textwidth]{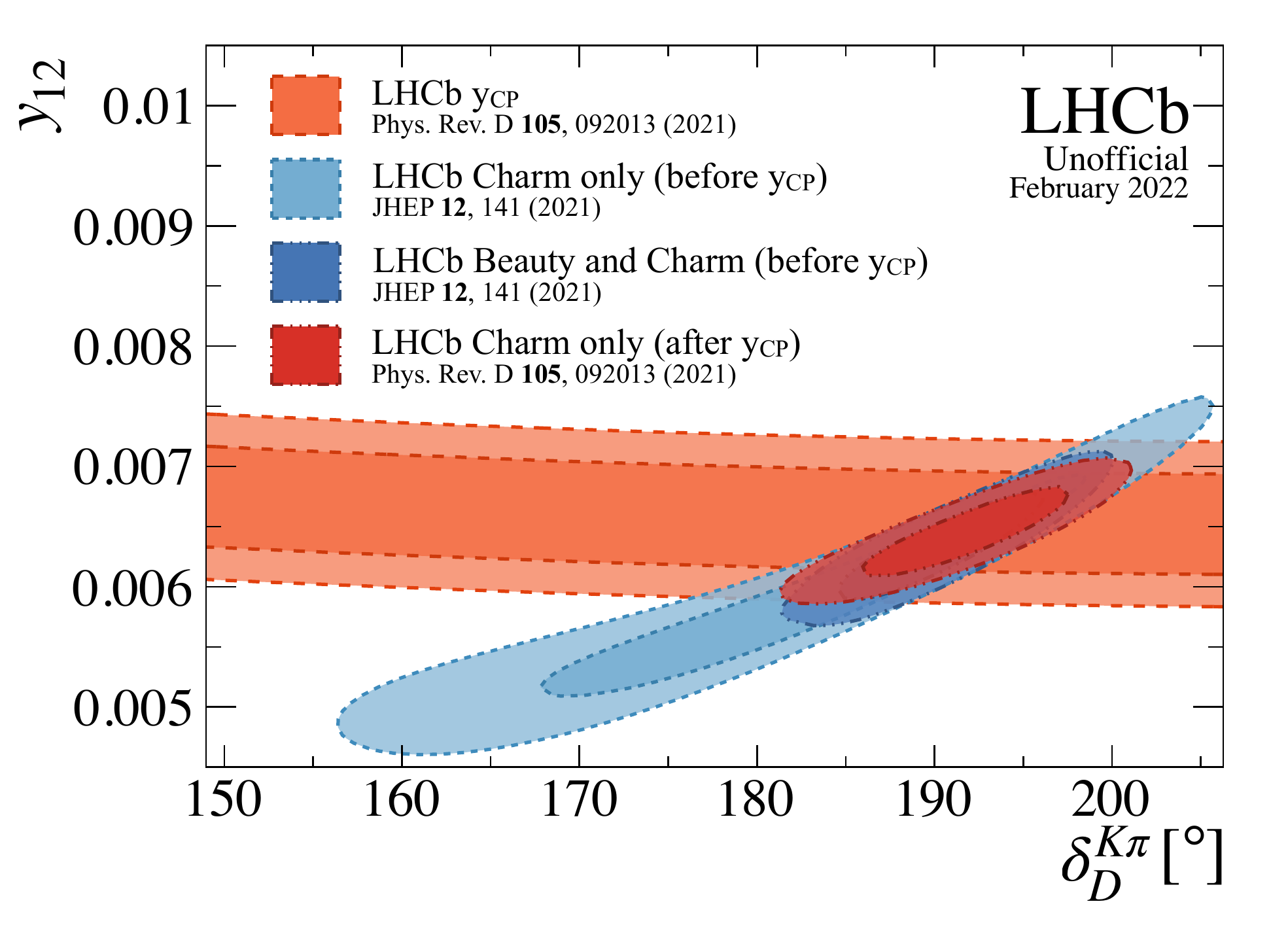}
    \end{center}
    \vspace*{-4mm}
    \caption{
        (Left) Two-dimensional confidence regions for the CKM angle $\gamma$ and for the strong-phase difference $\delta_B^{\D K}$, for different \D decay channels and for their combination.
        (Right) Two-dimensional confidence region for the mixing parameter $y_{12}$ and for the strong-phase difference \deltakpi.
        The correlation in the blue and red contours is due to the measurement of $y^\prime$ in \DzWS decays~\cite{LHCb-PAPER-2017-046}.
        Both the simultaneous combination of $\gamma$ and charm measurements, and the inclusion of the \ycp{} measurement discussed in \cref{sect:ycp}, improve the precision on $y_{12}$ and on \deltakpi by around a factor of two.
        Figures taken from refs.~\cite{LHCb-PAPER-2021-033,LHCb-PAPER-2021-041}.
    }
    \label{fig:beauty}
\end{figure}

The precision on the mixing parameters can be further improved by combining charm measurements with those of the angle $\gamma$ of the CKM unitarity triangle~\cite{Rama:2013voa,LHCb-PAPER-2020-036,LHCb-PAPER-2021-033}.
The most precise measurements of $\gamma$ to date employ \decay{\Bp}{\D\Kp} and \decay{\Bp}{\D\pip} decays, where \D stands for either of \Dz or \Dzb mesons reconstructed in a final state that is shared by the two mesons.
The final states $\KS\pip\pim$ and $\Km\pip$ are the most relevant~\cite{LHCb-PAPER-2020-019,LHCb-PAPER-2020-036}; see \cref{fig:beauty} left.
In particular, the rate of the decays where the \D meson is reconstructed in the $\Km\pip$ final states is equal to
\begin{equation}
    \label{eq:gamma}
    \begin{aligned}
    \Gamma(B^\pm \to (\Kmp\pipm)_D\, \hpm) \propto & \lvert r_D^{K\pi} e^{-i\deltakpi} + r_B^{Dh} e^{i(\delta_B^{Dh} \pm \gamma)}\rvert^2 \\
                                          = & (r^{K\pi}_D)^2 + (r^{Dh}_B)^2 + r_D^{K\pi} r_B^{Dh} \cos(\deltakpi + \delta_B^{Dh} \pm \gamma),
    \end{aligned}
\end{equation}
where $r_B^{Dh} e^{i(\delta_B^{Dh} + \gamma)}$ is the ratio of the \decay{\Bp}{\Dz\hp} to \decay{\Bp}{\Dzb\hp} decay amplitudes, $r_D^{K\pi} e^{-i\deltakpi}$ is the ratio between the \decay{\Dzb}{\Km\pip} and \decay{\Dz}{\Km\pip} decay amplitudes, and subleading effects from charm mixing~\cite{Rama:2013voa} are neglected for the sake of simplicity.\footnote{Note that a rigorous definition of the strong phases should be based on the full \decay{\Bp}{D(\decay{}{f})\hp} decay chain, as the individual ratios of amplitudes are not quark- nor meson-rephasing invariant.}
However, the measurement of these decay rates does not allow for an unambiguous determination of $\gamma$ nor of the strong phases, since the cosine function in \cref{eq:gamma} is not injective~\cite{Atwood:2000ck}; see \cref{fig:beauty} left.
By contrast, multibody \Dz decay channels such as $\decay{\D}{\KS\pip\pim}$ allow, in combination with external inputs for the charm hadronic parameters~\cite{CLEO:2010iul,BESIII:2020khq}, for an unambiguous measurement of $\gamma$.
In fact, the variation of the strong phase, $\delta_{\D}^{\KS\pip\pim}$, as a function of the phase space allows the trigonometric degeneracy to be resolved; then not only $\gamma$, but also $r_B^{Dh}$ and $\delta_B^{Dh}$, which are independent of the \D final state, can be measured~\cite{Giri:2003ty}.

The combination of the measurements of the \RS and \KS\pip\pim final states improves the precision not only for $\gamma$ and $\delta_B^{Dh}$ (see \cref{fig:beauty} left) but, through \cref{eq:gamma}, also for $\deltakpi$.
This has a beneficial effect on the knowledge of charm mixing parameters, since one of the most precise measurements of mixing is that of the observable $y^\prime \equiv -y_{12}\cos\deltakpi + x_{12}\sin\deltakpi$ in \decay{\Dz}{\WS} decays~\cite{LHCb-PAPER-2017-046},
and the uncertainty on $\deltakpi$ has limited the precision of its interpretation in terms of the mixing parameters for a long time.

A new simultaneous combination of $\gamma$ and charm measurements has recently been performed that allows the precision on \deltakpi to be improved by a factor of two~\cite{LHCb-PAPER-2021-033}.
Since the $U$-spin breaking difference of \deltakpi from $\pi$ is small, $y^\prime$ is mostly sensitive to $y_{12}$.
Therefore, the precision on $y_{12}$ is also improved by around a factor of 2.
This improvement is comparable to that allowed by the \ycp{} measurement that was published shortly after the combination, and the two results show excellent agreement; see \cref{fig:beauty} right.

The benefits of a simultaneous combination could be even larger in future studies of multibody decays such as \decay{\Dz}{\Kpm\pimp\pip\pim}, whose contribution to the determination of both the $\gamma$ angle and charm mixing and \CP violation is limited by the knowledge of the relevant \Dz hadronic parameters (strong-phase differences and amplitude ratios) as a function of phase space~\cite{BESIII:2021eud}.
The combination would indirectly improve the knowledge of these parameters, thanks to the different way in which they enter the mixing and \CP violation observables measured in the \decay{\Bp}{\D\hpm} and \Dz decays into the same phase-space regions.
In this case, the combination may improve the precision not only for the charm sector, but also for the angle $\gamma$~\cite{Harnew:2014zla,Evans:2019wza,BESIII:2021eud}.

\section{Conclusions and prospects}
\label{sect:conclusions}
The study of charm mixing and \CP violation constitutes a unique tool to test the up-type quark sector of CKM paradigm.
After many years of frustrated experimental searches, this field is eventually entering the era of precision studies.
This implies new challenges in the theoretical interpretation of the results, whose precision is limited by the understanding of strong interactions at the energy scale of the charm mass.
In particular, the compatibility of the historic first observation of \CP violation in \DzKK and \DzPP decays with the SM constitutes an open puzzle.
A parallel progress of the theoretical tools and of the experimental studies of \CP violation and rescattering is needed to shed light on this issue.
This review presented the most recent measurements of mixing and \CP violation performed with the data collected at the \lhcb experiment --- the major player of this experimental endeavour --- during its \runtwo (2015--2018).

Apart from the non-zero value of $\Delta A_{\CP}$, all other searches for \CP violation have so far yielded null results.
However, a very recent measurement of $A_{\CP}(\decay{\Dz}{\KK})$ indirectly establishes the first evidence for of \CP violation in a single decay channel, \decay{\Dz}{\PP}.
In addition, significant improvements in precision are obtained in decay channels with neutral particles such as \decay{\Dz}{\KS\KS} and \decay{\DpOrDsp}{\hzero\hp}.
Many of these measurements achieve world-leading precisions, at the per cent level or below, despite the challenges posed by their study at hadron colliders.
The potential of the \runtwo dataset is not exhausted yet, and many new measurements are expected in the next few years, especially for multibody decays.

Even larger improvements in precision have been obtained in time-dependent measurements, which determined the value of the mixing parameters $x_{12}$ and $y_{12}$ with 12\% and 4\% relative uncertainty, respectively.
The weak phases responsible for \CP violation in the mixing are still compatible with zero (though in slight tension at the level of around two standard deviations), within uncertainties of around 25\mrad.
This value is one order of magnitude larger than the SM estimates.
The prospect of using multibody decays such as \decay{\Dz}{\Kpm\pimp\pip\pim} to improve the precision on these phases and on the mixing parameters looks particularly attractive.
The improvement of the trigger between \runone and \runtwo significantly increased the yield per integrated luminosity of multibody decays; see for example ref.~\cite{LHCb-PAPER-2021-009}.
Moreover, while their analysis is complicated by the five-dimensional phase space of the final state, the interplay with the measurements of \decay{\B}{(\Kpm\pimp\pip\pim)_\D \hp} decays might indirectly improve the precision also on the angle $\gamma$ of the CKM unitarity triangle~\cite{Harnew:2014zla,Evans:2019wza}.

In the end, it is likely that larger data samples will be needed to yield the first observation of \CP violation in a single decay channel and of \CP violation in the mixing.
All current measurements are statistically limited, and no irreducible systematic uncertainties have been pinpointed yet.
Therefore, several experiments are planned to improve the statistical precision of these measurements by increasing the size of the collected samples.
In the near future, the \belletwo experiment is expected to contribute significantly to the studies of final states with neutral particles~\cite{Kou:2018nap}, whereas the upcoming \lhcb \upgradeone will yield the best precision for all other final states.
The latter plans to increase the collected integrated luminosity to $25 \invfb$ ($50\invfb$) by the end of \runthree (\runfour), scheduled to take place from 2022 to 2025 (from 2029 to 2032).
This will be obtained by increasing the instantaneous luminosity fivefold, reaching a value of $2 \times 10^{33} \cm^{-2} \sec^{-1}$, and the centre-of-mass energy to $14\tev$~\cite{LHCb-TDR-012,LHCb-TDR-016}.
Another crucial improvement of this upgrade will be the removal of the hardware trigger, whose thresholds were often tighter than those affordable in the software trigger and in the offline analysis of charm decays.
The possibility to reconstruct all collision events in the software will not only improve the collection efficiency, but also reduce the detection asymmetry due to triggering with the hadronic calorimeter.
In addition, the greater flexibility in the design of the trigger selections will be beneficial in reducing the correlations between decay time and kinematics introduced by the trigger requirements, and to keep systematic uncertainties such as those described in \cref{sect:dy,sect:kspipi} at a low level.
Finally, the introduction of new trigger lines dedicated to displaced \KS mesons in the first level of the software trigger will significantly benefit the collection efficiency of decays including these particles in their final state.
To take full advantage of the increased collected yields also in the study of multibody decays, improving the knowledge of the charm hadronic parameters of decays such as \decay{\Dz}{\KS\pip\pim} and \decay{\Dz}{\Kpm\pimp\pip\pim} will be essential.
The existing BESIII charm factory should therefore be upgraded, or new ones should be built~\cite{BESIII:2020nme,Eidelman:2015wja,Luo:2019xqt}.

In the long term, an \upgradetwo of the \lhcb experiment has been proposed to increase the integrated luminosity to about 300\invfb~\cite{LHCb-PII-EoI,LHCb-PII-Physics}.
While the experimental and computational challenges of operating a detector at an instantaneous luminosity larger by a factor of ten than that of the \upgradeone, $2 \times 10^{34} \cm^{-2} \sec^{-1}$, are formidable, this would be a unique opportunity to reduce the current statistical uncertainties by nearly one order of magnitude.
This might allow to eventually detect \CP violation in \Dz mixing, as well as to measure the \CP asymmetries of a variety of decay channels with precision ranging from below $10^{-4}$ to $10^{-3}$.
Sixty years after the November revolution~\cite{E598:1974sol,SLAC-SP-017:1974ind,Bianco:2003vb}, a complete picture of the phenomenology of charm-quark decays may thus eventually be achieved.

\appendix

\section*{Acknowledgments}
I am grateful to the editors of Modern Physics Letters A for inviting me to write this review, and to the participants to the 2022 MIAPP workshop ``Charming clues for existence'' for stimulating discussions.
I am especially indebted to Alex Gilman and to Guy Wilkinson for proofreading the manuscript and for providing valuable comments.
I thank Adam Davis for pointing out some typos in \cref{eq:gamma_uspin} and its analogue for $M_{1,0}$ in the manuscript.
This work has been supported in part by the Science and Technology Facilities Council [grant number ST/S000933/1], and has benefitted from the support of the Munich Institute for Astro- and Particle Physics (MIAPP), which is funded by the Deutsche Forschungsgemeinschaft (DFG, German Research Foundation) under Germany Excellence Strategy [EXC-2094/390783311].

\bibliographystyle{JHEP-tpajero}
\bibliography{main,standard,LHCb-PAPER,LHCb-TDR,LHCb-DP}

\end{document}